THE ROLE OF STRESS AND DIFFUSION IN STRUCTURE FORMATION IN
SEMICONDUCTORS

by

Mathieu Bouville

A dissertation submitted in partial fulfillment
of the requirements for the degree of
Doctor of Philosophy
(Materials Science and Engineering)
in The University of Michigan
2004

Doctoral Committee:

    Assistant Professor Michael Falk, Chair
    Professor Bradford G. Orr
    Assistant Professor Krishnakumar R. Garikipati
    Assistant Professor Joanna Mirecki-Millunchick

# TABLE OF CONTENTS

# LIST OF FIGURES

# LIST OF TABLES

# ABSTRACT

This dissertation addresses two aspects of the theory and simulation of stress-diffusion coupling in semiconductors. The first part is a study of the role of kinetics in the formation of pits in stressed thin films. The second part describes how atomic-scale calculations can be used to extract the thermodynamic and elastic properties of point-defects. For both aspects, there exists an interaction between phenomena at the atomic and macroscopic scales and the formation of both point-defects and surface features depends on the stress state of the system.

Recently, pit nucleation has been observed in a variety of semiconductor thin films. We present a model for pit nucleation in which the adatom concentration plays a central role in controlling the morphological development of the surface. Although pits relieve elastic energy more efficiently than islands, pit nucleation can be prevented by a high adatom concentration. Three-dimensional islands act as adatom sinks and the lower adatom density in their vicinity promotes pit nucleation. Thermodynamic considerations predict several different growth regimes in which pits may nucleate at different stages of growth depending on the growth conditions and materials system. When kinetics are taken into account, the model predicts a wide range of possible morphologies: planar films, islands alone, island nucleation followed by pit nucleation, and pits alone. The model shows good agreement with experimental observations in III-V systems given the uncertainties in quantifying experimental parameters such as the surface energy.

The same stresses which lead to the nucleation of surface features can have a significant effect on the stability of dopant profiles by altering diffusivities and by inducing chemical potential gradients. We perform empirical calculations regarding a simple model point-defect, a vacancy in the Stillinger Weber model of silicon. In the context of these calculations we devise a method to extract the strength of the elastic relaxation in the vicinity of the defect. This quantity is extracted from the leading order term which must be evaluated sufficiently far from the defect and the boundaries. It is also directly related to the formation volume, the thermodynamic quantity that couples the defect free energy to the externally applied stress.

# CHAPTER 1 — INTRODUCTION

## 1) Stress and diffusion in semiconductors

Semiconductors were the basis of a technological revolution in the 20$^{th}$ century. Electronic and optoelectronic devices are ubiquitous nowadays. They are found in computers, cell phones, lasers and solar cells. In the 21$^{st}$ century new devices are expected to play a central role in technological progress, e.g. to develop alternative energy sources or enable totally new technologies, such as nanomachines for biomedical applications. Such technologies are very demanding and the need for further miniaturization and integration of electronic and micromechanical systems necessitates very high-quality synthesis and control on the nanometer scale. The presence of defects such as dislocations and grain boundaries often results in the degradation of the electronic properties of devices [Ohring 1992]. Nanometer-scale defects can cause major reliability issues for MEMS (Micro Electro-Mechanical Systems) and NEMS (Nano Electro-Mechanical Systems). Such next-generation technology will exploit physics at a length scale where continuum predictions break down. While lithography can be used to create patterns on the micron-scale, on the nano-scale self-assembly is seen as a more suitable solution for patterning. Features spontaneously form patterns for physical reasons, related to stress or surface energy, not because patterns are directly imposed by man. Self-assembly is very promising but it requires a thorough understanding of the physics at this small scale. The work presented in this dissertation contributes to the understanding of one important aspect of the physics relevant to nanoscale features: the interaction of stress and diffusion.

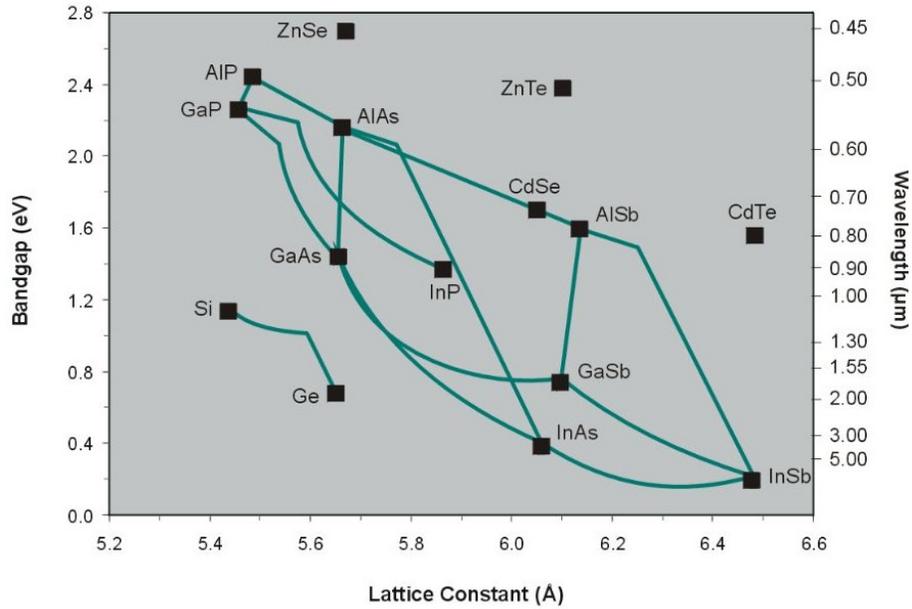

FIG. 1.1: Semiconductor band gap and lattice constant for elemental and compound semiconductors.

The primary processing technologies that enabled the technological revolution in semiconductor electronics were techniques to grow single crystal films, also called epitaxy. The word "epitaxy" comes from the Greek επι, "on", and ταξισ, "in order" [Markov 1995], meaning "ordered on". The basic principle of epitaxy — whether by molecular beam epitaxy (MBE), chemical vapor deposition (CVD) or any other technique — is to deposit atoms on a crystalline substrate. Since the substrate is a perfect crystal, the lowest energy state occurs when atoms occupy lattice sites and a crystallographically perfect flat film results.

For electron confinement, a low band gap material must be deposited on a high band gap material. Hence to produce (opto)electronic devices, different materials must be grown next to each other. Figure 1.1 shows the band gaps and lattice parameters of group IV, III-V and II-VI semiconductors. It shows that materials with different band gaps also have different lattice parameters. In this case — unless prevented by wetting issues — the film will adopt the lattice constant of the substrate rather than its natural lattice parameter, as shown in Fig. 1.2. This difference in lattice parameters is quantified by the lattice mismatch (also called misfit) defined as

$$f = \frac{a_s - a_f}{a_s} \qquad (1.1)$$

where $a_s$ and $a_f$ are the natural lattice parameters of the substrate and of the film respectively.

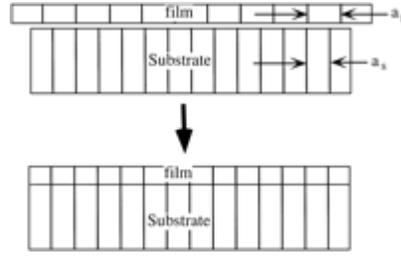

FIG. 1.2: Illustration of misfit strain (from Gao and Nix 1999)

This dissertation focuses on two important consequences of the misfit strain. The first consequence arises because the strain energy is proportional to the thickness. As growth proceeds the energy of the system increases and the crystallographically-perfect film becomes less and less stable. One possible mechanism to release elastic energy is the creation of misfit dislocations [Matthews and Blakeslee 1974, Matthews 1975]. These dislocations are located at the interface between the substrate and the film. If the film is tensile for instance, the addition of extra half planes can reduce the tension and thus the elastic energy. Strain energy can also be released by surface features such as islands [Snyder *et al.* 1991, Orr *et al.* 1992] or by the Asaro-Tiller mechanism [Asaro and Tiller 1972, Grinfel'd 1986, Srolovitz 1989]. As Fig. 1.3 shows, the strain is lower at the top of the feature, hence the elastic energy is lower there. This mechanism is detailed in the second section of this chapter. These strain relaxation mechanisms have important consequences for devices: dislocations can destroy the electronic properties of the devices [Ohring 1992] and surface features can make the surface very rough which is detrimental to electron mobility [Roblin *et al.* 1996, Kitabayashi *et al.* 1997, Lew *et al.* 1998, Yang *et al.* 1998]. On the other hand nanometer scale surface features can potentially be used as quantum dots or wires in future quantum devices, but accomplishing this will require us to understand and control the formation of nanometer-scale structures on the surface.

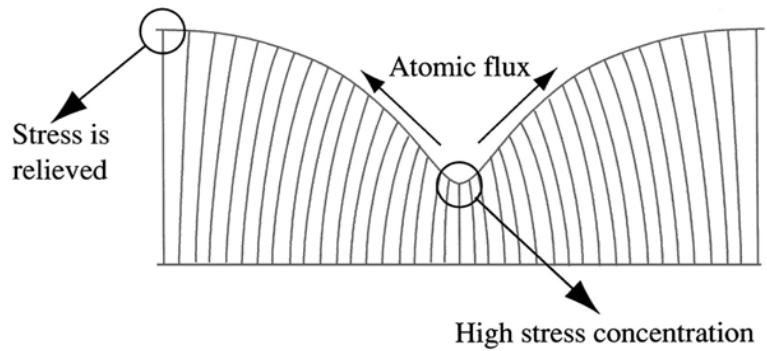

FIG. 1.3: Stress is relieved at the top of islands/ripples due to lattice relaxation (from Gao and Nix 1999)

While the first consequence of the stress described above pertains to the surface, strain is also important sub-surface. In the bulk, stress arising from the misfit or other origins can influence the energetics of the formation and migration of point defects. For instance the relative populations of vacancies and interstitials are different under tension and under compression. Also stress gradients can drive the diffusion of dopants resulting in inhomogeneities in their concentration and poor electronic properties. This is described in more detail in chapter 3.

Equally important to the stress and strain arising during epitaxy is the way the material responds to stress *via* diffusion. Diffusion is of importance in semiconductors both during growth and post-processing. During epitaxial growth, atoms are deposited on the surface at random locations. If these adatoms cannot diffuse, the growth is essentially stochastic and the end result is not governed by thermodynamics. When adatoms can diffuse they are able to reach lower energy states and growth takes place closer to equilibrium. Since, for a given materials system, surface diffusion is governed mainly by the growth temperature low-temperature films are rough with a high point defect concentration while higher-temperature films are smoother. There have been extensive studies of growth morphology as a function of the ratio of temperature to deposition flux. Figure 1.4 shows the effect of the ratio of the diffusivity to the deposition rate. If the ratio is low atoms form small diffusion-limited aggregates (DLA) close to where they landed, while for a higher diffusivity they can diffuse longer distance to form larger aggregates. At higher diffusivities and where edge diffusion is important compact islands would be expected rather than DLA clusters.

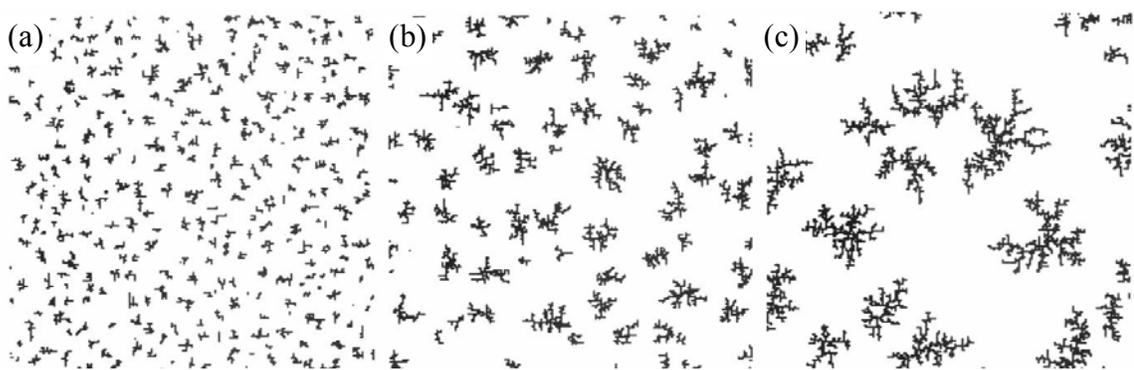

FIG 1.4: Surface morphology studied by Monte Carlo as a function of the ratio of the diffusivity, D, to the deposition rate, F, for three cases: $D/F = 10^5$ (a), $10^7$ (b) and $10^9$ (c). From Amar et al. 1994.

After the film is grown it must be doped. Controlling the dopant concentration profile requires controlling dopant diffusion. Point defects — vacancies, interstitial and substitutional atoms — play a central role in diffusion. Thus an understanding of the formation of these defects is a first step towards controlling dopant diffusion.

**2) Surface features in heteroepitaxy**

A flat crystallographically-perfect film minimizes the surface area (and hence the surface energy) and has no energy associated with defects such as dislocations. However in the presence of misfit it may have a large strain energy. Since the strain energy is proportional to the thickness, the thicker the film the higher its energy. When the film is thick enough, it is not energetically favorable that the film remain flat. Figure 1.3 shows a rippled film: at the top of the ripple, the system is less constrained and some relaxation is possible, decreasing the strain energy. This happens at the cost of surface energy.

Asaro and Tiller (1972), Grinfel'd (1986) and Srolovitz (1989) independently showed that the strain energy can lead to a surface instability with a critical wavelength controlled by the surface energy (Fig. 1.5). As an instability this process does not involve any energetic barrier. This model has been successful in predicting the periodicity of features at low misfit. Several studies have used an instability formalism to study thin film growth. Golovin and coworkers (2003) analytically studied the dynamics of island formation through a time-evolution equation for the height of the film. Island formation has also been studied by computer simulations [Liu *et al.* 2003, Zhang *et al.* 2003]. Numerical treatments have the possibility to account for anisotropy while the original papers and most analytical studies assume that the surface energy is isotropic. Eggleston and Voorhees studied the growth of self-assembled islands on patterned substrates using a phase field model [Eggleston and Voorhees 2002]. The optimal shape of isolated islands has been studied using an instability model [Shanahan and Spencer 2002].

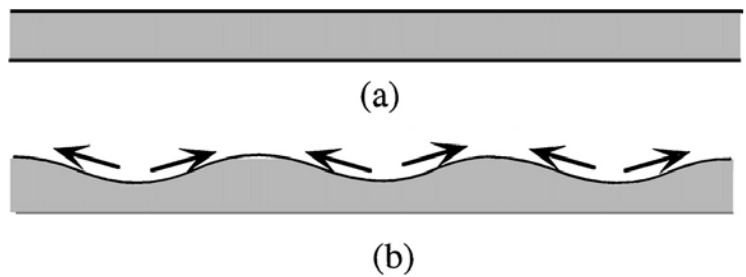

FIG. 1.5: A flat film (a) can form ripples (b) through an instability. The arrows show the direction of mass transfer. From Gao & Nix 1999.

These models do not take into account the inherent discreteness of the surface: when the instability starts growing, the amplitude is small compared to the size of the atoms, and the continuum assumption is not strictly applicable. More importantly, they predict smooth undulations on the surface, inconsistent with experimental observations in systems where pitting occurs which do not show ripple patterns in the early stages of growth. Instead they show isolated features which we interpret as indicative of nucleation (Fig. 1.6).

An alternate picture to the instability model for releasing strain energy is that of the nucleation of 3D islands. Nucleated islands and pits are distinct from features formed due to instabilities in that they are localized features which arise isolated on the surface. Islands can nucleate if fluctuations of the island size due to the motion of atoms on the surface are sufficient to overcome the energetic barrier caused by surface energy. Figure 1.6 shows islands (white dots) in a $In_{0.27}Ga_{0.73}As$ film on GaAs. Islands such as these have been seen experimentally [Guha *et al.* 1990, Eaglesham and Cerrullo 1990] and studied theoretically [Frankl and Venables 1970, Venables 1973, Venables 1984, Tersoff *et al.* 1994]. Elastic calculations [Vanderbilt and Wickham 1991] indicate that pits can also release strain energy. However pit nucleation is experimentally much less commonly observed.

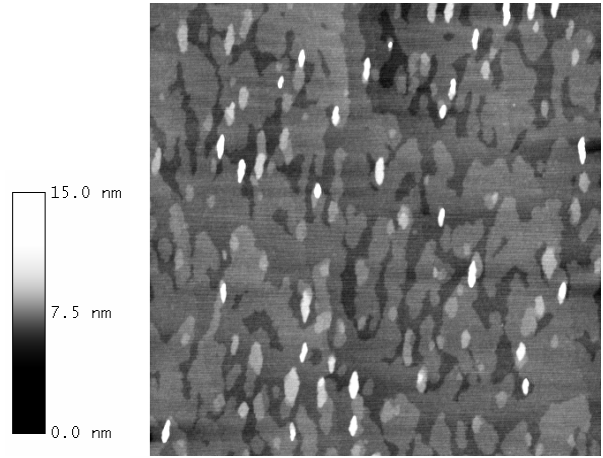

FIG. 1.6: 2 μm x 2 μm AFM image of a 25 ML thick $In_{0.27}Ga_{0.73}As$ film grown on GaAs at 490°C and an AsBEP of $10 \times 10^{-6}$ torr. Courtesy of A. Riposan.

The formation of surface features is dependent upon the quantity of adatoms which are available to form them. The effect of adatom concentration is therefore central to the work presented in chapter 2. Experimental observations [Johnson *et al.* 1997] and Monte Carlo simulations [Zhang and Orr 2003] have revealed adatom concentrations in GaAs significantly higher than was generally expected. This has been shown to be due to the role of As overpressure on the surface thermodynamics [Tersoff *et al.* 1997]. Experiments in SiGe found that, although the adatom concentration was almost uniform across the surface, small inhomogeneities could lead to localization of the nucleation of islands [Theis and Tromp 1996].

Commonly pits that form during thin film epitaxy arise due to heterogeneous nucleation [Chen *et al.* 1995, Weil *et al.* 1998, Li *et al.* 2001, Radhakrishnan *et al.*

2004]. For example, SiC or SiO$_2$ impurities have been identified as nucleation sites for pits during growth on silicon substrates. These pits are of interest to thin film growers because of the effect they can have on the film properties, but the mechanism of nucleation is sensitive to the concentration and chemistry of the contaminants. Heterogeneously nucleated pits are not considered in this work. Rather this work will focus on a number of experimental investigations in silicon-germanium on silicon and in III-V semiconductors that revealed that pits can arise in the absence of contaminants (Table 1.1). In these cases pit nucleation appears to be homogeneous. The mismatch in these systems ranges from 1.2 % [Gray *et al. 2001,* 2002, Vandervelde *et al.* 2003] to 4 % [Jesson *et al.* 2000] to 7 % [Seshadri and Millunchick 2000] and growth conditions vary greatly; for example temperatures range from 300°C to 600°C.

| reference | system | misfit | T | islands and pits? |
|---|---|---|---|---|
| Jesson 1996 | Si$_{0.5}$Ge$_{0.5}$/Si | 2% | "low"* | islands and pits form ripples |
| Jesson 2000 | Ge/Si | 4% | 300°C | pits far from islands |
| Gray 2001, 2002; Vandervelde 2003 | Si$_{0.7}$Ge$_{0.3}$ /Si | 1.2% | 550°C | pits without islands |
| Chokshi 2002 ; Riposan 2003 | In$_{0.27}$Ga$_{0.73}$As/GaAs | 1.9% | 500°C | pits close to islands |
| Lacombe 1999 | In$_{0.8}$Ga$_{0.2}$As/InP | 1.8% | 540°C | pits without islands |
| Seshadri 2000 | InSb/InAs | 7% | 400°C | pits far from islands |

Table 1.1: Observations of pits in semiconductors. *: the samples were annealed 5 minutes at 590°C

Elastic calculations indicate that pits can release strain energy and can therefore participate to the strain relaxation process [Vanderbilt and Wickham 1991]. But, in the systems studied here, pits are generally seen only after islands nucleated and close to islands. Gray *et al.* have observed pits in the absence of islands in $Si_{0.7}Ge_{0.3}$ films grown on Si at 550°C, as can be seen in Fig. 1.7(a). Jesson *et al.* have observed pits appearing simultaneously with islands in $Si_{0.5}Ge_{0.5}$ films grown on Si and annealed at 570°C for 5 min, as shown in Fig. 1.7(b).

**3) Defects in silicon**

This dissertation will also explore some studies related to point defects in semiconductors. Point defects — vacancies, interstitial and substitutional atoms — play an important role in dopant diffusion in semiconductors. The formation and migration of defects are stress-dependent. The stress field resulting from the contraction of the crystal around a vacancy interacts with the stress fields of other vacancies and with the externally applied stress. If the stress is not uniform (due to manufacturing for instance), defects are more likely to form in regions of high/low stress or to migrate to these regions. Whether they are attracted to high or low stress regions depends on their formation volume which will be defined and discussed in chapter 3.

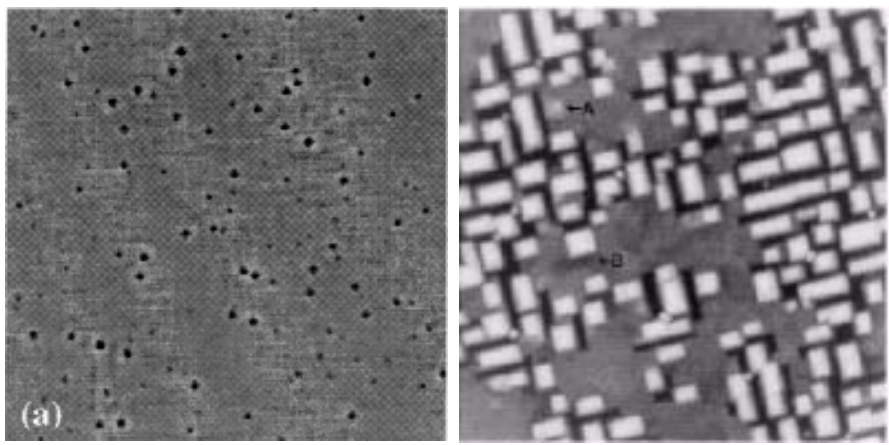

FIG. 1.7: 2 µm x 2 µm AFM image of (a) a 15 nm thick $Si_{0.7}Ge_{0.3}$ film grown on Si at 550°C [Gray et al. 2001], (b) a 5 nm thick $Si_{0.5}Ge_{0.5}$ film grown on Si and annealed at 570°C for 5 min [Jesson et al. 1996] (islands (A) and pits (B) nucleate next to each other.)

Semiconductors have to be doped to be of technological use and the dopant profile must be controlled. In SiGe wafers, for instance, the widely disparate diffusive behaviors of boron and arsenic result in very complex process steps to create the desired dopant profiles. Exploiting stress-mediated diffusion in SiGe will hopefully lead to simpler process steps, more reliable profiles and higher yield. Another consequence of stress inhomogeneities is that they can result in inhomogeneities of the formation energy of vacancies. Vacancies can cluster and form voids or they may favor the initiation of cracks. This is an important issue for the reliability of devices. What is true of dopants can also be applied to contaminants. Metallic impurities are considered a major issue in silicon wafers [*International technology roadmap for semiconductors* 2002]. Transition metals (mostly Fe, Cu and Ni) generally reside on interstitial sites in silicon [Graff 2000]. The fabrication and processing of Si wafers can result in metal concentrations up to $10^{13}$ atoms/cm$^3$. Since they can diffuse fairly fast, even at low temperature, these impurities can gather and form metal silicides since their solubility in Si is very low at room temperature. As the solubility of these contaminants depends on stress the location of the precipitation can be stress dependent if the stress field is inhomogeneous. Precipitation near contacts, for example, can have consequences for the reliability of devices.

The stress field in the crystal arises due to external stress plus the superposition of the individual stress fields due to the defects. The resulting stress gradients can in turn drive the diffusion of the defects leading to two-way coupling of stress and defect populations [Garikipati *et al.* 2001, Garikipati *et al.* 2004]. Although the analytical form of the stress field around a center of contraction can be calculated by linear elasticity, atomic-scale information is needed to calculate the strength of the contraction/dilation.

To predict diffusion it is important to connect the diffusive process to atomic mechanisms. Figure 1.8 shows the atomistic mechanism for vacancy-assisted diffusion. The intermediate step is the saddle point: it is the highest free energy state along the minimum energy diffusion path. The difference in free energy between the intermediate step and the initial step is called the migration energy. In transition state theory, this is the free energy fluctuation necessary for the vacancy to migrate. The excess free energy of a crystal with a vacancy compared to a perfect crystal is called the formation energy of the vacancy. Diffusion depends on both energies since diffusion requires that there be a vacancy (formation) and that the vacancy migrate (migration). Therefore the activation energy for diffusion is the sum of the formation energy and the migration energy.

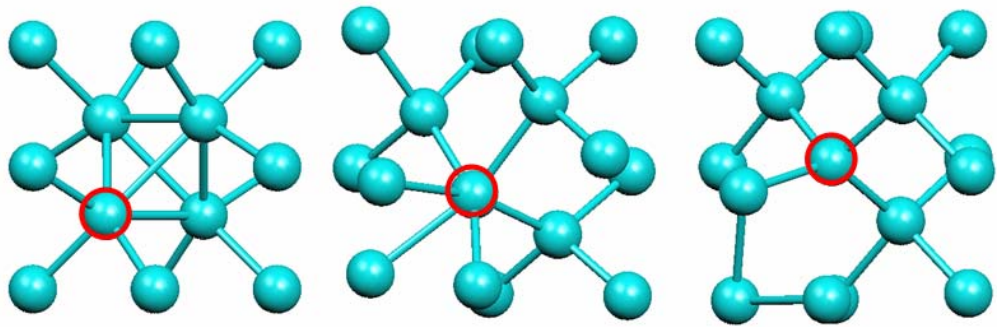

FIG. 1.8: Vacancy-assisted diffusion. The circled atom exchanges place with the vacancy.

The energy of formation of dopants is dependent upon the local stress state. Stress gradients can therefore drive their diffusion. As the formation energy of point defects is stress-dependent, the number of vacancies changes with stress. Since vacancies participate in diffusion, the mobility of dopants can be affected by stress through a change in vacancy concentration.

The diffusivity can be expressed as a function of $G^*$, the Gibbs free energy of activation [Vineyard 1957, Aziz 2001]

$$D = \frac{f \nu d^2}{6} \exp\left(-\frac{G^*}{kT}\right) \tag{1.2}$$

where $\nu$ is an attempt frequency, d is the hop distance, f is a correlation factor, k is Boltzmann's constant, T is the temperature and the factor of 6 arises from the dimensionality of the system. As discussed above, $G^*$ can be split into two contributions: a free energy to create the defect, $G^f$, and a migration free energy, $G^m$, such that

$$G^* = G^f + G^m. \tag{1.3}$$

These three energies depend upon stress through three quantities with units of volume: an activation volume, $\mathbf{V}^*$, a formation volume, $\mathbf{V}^f$, and a migration volume, $\mathbf{V}^m$,

$$V_{ij}^f = -\left(\frac{\partial G^f}{\partial \sigma_{ij}}\right), \quad V_{ij}^m = -\left(\frac{\partial G^m}{\partial \sigma_{ij}}\right), \quad V_{ij}^* = V_{ij}^f + V_{ij}^m. \tag{1.4}$$

The volumes are in boldface because they are tensors. If the defect is isotropic and the stress field is hydrostatic, these tensors can be expected to be proportional to identity, i.e. they can be treated as scalars. A di-vacancy, for example, has an orientation and is therefore intrinsically anisotropic; in this case, the activation, formation and migration volumes must be considered as tensors.

Equation (1.2) is sufficient for instances in which a single atomic mechanism dominates diffusion and acts isotropically in the crystal. However, if multiple atomic mechanisms operate, if diffusion is anisotropic, and in instances where externally applied stress breaks crystalline symmetry, more complex descriptions are required to incorporate the anisotropy of the atomic level processes. An early theory by Dederichs and Schroeder (1978) elucidated this relationship for simple defects and a more detailed theory has been worked out recently by Daw and coworkers (2001). Using this formalism it is possible, at least in principle, to extract the isotropic or anisotropic diffusivity and the dependence of the diffusivity on stress by a careful enumeration and calculation of energetic barriers associated with formation and migration of defects.

Due to its technological importance, dopant diffusion in silicon has received most of the attention of experimentalists. The most studied dopants in silicon are boron (B), phosphorous (P) and antimony (Sb). Two mechanisms exist for the diffusion of these atoms: Sb diffuses by a vacancy mechanism while B and P use an interstitial mechanism [Fahey *et al.* 1989, Ahn *et al.* 1988, Osada *et al.* 1995, Zaitsu *et al.* 1998, Chaudhary and Law 1997, Shimizu *et al.* 1998, Rao and Wosik 1996]. Figure 1.9 sketches vacancy (top) and interstitial (bottom) formation and migration.

At the time of this writing there are only two cases in which the trace of the activation volume tensor has been measured experimentally, for the diffusion in Si of B [Zhao *et al.* 1999A] and of Sb [Zhao *et al.* 1999B]. The diffusivities were measured at constant temperature (810ºC for B and 860ºC for Sb) and various hydrostatic pressures and the activation volume was extracted. In the case of boron the activation volume is negative and it is positive in the case of antimony.

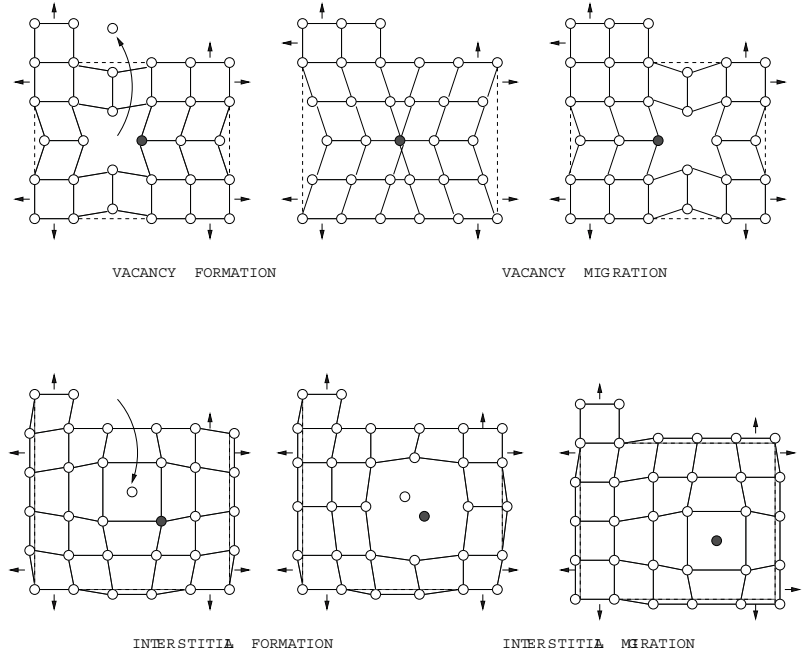

FIG. 1.9: Lattice distortion due to vacancy (top) and interstitial (bottom) formation and migration. From Aziz 2001.

The tensile/compressive biaxial stress in strained Si and $Si_{1-x}Ge_x$ has been used to study the effect of stress on dopant diffusivity as a function of mechanism by exploiting the fact that P and B diffuse primarily by an interstitial mechanism and Sb by a vacancy mechanism in both Si and $Si_{1-x}Ge_x$ [Kringhøj *et al.* 1996, Larsen and Kringhøj 1997, Kuznetsov *et al.* 1999, Fedina *et al.* 2000, Kuznetsov *et al.* 2001]. Kringhøj and coworkers studied the diffusivity of Sb in tensile Si compared to unstrained Si and in compressive $Si_{0.91}Ge_{0.09}$ compared to unstrained $Si_{0.91}Ge_{0.09}$. In both materials systems they found that the diffusivity is higher at higher stress. Using the relationship between the activation volume under hydrostatic and biaxial loadings [Aziz 1997], the results by *Zhao et al.* are in agreement with those of Kringhøj *et al.* These results are in qualitative agreement with retardation of B diffusion by compressive biaxial stress and enhancement by both compressive hydrostatic stress and tensile biaxial stress, and retardation of Sb diffusion by tensile biaxial stress and compressive hydrostatic stress [Kuo *et al.* 1995, Moriya *et al.* 1993, Cowern *et al.* 1994].

It is also possible to experimentally measure the formation volume instead of the activation volume. Simmons and Balluffi have measured the concentration of vacancies in aluminum [Simmons and Balluffi 1960]. They measured simultaneously and independently the change in length, $\Delta L$, and the change in lattice parameter, $\Delta a$, of the samples as a function of temperature, from 229ºC to 415ºC. If the vacancy concentration is constant, $\Delta L/L = \Delta a/a$. A deviation from this equality is indicative of a change in vacancy concentration. Performing this kind of experiment for various pressures would give a measurement of the formation volume of vacancies. However this has not been done yet due to the difficulty in obtaining precise enough concentrations for the pressure dependence to overcome the noise.

Since experimental studies like those described above are difficult there has been significant interest in using simulations to calculate $\mathbf{V}^f$ and $\mathbf{V}^m$. Formation energies have been computed by *ab initio* and tight-binding simulations [Sugino and Oshiyama 1992, Antonelli and Bernholc 1989, Tang *et al.* 1997, Antonelli *et al.* 1998, Puska *et al.* 1998, Zywietz *et al.* 1998, Lenosky *et al.* 1997, Zhu *et al.* 1996]. The purpose of these studies has been to calculate the formation energy of neutral vacancies. Little work has been concerned with the stress dependence of the energy. Antonelli and coworkers (1998) found two possible geometries for the neutral Si vacancy with similar formation energies but different formation volumes. An apparent consequence of this result is that the concentration of vacancies should increase both under tension and compression, the concentration would be at its lowest in a stress-free crystal. This lack of monotonicity is not intuitive but it cannot be discarded since no experiment has been carried with enough accuracy to measure this subtle change. Table 1.2 presents the results of several computational investigations along with an experimental result. It shows that results vary greatly and that there is no agreement upon the values of the activation, formation and migration volumes.

| technique | reference | tr($V^*$) | tr($V^f$) | tr($V^m$) |
|---|---|---|---|---|
| *ab initio* | Sugino and Oshiyama 1992 | 0.06 Ω | 0.468 Ω | 0.534 Ω |
| *ab initio* | Antonelli and Bernholc 1989 | | 0.75 Ω | - 0.68 Ω |
| MD/TB | Tang *et al.* 1997 | | 0.03 Ω | 0.04 Ω |
| *ab initio* | Antonelli *et al.* 1998 | | - 0.09 Ω | 0.16 Ω |
| experiment | Zhao *et al.* 1999 | 0.07 Ω | | |

Table 1.2: Theoretical and experimental results for activation, formation and migration volumes. Ω is the atomic volume.

One issue with such numerical simulations of defects is the system size. Quantum mechanics calculations provide good accuracy but they are limited to small systems, typically smaller than 256 atoms per super cell. Probert and Payne (2003) studied their convergence and found that the speed of the convergence depends on the organization of the supercells — they can be on a simple cubic lattice, face-centered cubic or body-centered cubic. Mercer and coworkers (1998) showed that the volume converges more slowly with supercell size than the energy. This is an indication that elastic interactions have a strong influence upon the formation volume. Because periodic boundaries are generally used in atomistic simulations the vacancy interacts with its periodic images. Thus a simulation with few atoms is equivalent to a high concentration of defects and the properties obtained are not those of an isolated vacancy but of a dense network of vacancies. *Ab initio* calculations for instance compute the formation energy and the structure of the vacancy under a stress state partly due to the periodic images of the defect, i.e. the stress state is in part an artifact coming from the small system size. Since the elastic interaction between a point defect and its periodic images is long-ranged, even the largest of these *ab initio* calculations is too small. One goal of this dissertation is to present new methodologies using continuum mechanics to handle the long-ranged elastic interactions.

The marriage of the atomic and macroscopic simulation and theory can potentially result in new numerical tools for the design of production processes of semiconductors. This will make it possible to use results on the atomic scale to provide data to link stresses calculated *via* continuum elasticity to changes in material composition calculated through continuum diffusion. In the long-term, this will enable the accurate simulation of stress-induced changes in semiconductor growth and failure processes.

# CHAPTER 2 — PIT NUCLEATION IN HETEROEPITAXIAL GROWTH

## 1) Introduction

Morphological features such as three-dimensional islands and strain-induced ripples [Guha *et al.* 1990, Cullis *et al.* 1992] arise spontaneously during the heteroepitaxial growth of III-V compound semiconductor alloys. While models exist that can explain nucleation of 3D islands or development of surface instabilities during growth [Frankl 1970, Venables 1973, Venables *et al.* 1984, Tersoff *et al.* 1994], the spontaneous formation of more complex morphologies is a matter of current investigation. Better prediction and control of such structures could improve the performance of devices and provide processing routes for the manufacture of new devices that exploit the unique physics of features near the atomic scale. It has been shown, for instance, that pits may act as nucleation sites for quantum dots resulting in quantum dots (QD) distributions and densities suitable for cellular automata applications [Gray *et al.* 2001, 2002, Vandervelde *et al.* 2003].

Previous work [Chokshi *et al.* 2000, 2002, Riposan *et al.* 2002, 2003] revealed the onset of pits subsequent to the nucleation of 3D islands during the heteroepitaxial growth of InGaAs on GaAs. Pits have also been observed to play an important role in morphological development in silicon-germanium on silicon [Jesson *et al.* 1996, 2000, Gray *et al.* 2002, Vandervelde *et al.* 2003], InGaAs on InP [Lacombe 1999, Lacombe *et al.* 1999] and InSb on InAs [Seshadri *et al.* 2000]. In these systems, the process of island and pit nucleation leads to surface patterning in which features on the surface are correlated to approximately 150 nm. It has been shown that pits may act as nucleation sites for islands [Gray *et al.* 2001, 2002, Vandervelde *et al.* 2003, Songmuang *et al.*

2003], resulting in island distributions and densities suitable for application such as cellular automata [Lent *et al.* 1993].

These experimental observations form the basis of the theoretical analysis presented here, in which a secondary feature, in this case a pit, nucleates on a surface upon which primary features, 3D islands, have already nucleated. A nucleation model may be more appropriate for this material system than a linear instability model because the initial 3D island features are observed to emerge isolated on an otherwise nearly flat substrate (Fig. 1.6). However it is, in part, this conjecture that will be evaluated by providing a detailed account of the logical consequences of such an assumption.

To that end this chapter details a model in which pit nucleation in thin films is considered to arise from a near-equilibrium nucleation process. Central to this model is the assumption that the adatom concentration plays a critical role in controlling the morphological development of the surface. One consequence of this is that although pits relieve elastic energy more efficiently than islands, pit nucleation can be prevented by a high adatom concentration. However three-dimensional islands act as adatom sinks and the lower adatom density in their vicinity promotes pit nucleation. Thermodynamic considerations predict several different growth regimes in which pits may nucleate at different stages of growth depending on the growth conditions and materials system. However direct comparisons to experimental observations require that kinetics be taken into account as well. The model predicts a wide range of possible morphologies: planar films, islands alone, islands nucleation followed by pit nucleation, and pits alone. The model shows good agreement with experimental observations in III-V systems given the uncertainties in quantifying experimental parameters such as the surface energy.

The central question addressed here is why and how pits nucleate only at later stages of growth and near islands. A stress concentration near three-dimensional islands is more often considered as the origin for this effect. This has been studied using the finite element method [Benabbas *et al.* 1999, Meixner *et al.* 2001], but no analytical form of the strain in the film exists, which makes the integration of these results into the adatom-based model difficult. The effect of adatom concentrations has not been addressed in detail and provides the subject of this analysis. Therefore in what follows a supposition is made of a uniform strain across the surface. As will be discussed in section 2 one of the motivations for focusing on the role of adatom concentration is that direct experimental observations of adatoms on GaAs surfaces have revealed adatom concentrations significantly higher than was generally expected [Johnson *et al.* 1997]. Monte Carlo simulations also found high Ga densities in GaAs [Zhang and Orr 2003]. It has been theorized that this occurs due to the role of As overpressure on the surface thermodynamics [Tersoff *et al.* 1997]. Other experiments in SiGe found the adatom concentration to be almost uniform across the surface, yet small inhomogeneities in the adatom concentration were observed to lead to localization of the nucleation of islands [Theis and Tromp 1996]. Using a nucleation model, we show that under a variety of conditions such small inhomogeneities could account for pit nucleation, particularly in the presence of 3D islands.

**2) The energetics of island and pit nucleation**

*a) Analogy with classical nucleation theory*

Classical nucleation theory was originally devised to describe solidification and other first order phase transitions. An analogy is often drawn between nucleation during a first order phase transition and the nucleation of islands and pits. When a cluster (or island or pit) is large enough, i.e. when finite size effects can be neglected, its energy can be expressed as the energy released per unit volume plus an energetic cost associated with the surface energy [Volmer and Weber 1925, Volmer 1929, Becker and Döring 1935]:

$$E_n = -\mathcal{E} n + \gamma n^{2/3} \qquad (2.1)$$

where $\mathcal{E}$ is an energy released per volume, $\gamma$ is a surface energy, and n is the number of atoms in the cluster (or the number of cation-anion pairs in the case of compound semiconductors). n is proportional to the volume and $n^{2/3}$ to the surface area (Fig. 2.1). For the sake of simplicity (to avoid carrying atomic volume factors), $\mathcal{E}$ and $\gamma$ have units of energy. $\mathcal{E}$ and $\gamma$ depend on the shape of the feature. Even though the functional form is the same for islands and for pits, the values of $\mathcal{E}^{isl}$ and $\mathcal{E}^{pit}$ and the values of $\gamma^{isl}$ and $\gamma^{pit}$ may differ due to differences in geometry between islands and pits. In general $\mathcal{E}$ is a function of the strain, and near islands the strain is enhanced [Benabbas *et al.* 1999,

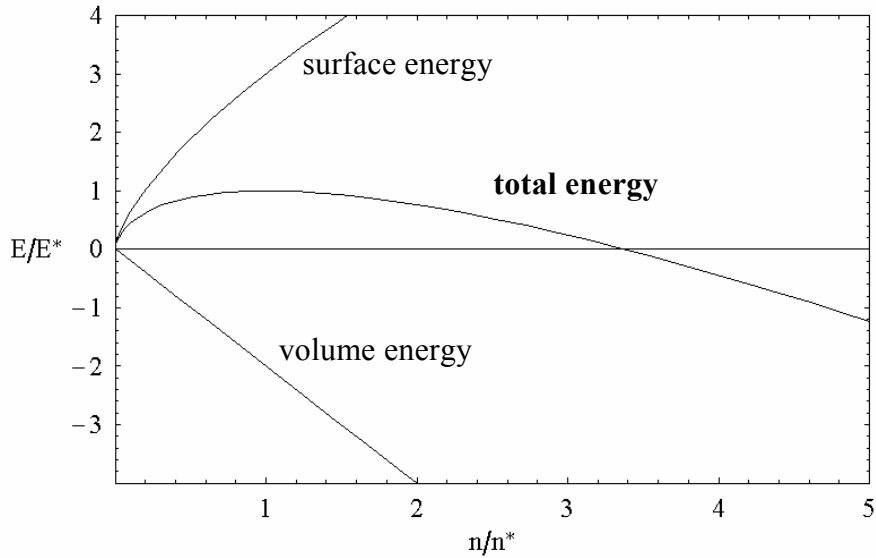

FIG. 2.1: surface, volume and total energies as a function of size n. $n^*$ is defined in equation (2.2)

Meixner *et al.* 2001]. However, as discussed previously, the analysis presented here will focus on the effect of adatom concentrations. Therefore, the value of $\mathcal{E}$ used here will taken to be uniform and will be determined from the nominal mismatch of the film only.

The maximum energy in equation (2.1) occurs at n*, the critical cluster size:

$$n^* = \left(\frac{2\gamma}{3\mathcal{E}}\right)^3. \qquad (2.2)$$

For a subcritical cluster, i.e. a cluster of size n < n*, an increase in the size of the cluster results in an increase in energy, and the cluster tends to shrink. If n > n* (supercritical cluster), an increase in the size of the cluster results in a decrease of the energy, and the cluster tends to grow. However adatoms always have finite probability to reach the island and atoms in the island have a finite probability to leave the island. Such fluctuations may cause a subcritical cluster to grow or a supercritical cluster to shrink. The energy barrier is the energy at the critical size, $E^* = E(n^*)$:

$$E^* = \frac{\gamma\, n^{*2/3}}{3}. \qquad (2.3)$$

Equation (2.1) can be rewritten in terms of critical size n* and nucleation barrier $E^*$ instead of $\gamma$ and $\mathcal{E}$:

$$E = E^*\left[3\left(\frac{n}{n^*}\right)^{\frac{2}{3}} - 2\frac{n}{n^*}\right]. \qquad (2.4)$$

The difference in energy between a cluster of size n and a cluster of size n+1 will be called $\Delta E_n$. For a large enough cluster, a continuum approximation can be used:

$$\Delta E_n \approx \frac{\partial E_n}{\partial n} = \frac{2E^*}{n^*}\left[\left(\frac{n}{n^*}\right)^{-\frac{1}{3}} - 1\right]. \qquad (2.5)$$

*b) Molecular dynamics simulations of islands*

Islands of sizes ranging from fewer than 10 atoms to around 500 atoms were simulated for silicon using molecular dynamics to determine their critical size $n^*$ and energy barrier $E^*$. In the work that follows, *i* will be used to denote the size of an island (in atoms), *p* the size of a pit and *n* the size of any feature when writing general equations applying to both islands and pits.

The potential used is the one published by Stillinger and Weber [Stillinger and Weber 1985]. Simulations were run at 10 K, the temperature was kept constant using a Nosé-Hoover thermostat [Nosé 1984, 1986, Hoover 1985, 1986]. Using higher temperatures would result in more thermal fluctuations which would require longer simulation times. Periodic boundaries were imposed in the x-y plane, the surface perpendicular to the z axis was free and the atoms at the bottom were immobile. Each island was constructed by keeping all atoms inside a hemisphere of a given radius. Then atoms of coordination one were removed; such atoms would have had to diffuse to reach a stable state, increasing the simulation time and decreasing the reproducibility of the results. A typical island is shown in Fig. 2.2; it looks somewhat faceted because of finite size effects. A (2x1) reconstruction was imposed on the free surface around the island.

Two series of simulations were run simulating a matched film and a film under 5 % compression. This choice of the strain is arbitrary — experiments generally take place anywhere between 2 % and 7 % — but results for other mismatches can be extrapolated using equations (2.1) through (2.4). In each simulation, the difference in energy between an island of size *i* and *i* bulk atoms (see Fig. 2.3) was calculated.

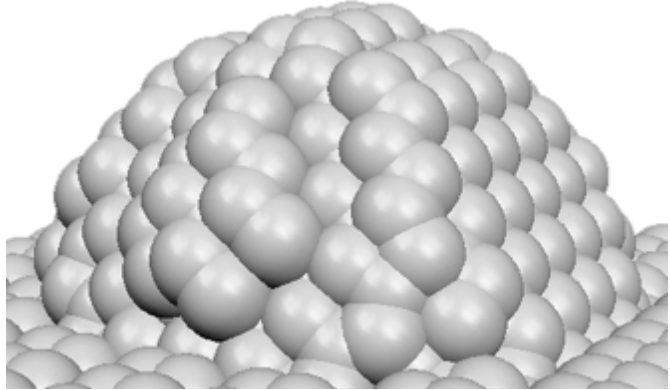

FIG. 2.2: Example of an island of 477 atoms used in the simulations. The island looks somewhat faceted due to finite size effects.

For the case where there is no mismatch, $\varepsilon = 0$, the continuum expression for the energy reduces to $E_i = \gamma i^{2/3}$. So the plot of $E_i\, i^{-2/3}$ as a function of island size $i$ should be constant and give the surface energy $\gamma$. Figure 2.4 is a plot of $\gamma$ obtained from the simulations as a function of the island size. For island sizes greater than 150 atoms, $\gamma$ is indeed approximately constant with a value of 1.5 eV/atom. For sizes less than 150 atoms the continuum assumption does not apply and $\gamma$ is strongly influenced by finite size effects. This value of $\gamma$ appears to be meaningful as it gives a surface energy for the lateral surface of the island within 15 % of published values for the (001) surface [Guyer and Voorhees 1996].

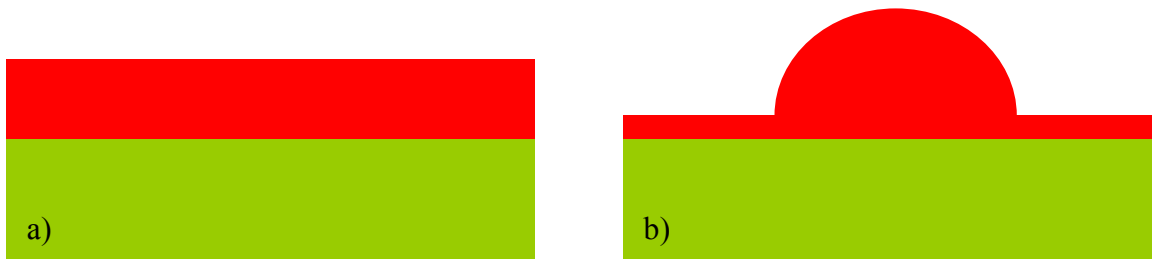

FIG. 2.3: the energies of a flat film (a) and that of an island (b) having the same volume are compared using molecular dynamics.

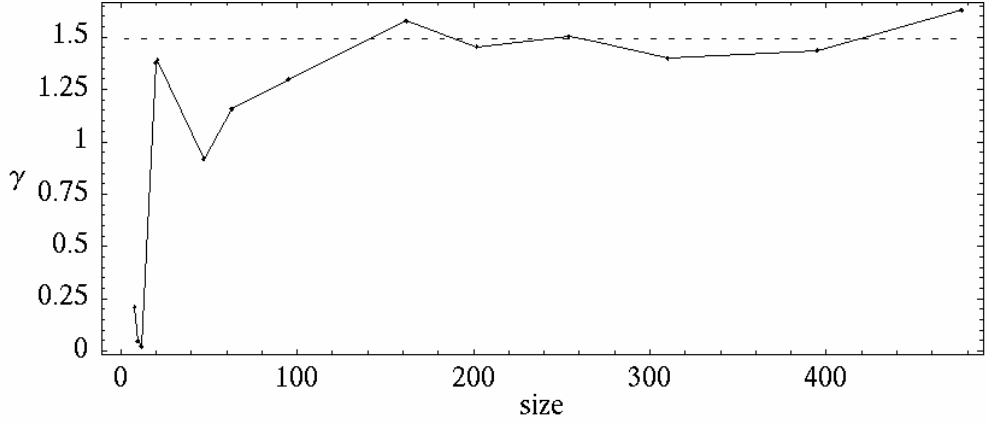

FIG. 2.4: surface energy (in eV/atom) as a function of the island size in Stillinger-Weber silicon. The dashed line is the average over islands larger than 150 atoms.

This surface energy γ is assumed to be independent of the strain, so the difference in energy between an unstrained island $E_{unstrained}$ and an island under 5 % compressive strain $E_{5\%}$ provides a measure of the volume energy:

$$E_{unstrained} - E_{5\%} = \gamma i^{\frac{2}{3}} - \left(\gamma i^{\frac{2}{3}} - \mathcal{E} i\right) = \mathcal{E} i \quad (2.6)$$

Plotting $\dfrac{E_{unstrained} - E_{5\%}}{i}$ as a function of the island size, as done in Fig. 2.5, therefore gives the strain energy prefactor $\mathcal{E}$. Again, for small islands the continuum assumption does not apply. For larger islands, the volume energy is 50 meV/atom. This is within 10 % of the elastic energy as given by continuum elasticity. This does not mean that the strain is entirely relieved in the island, a part of the energy released comes from the atoms in the substrate underneath the island that are also allowed to relax.

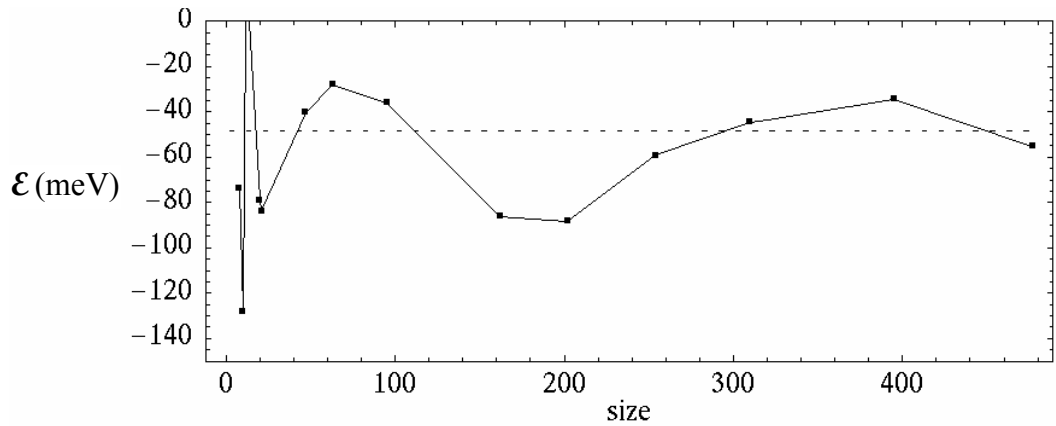

FIG. 2.5: $\mathcal{E}$, the volumetric contribution to the energy released per atom, at 5% strain as a function of the island size in Stillinger-Weber silicon. The dashed line is the average.

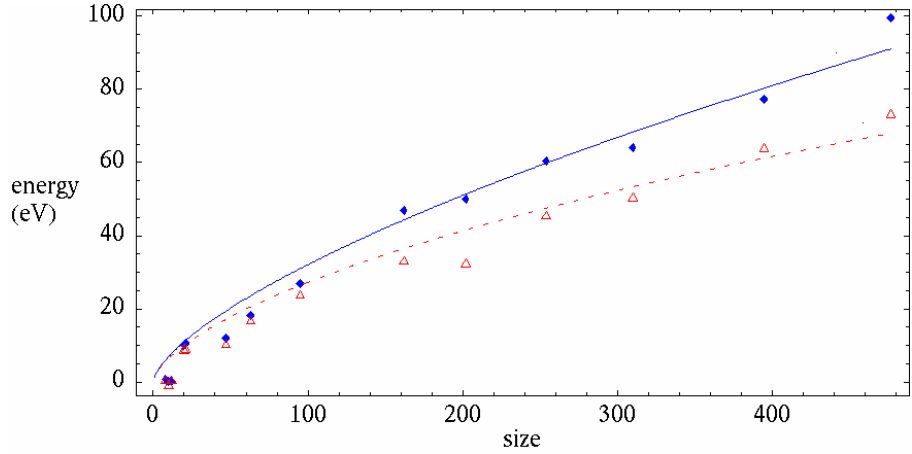

FIG. 2.6: Energy as a function of island size. The filled diamonds and continuous lines are simulation results without strain and interpolation using Eq. (2.1) respectively. Open triangles and dotted lines are the same at 5 % compression.

Knowing the two parameters for Eq. (2.1), it is possible to plot Eq. (2.1) and simulation data on the same graph. This is done in Fig. 2.6, where the difference in energy between the feature and a flat film is shown as a function of feature size. The form of Eq. (2.1) provides a reasonable fit for the data.

Critical sizes and energy barriers for 5 % compression can now be extrapolated when the reference state is taken to be the bulk energy. It is found that the barrier $E_i^* \approx 210$ eV and the critical size $i^* \approx 8\,700$ atoms. The order of magnitude of these critical sizes and energies seems to make it impossible to nucleate an island in a reasonable amount of time.

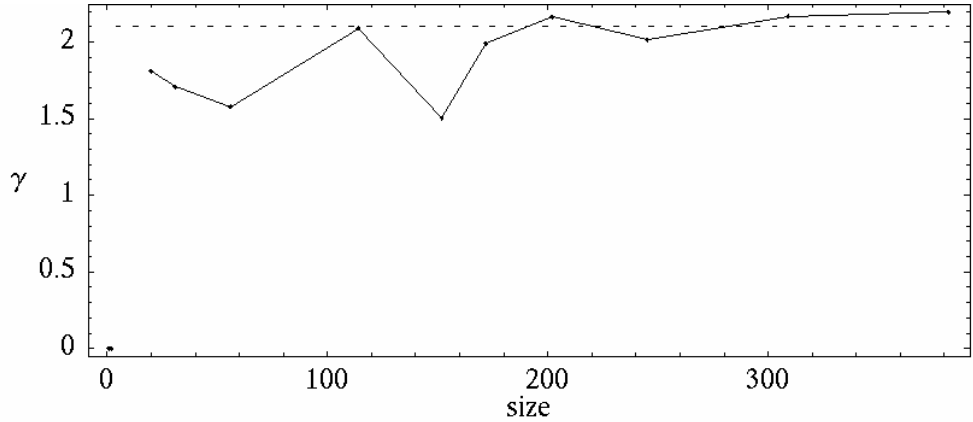

FIG. 2.7: surface energy (in eV/atom) as a function of pit size in Stillinger-Weber silicon. The dashed line is the average over pits larger than 150 atoms.

*c) Molecular dynamics simulations of pits*

Simulations show that pit energy obeys Eq. (2.1) as islands do but with slightly different values for the critical size and energy barrier. These values have been obtained by repeating the analysis in the previous section for pits. For the case where there is no mismatch, the continuum expression for the energy reduces to $E_p = \gamma p^{2/3}$. So the plot of $E_p \, p^{-2/3}$ as a function of pit size $p$ should be constant and give the surface energy $\gamma$. Figure 2.7 is a plot of $\gamma$ obtained from the simulations as a function of the pit size. The surface energy of a pit is constant for large enough pits with a value of 2.1 eV per atom, slightly higher than the surface energy of islands[1].

---

[1] This difference between islands and pits is most likely a curvature effect that will diminish at very large island/pit size.

The plot of $\frac{E_{unstrained} - E_{5\%}}{p}$ as a function of the pit size *p*, as shown in Fig. 2.8, gives the strain energy prefactor $\mathcal{E}$. As for islands, for sizes less than around 100 atoms the continuum assumption does not apply. And for large pits, the strain energy relieved by pits is slightly higher than that of islands, 90 meV/atom *vs.* 50 meV/atom. This is consistent with elastic calculations by Vanderbilt and Wickham (1991) who found that pits could relieve more elastic energy per unit volume than islands.

Figure 2.9 shows the difference in energy between the pit and a flat film is shown as a function of pit size *p* from Eq. (2.1) and from simulation data. This gives an energy barrier $E_p^* \approx 160$ eV and the critical size $p^* \approx 3\,500$ atoms. These values are on the same order as those calculated for islands.

*d) Difficulties with the analogy to classical nucleation*

The values of the critical sizes and barrier energies found in the previous sections appear to preclude the nucleation of islands and pits, inconsistent with experimental results. It is therefore necessary to find an alternative nucleation model providing more reasonable critical size and barrier energy.

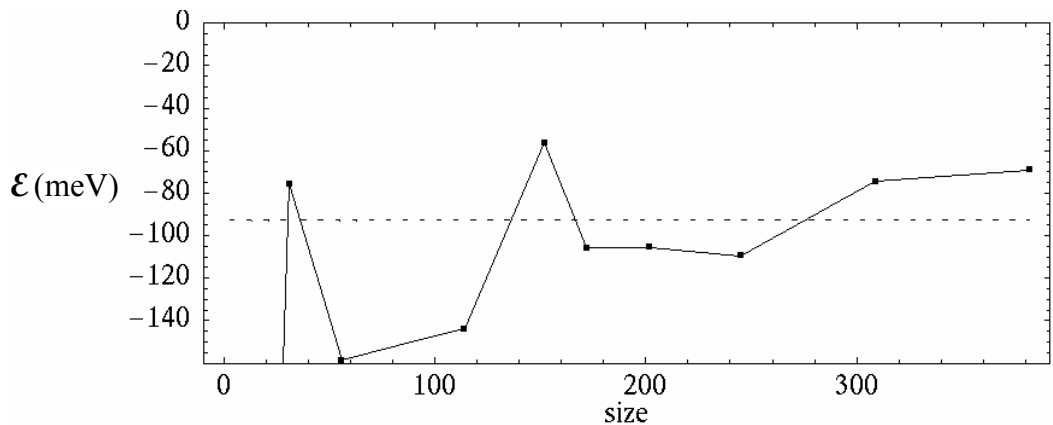

FIG. 2.8: volume energy as a function of pit size in Stillinger-Weber silicon. The dashed line is the average over pits larger than 150 atoms.

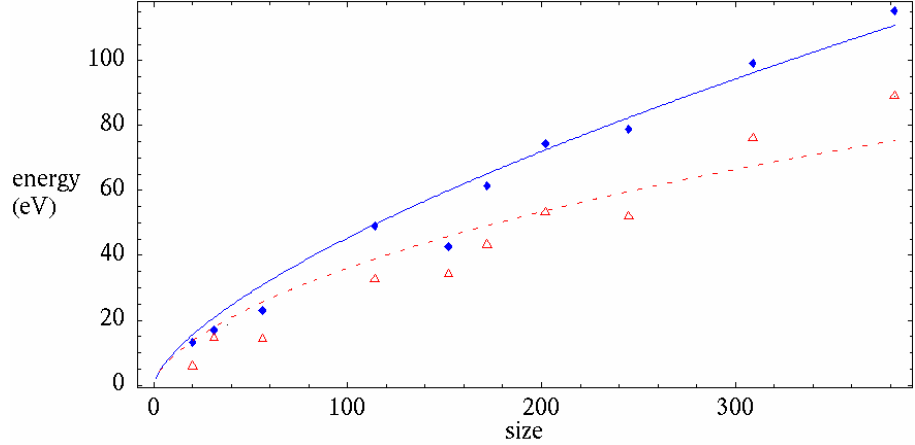

FIG. 2.9: Energy as a function of pit size. The filled diamonds and continuous lines are simulation results without strain and interpolation using Eq. (2.1) respectively. Open triangles and dotted lines are the same at 5 % compression.

This apparent discrepancy in the barrier energy and critical size arises due to the implicit assumption that there exists an equilibrium between the island or the pit and the bulk. However when the surface is held in contact with a reservoir, e.g. arsenic overpressure during the growth of arsenides, the underlying thermodynamics of the system must reflect the exchange of adatoms with this reservoir through the chemical potential of the adatoms µ. The formation energy of an island is then expressed as

$$V_i = E_i - i\,\mu. \qquad (2.7)$$

In the case of III-V semiconductors the chemical potential of a group-III adatom in the presence of an arsenic overpressure – assuming an equilibrium between island, adatoms and vapor – can be expressed as [Tersoff *et al.* 1997]

$$\mu = E_x + kT \ln(P/P_0)/m, \qquad (2.8)$$

where P is the pressure of the $As_m$ vapor; $P_0$ is a reference pressure, and $E_x = 2.7 \pm 0.6$ eV is the formation energy of GaAs when the arsenic is initially in the vapor and Ga is an adatom.

The effective critical size for island nucleation $i_{crit}$ is determined by the maximum of $V_i$,

$$i_{crit} = \frac{i^*}{(1+\mu/\mathcal{E})^3}. \tag{2.9}$$

If $\mu$ is larger than $\mathcal{E}$ the effective critical size can be much lower than $i^*$, which could account for the nucleation of islands on observable time scales. The critical size, $i_{crit}$, given by Eq. (2.9) and $i^*$ are apparent in Fig. 2.7 which shows the energy of an island as a function of the island size. In Fig. 2.7, the continuous line does not take the chemical potential into account and the dotted line does. The parameters used to draw Fig. 2.10 are slightly different from the ones found in the simulations to make the curves appear more clearly. Indeed the maximum of the dotted curve should occur at a value of $i/i^*$ so

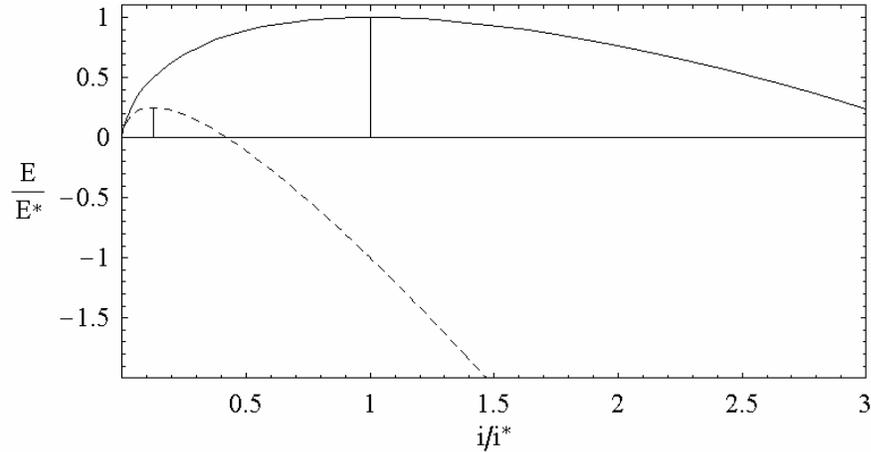

FIG. 2.10: Energy of an island as a function of the island size.

The continuous line is for $E_i$, the dotted line is for $V_i = E_i - i\mu$.

Drawn with $i^* = 1000$ atoms, $E^* = 100$ eV and $\mu = 0.2$ eV.

small that the dotted line would appear as a straight line from the origin. Although the value of $\mu$ is not known, $\mu$ can be expected to be larger than $\varepsilon$ since it involves the creation of bonds while $\varepsilon$ includes only the stretching of the bonds. The critical size as defined in Eq. (2.9) is then much smaller than $i^*$. On its face this would mean that as soon as an island becomes a few atoms big, the system can decrease its energy by an adatom joining this island. But because the equations we use are all continuous, this result must be considered with care. Indeed for very small features finite size effects can be large enough to make the continuum picture inapplicable. All we can say is that the critical size is on the order of a few atoms, without being able to say more precisely how many. Also, these results for the energy barrier and the critical size were obtained using Stillinger-Weber potential which is only an approximate representation of silicon. The finite substrate used in simulations too may have an effect on the energies obtained.

This result is closer to published experimental values of critical sizes [Venables *et al.* 1984] than the thousands of atoms obtained from Stillinger-Weber and Eq. (2.1). This does not mean that this new picture contains all the physics necessary to explain island nucleation. It simply shows that the critical size and energy may not be very large if the local adatom density is not in equilibrium with the bulk crystal. One should also keep in mind that this is obtained using a continuum approximation. Any quantitative match between these numbers and experimental critical sizes is fortuitous, but these results seem to indicate that the system either produces these features through a non-nucleation process or the chemical potential (and hence the adatoms) plays a central role in determining island nucleation. We will explore the latter possibility.

As with islands the chemical potential must be taken into account. In the case of pits, bonds are broken, not created, when atoms leave a pit; Eq. (2.3) becomes

$$V_p = E_p + p\,\mu. \tag{2.10}$$

This leads to an effective critical size of

$$p_{crit} = \frac{p^*}{(1-\mu/\mathcal{E})^3}. \tag{2.11}$$

Thus while the critical island size decreases when the chemical potential is taken into account the critical size for pit nucleation increases with the chemical potential. If $\mu$ is larger than $\mathcal{E}$ pits cannot nucleate as Eq. (2.11) has no solution. This difference in behavior between islands and pits can be seen by comparing Fig. 2.7 and 2.11. Nevertheless, pits are experimentally observed indicating that other factors come into play in the nucleation and growth of pits. In particular, this result does not take the adatom concentration and the related entropic effects into account.

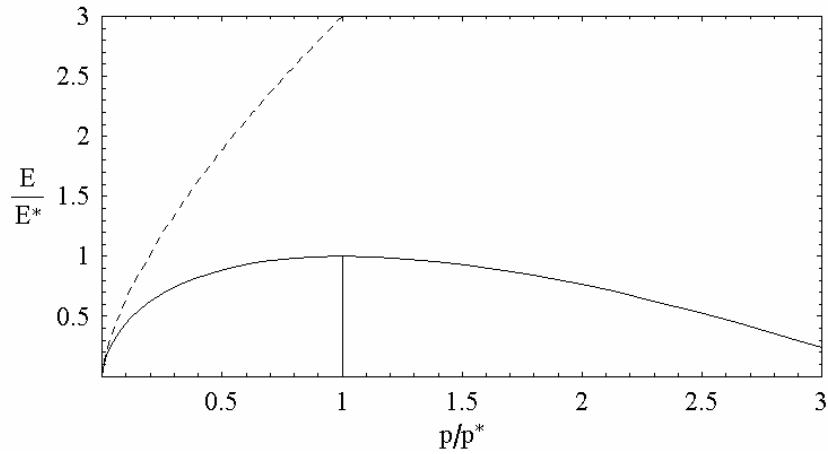

FIG. 2.11: Energy of a pit as a function of the feature size. The continuous line is for $E_p$, the dotted line is for $V_p = E_p + p\mu$. Drawn with $p^* = 1000$ atoms, $E^* = 100$ eV and $\mu = 0.2$

### 3) Rate equations for islands and pits

*a) Rate equations for islands*

A mean-field kinetic model will be developed for the nucleation of islands on a surface with a reservoir upon which the adatom concentration may vary. The purpose is to use the adatom density as a relevant local thermodynamic variable and characterize the growth of islands as a function of this parameter.

In order to extend the picture of island and pit nucleation and growth beyond the near-equilibrium version presented above some simple kinetic models will be discussed that are often used to describe the nucleation process. These models, by explicitly including the adatom-island and adatom-pit kinetics capture some of the entropic contributions to the nucleation process that are ignored by only considering island and pit energetics. In these models the growth or decay of an island will result from the fluxes of atoms to and from the island as illustrated in Fig. 2.12. The first flux in Fig. 2.12 is the flux of adatoms going from the surface to the island, the second is the flux of atoms leaving the island and going to the surface and the third flux is the direct impingement of atoms of the beam onto the island.

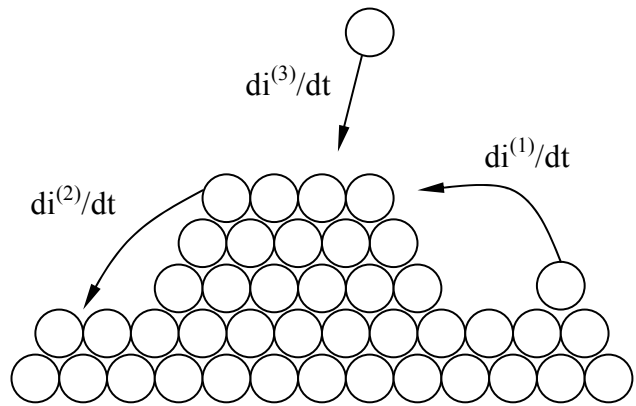

FIG. 2.12: Fluxes from the film to an island (1), to the film (2) and from the beam (3).

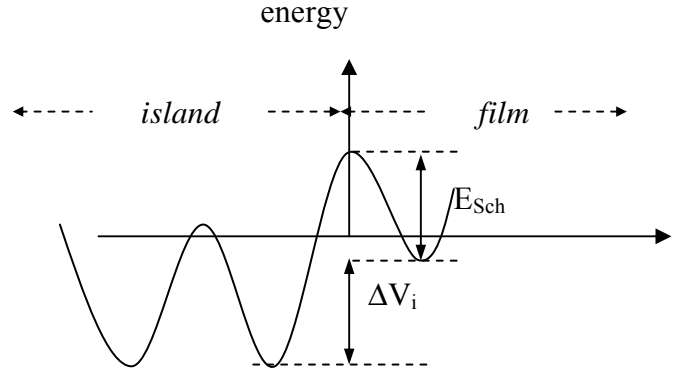

FIG. 2.13: $\Delta V_i$ and the Schwoebel barrier, $E_{Sch}$.

In a mean-field treatment of these processes applicable to a single supercritical hemispherical island, the expressions for the three fluxes are [Frankl and Venables 1970, Venables 1973, Venables *et al.* 1984]:

$$\frac{di^{(1)}}{dt} = D\,\sigma\,\eta\,\exp\left(\frac{-E^i_{Sch}}{kT}\right) \qquad (2.12\ a)$$

$$\frac{di^{(2)}}{dt} = -\nu\,(1-\eta)\exp\left(-\frac{E^i_{Sch} - \Delta E_i + \mu}{kT}\right) \qquad (2.12\ b)$$

$$\frac{di^{(3)}}{dt} = F\left(\pi\,r_i^2\right) \qquad (2.12\ c)$$

The above expressions use the following definitions: i is the number of atoms in the island (atom), $\eta$ is the density of adatoms (atom/site), $\nu$ is an attempt frequency (s$^{-1}$), $\sigma$ is a capture number (dimensionless), D is the diffusivity (sites/s), $r_i$ is the radius of the island (Å) and F is the deposition rate (atom/Å$^2$/s).

As the island is supercritical, the only barrier energy for an adatom to join the island — process 1 in Fig. 2.12 and Eq. (2.12 a) — is a local Schwoebel-like [Schwoebel 1969] barrier $E_{Sch}$ shown in Fig. 2.13. $\mu - \Delta E_i$ is the difference between the energy of an island of size i+1 and the energy of an island of size i plus one adatom.

In the dilute limit ($\eta \ll 1$), the summation of the three terms above leads to:

$$\frac{di}{dt} = D\sigma e^{-\frac{E^i_{Sch}}{kT}} \left[\eta - \frac{\nu}{D\sigma} \exp\left(\frac{\Delta E_i - \mu}{kT}\right)\right] + F\pi r_i^2. \qquad (2.13)$$

If we focus on $\eta$ we see that this equation is affine and can be rewritten in the form:

$$\frac{di}{dt} = \frac{\eta - \eta_i}{\tau_i} \qquad (2.14\text{ a})$$

where

$$\eta_i = \frac{\nu}{D\sigma} \eta_e\, e^{\Delta E_i/kT} - \pi r_i^2\, F\tau_i \qquad (2.15)$$

and

$$1/\tau_i = D\sigma e^{-E^i_{Sch}/kT}. \qquad (2.16)$$

$\eta_e$ is the equilibrium adatom concentration in the absence of islands and pits

$$\eta_e = \exp\left(-\frac{\mu}{kT}\right). \qquad (2.17)$$

Previous reports [Johnson *et al.* 1997] indicated that at 590ºC and an arsenic overpressure of $10^{-6}$ torr the equilibrium adatom concentration on GaAs(001) in the absence of 3D features and of deposition flux, $\eta_e$, is close to 0.1 atom per site.

We can notice that if the adatom density $\eta$ is smaller than $\eta_i$, islands can decay (even islands which appear to be supercritical from a solely energetic analysis.) This phenomenon of islands existing only when the adatom density $\eta$ is higher than $\eta_i$ can be understood in analogy to the equilibrium between a gas and a condensed phase where the latter exists only when the pressure is higher than some saturation pressure $P_{sat}$. If $\eta$ is higher than $\eta_i$, the island can be in equilibrium with the sea of adatoms as will be seen now.

*b) Equilibrium*

From the rate equations of the previous section we can calculate the equilibrium size of an island as a function of experimental conditions. For simplification we restrict ourselves to the case in which the flux of adatoms directly to the islands *via* the beam is negligible. At the equilibrium, given by $\frac{di}{dt} = 0$, $\eta = \eta_i$ and

$$\ln \eta = \Delta E_i / kT + \ln \frac{\nu}{D\sigma} + \ln \eta_e \tag{2.18}$$

Using Eq. (2.5), the equilibrium can be written as

$$\ln \eta = \frac{\varepsilon}{kT}\left[\left(\frac{i}{i^*}\right)^{-\frac{1}{3}} - 1\right] + \ln \frac{\nu}{D\sigma} + \ln \eta_e \tag{2.19}$$

This finally gives an expression for the supercritical island size in terms of the value of $i^*$ and the adatom density

$$i_{crit} = \frac{-\gamma'^3}{\ln^3(\eta^{floor}/\eta)} \tag{2.20}$$

where

$$\gamma' = \frac{2\gamma}{3kT} \tag{2.21}$$

and

$$\eta^{floor} = \frac{\nu}{D\sigma}\eta_e\, e^{-E'_i} \tag{2.22}$$

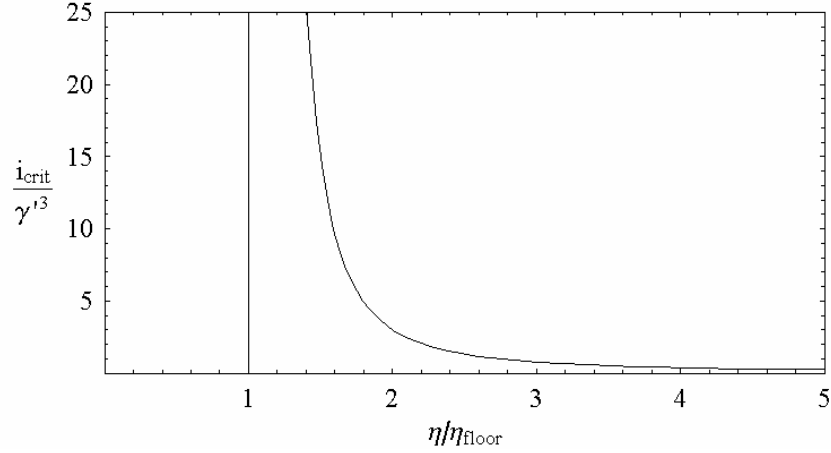

FIG. 2.14: equilibrium size as a function of the adatom density. The vertical asymptote shows that no island can be stable if $\eta < \eta_{floor}$

We can notice that $\lim_{\eta \to \eta^{floor}} \ln(\eta/\eta^{floor}) = 0$ which implies that $\lim_{\eta \to \eta^{floor}} i_{crit} = +\infty$. So $\eta^{floor}$ is the adatom concentration below which no stable island can exist (Fig. 2.14).

*c) Subcritical islands* vs. *supercritical islands*

Equations (2.12) is given for a supercritical island, that is when the island is at a lower energy than the surface. In this case there is an energy cost to leave the island but not to go to the island. For subcritical islands the opposite is true and, in the dilute limit and without deposition, Eqs. (2.12) become

$$\frac{di^{(1)}}{dt} = D\,\sigma\,\eta \exp\left(-\frac{E^i_{Sch} + \Delta E^{adatom}_i}{kT}\right) \quad (2.23\ a)$$

$$\frac{di^{(2)}}{dt} = -\nu \exp\left(-\frac{E^i_{Sch}}{kT}\right) \quad (2.23\ b)$$

Equation (2.14) becomes

$$\frac{di}{dt} = \frac{\eta - \eta_i}{\tau_i} \quad (2.24)$$

where

$$\eta_i = \frac{\nu}{D\sigma}\eta_e\, e^{\Delta E_i/kT} \qquad (2.25)$$

and

$$1/\tau_i = D\sigma\, e^{-E^i_{Sch}/kT}\, e^{-\Delta E_i/kT}. \qquad (2.26)$$

The rate equations used for the supercritical island — equations (2.14) — are almost the same as the ones for a subcritical island shown above. Moreover $\eta_i$ is the same in both cases and only the time scale $\tau$ changes.

*d) Statistical thermodynamics*

The equilibrium island size calculated through rate equations in the previous sections, Eq. (2.15), can also be obtained using statistical thermodynamics. The two approaches should give the same results for the equilibrium. In this section we calculate the internal energy and the entropy of a system constituted by $N - i$ adatoms and an island made of $i$ atoms. From these it is possible to find the minimum of the free energy which gives an equilibrium value for the island size of the kind of equation (2.15) obtained from rate equations.

The entropy of the island will be neglected compared to the one of the adatoms which, to a first approximation, is

$$S = k\ln\Omega = k\ln\frac{N_0!}{a!(N_0 - a)!} \qquad (2.27)$$

where $\Omega$ is the number of ways one can put $a$ adatoms on $N_0$ sites. We implicitly assume that the number of sites is independent of the size of the island, i.e. the surface of the island is small compared to the total surface (we are in the dilute limit $N \ll N_0$). Using that $N_0$, $a$ and $N_0-a$ are large numbers and Stirling approximation, one obtains

$$S \approx k\, a \left( \ln \frac{N_0}{a} + 1 \right) \tag{2.28}$$

The free energy of the system is then

$$G_i \approx E_i - i\mu - kT\,(N-i) \left( \ln \frac{N_0}{N-i} + 1 \right) \tag{2.29}$$

At the equilibrium one has

$$\frac{\partial G_i}{\partial i} = 0 \approx kT \left( \frac{\Delta E_i}{kT} - \frac{\mu}{kT} - \ln \eta \right) \tag{2.30}$$

which is of the same kind as equation (2.15) obtained from kinetics, however a term is different. The two equations are equivalent if $D\sigma/\nu$ is 1. $\sigma$ is qualitatively the number of sites around the island from which adatoms can hop to the island. $\nu$ is the frequency at which adatoms hop from the island. Both are proportional to the perimeter of the island, therefore the perimeter dependence cancels out and $D\sigma/\nu$ is independent of the island size. The process of hopping from the island is diffusion-like; it will be thermally activated with an activation barrier that should be the same as that which controls the diffusivity. So the temperature dependence should cancel out and $D\sigma/\nu$ be independent of T. Then $D\sigma/\nu = 1$ is possible. However, if $\sigma$ also depends on diffusion on the surface through Bessel functions [Venables 1973], it is impossible for $D\sigma/\nu$ to be equal to 1 as $\nu$ is purely local and does not depend on what happens on the surface. But in such a case the adatom concentration will not be homogeneous as required in steady-state and the statistical thermodynamics treatment does not apply.

    Using both kinetics and statistical thermodynamics we found that the equilibrium size of an island depends on the adatom concentration $\eta$. This size is small at high $\eta$ and

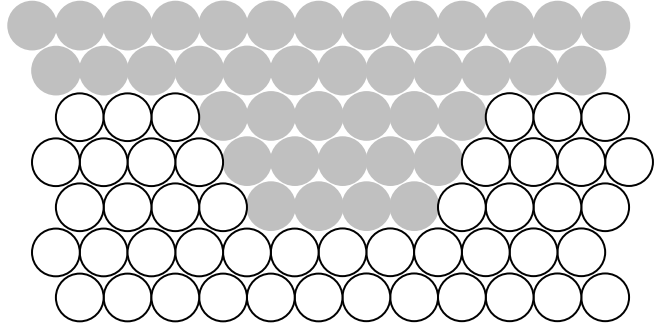

FIG. 2.15: a pit is shown as an island of

vacancies (atoms are white, vacancies are gray).

increases as the adatom concentration decreases. The most striking feature of this result is that when $\eta < \eta^{floor}$, no island can be stable.

*e) The case of pits*

Equations for the kinetics of pits can be derived from those for the island kinetics considering the pit as an island of vacancies as shown in Fig. 2.15, i.e. the density of adatoms $\eta$ has been replaced by the density of advacancies $1-\eta$. In what follows, we assume that $D\sigma/\nu = 1$. This leads to:

$$\frac{dp^{(1)}}{dt} = -\nu\,\eta\,\exp\left(\frac{\Delta E_p + \mu - E_{Sch}^p}{kT}\right) \qquad (2.31\text{ a})$$

$$\frac{dp^{(2)}}{dt} = \nu\,\exp\left(\frac{-E_{Sch}^p}{kT}\right) \qquad (2.31\text{ b})$$

$$\frac{dp^{(3)}}{dt} = -F\left(\pi\,r_p^2\right) \qquad (2.31\text{ c})$$

The above expressions use the following notations (equivalent to those for islands): p is the number of atoms in pit (atom), $\eta$ is thre density of adatoms (atom/site), $\nu$ is an attempt frequency (s$^{-1}$), $E_{Sch}$ is the Schwoebel barrier (eV), $r_p$ is the radius of the pit (Å) and F is the deposition rate (atom/Å$^2$/s)

The complete equation in the dilute limit ($\eta \ll 1$) is the sum of Eq. (2.31):

$$\frac{dp}{dt} = \nu \exp\left(\frac{\Delta E_p + \mu - E_{Sch}^p}{kT}\right)\left[\exp\left(-\frac{\Delta E_p + \mu}{kT}\right) - \eta\right] - F\pi r_p^2. \qquad (2.32)$$

$\Delta E_p$, the difference in energy between a pit of size p and a pit of size p+1, can be approximated as the derivative of the energy with respect to p:

$$\Delta E_p \approx \frac{dE_p}{dp} = -\mathcal{E} + \frac{2\gamma}{3} p^{-1/3}. \qquad (2.33)$$

Here, and in subsequent expressions, the elastic energy relieved per atom removed from a pit is labeled $\mathcal{E}$ instead of $\mathcal{E}^{pit}$ to simplify the notation.

In order to show the affine dependence on $\eta$, equation (2.32) can be rewritten as:

$$\frac{dp}{dt} = \frac{\eta_p - \eta}{\tau_p} \qquad (2.34)$$

with

$$1/\tau_p = \frac{\nu}{\eta_e} \exp\left(\frac{\Delta E_p - E_{Sch}^p}{kT}\right). \qquad (2.35)$$

Neglecting direct impingement into the pit, the concentration of adatoms in equilibrium with a pit of size p, $\eta_p$, is determined by the chemical potential and the energetics of the pit

$$\eta_p = \eta_e \exp\left(-\frac{\Delta E_p}{kT}\right). \qquad (2.36)$$

This equation is of the same form as equations (2.25) for islands. The equivalent of Eq. (2.20) for pits is

$$p_{crit} = \frac{-\gamma'^3}{\ln^3(\eta/\eta_{ceil})} \tag{2.37}$$

where $\eta_{ceil}$ is defined as

$$\eta_{ceil} = \eta_e \exp\frac{\varepsilon}{kT}. \tag{2.38}$$

When $\eta$ approaches $\eta_{ceil}$, the critical size goes to infinity. Thus pits cannot nucleate or grow at adatom concentrations above $\eta_{ceil}$. Similarly no islands can nucleate or grow below a minimal value of the adatom concentration, $\eta_{floor}$. This dependence of the nucleation of islands and pits on the adatom concentration accounts for the rarity of homogeneously-nucleated pits. As long as $\eta$ is above $\eta_{floor}$ and $\eta_{ceil}$, only islands can nucleate. Once islands form, however, they can act as sinks of adatoms, lowering the adatom density and, in some cases, allowing pits to nucleate and grow.

As in the case of islands, Eq. (2.37) can be obtained from statistical thermodynamics. The free energy is of the same form as for islands:

$$G_p \approx E_p^{bulk} + p\Delta E_s - kT(N+p)\left(\ln\frac{N_0}{N+p} + 1\right) \tag{2.39}$$

The equilibrium is given by $\frac{\partial G_p}{\partial p} = 0$. Which again agrees with the results from kinetics.

**4) Island-induced inhomogeneities in adatom concentration**

Our observations of pits [Chokshi *et al.* 2000, 2002, Riposan *et al.* 2002, 2003] indicate that they often nucleate close to islands or even surrounded by islands. This section considers the effect of the proximity of islands on pit nucleation, particularly the case of two islands with separation much smaller than their radii. In this case the islands are treated as two infinitely long parallel absorbing boundaries a distance 2 $\ell$ apart. The deposition rate F acts as a source of adatoms, and steps on the surface capture adatoms at a rate proportional to the difference between the local adatom concentration η and the equilibrium adatom concentration $\eta_e$. The proportionality constant $1/\tau$ is an indication of the time it takes to incorporate adatoms into steps. Fick's second law for this one-dimensional problem is

$$\frac{\partial \eta}{\partial t} = D\frac{\partial^2 \eta}{\partial x^2} - \frac{\eta - \eta_e}{\tau} + F \tag{2.40}$$

where x is the position between the islands and D is the diffusivity. In steady state, the adatom concentration at position x, η(x), is given by

$$\frac{\eta(x) - \eta_\infty}{\eta_{edge} - \eta_\infty} = \frac{\cosh x/L}{\cosh \ell/L} \tag{2.41}$$

where $L = \sqrt{D\tau}$ is a diffusion length, $\eta_{edge}$ is the adatom concentration at the island edge and $\eta_\infty = \eta_e + F\tau$ is the adatom concentration in the absence of islands. All distances are expressed in terms of distance between two surface sites, areas are in number of surface sites. Figure 2.16 shows how η varies between two islands separated by 2 $\ell$ as a function of position as given by Eq. (2.41). The adatom concentration is minimum at the edge of the island where $\eta = \eta_{edge}$ and reaches its maximum value midway between the islands, $\eta_{mid}$, which is given by

$$\frac{\eta_{mid} - \eta_\infty}{\eta_{edge} - \eta_\infty} = \text{sech}\frac{\ell}{L} \tag{2.42}$$

If $\eta_{ceil} < \eta_{edge}$, the adatom concentration is greater than $\eta_{ceil}$ everywhere on the surface and pitting is precluded. When $\eta_{edge} < \eta_{ceil} < \eta_{mid}$, pitting is localized near the islands.

When $\eta_{ceil} > \eta_{mid}$, the adatom concentration is lower than $\eta_{ceil}$ everywhere between the islands and pit nucleation is delocalized. Finally when $\eta_{ceil} > \eta_\infty$, the adatom concentration is lower than $\eta_{ceil}$ even in the absence of islands and pits can nucleate.

Whether or not pits may nucleate thus depends on the value of $\eta_{edge}$, which can be derived by considering the conservation of mass on the surface. As expressed in the following balance equation, the material from the beam (left hand side) will be exchanged with steps (first term of the right hand side) or with islands (second and third terms of the right hand side)

$$F\,2\ell = -\int_{-\ell}^{\ell} \frac{\eta(x)-\eta_e}{\tau}\,dx + 2v_{in}\eta_{edge} - 2\Phi_{out}. \qquad (2.43)$$

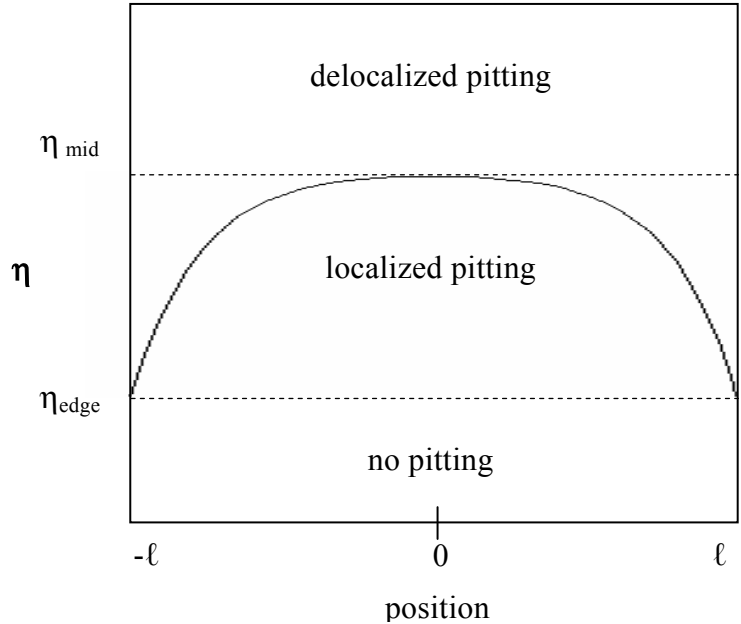

FIG. 2.16: The variation of the adatom concentration $\eta$ between two islands separated by a distance $2\ell$, as a function of position, as given by equation (2.41).

Here, the product $v_{in} \eta_{edge}$ is the rate of attachment to the island per boundary site and $\Phi_{out}$ is the rate of detachment from the island per boundary site. Equations (2.41) and (2.43) determine the functional form of $\eta_{edge}$

$$\eta_{edge} = \eta_\infty - \frac{\eta_\infty - \dfrac{\Phi_{out}}{v_{in}}}{1 + \dfrac{D}{L\, v_{in}} \tanh \dfrac{\ell}{L}} \qquad (2.44)$$

The diffusivity D can be related to atomic phenomena through Einstein's formula: $D = v_0/4$, where $v_0$ is the hop frequency associated with diffusion. The frequency for an adatom to attach to an island $v_{in}$ depends on the Schwoebel barrier at the island edge, $E_{Sch}$, such that $v_{in} = v_0 \exp{-E_{Sch}/kT}$. Also, in steady state, the detachment rate from the islands $\Phi_{out}$ equals the detachment rate $v_{in} \eta_e \exp{\Delta E_i /kT}$. From these $\eta_{edge}$ can be rewritten

$$\eta_{edge} \approx \eta_\infty - \frac{\eta_\infty - \eta_e \exp\left(\dfrac{\Delta E_i}{kT}\right)}{1 + \dfrac{1}{\Lambda} \tanh \dfrac{\ell}{L}} \qquad (2.45)$$

where $\Lambda = 4 L \exp{-E_{Sch}/kT}$.

**5) The thermodynamics of pit nucleation in the presence of islands**

The results detailed in the previous two sections allow us to present a detailed picture of when pitting is thermodynamically favored subsequent to three-dimensional islanding for a particular material system under given growth conditions. When pit nucleation is thermodynamically allowed pits may only be thermodynamically possible in the vicinity of islands or they may arise anywhere between the islands. The question of the kinetics of pit nucleation will be set aside until the next section.

The first step is to determine whether pitting is possible for a given materials system under a given set of growth conditions and at a stage of growth characterized by a particular island-island separation. This is accomplished by comparing the adatom concentration above which pits cannot nucleate, $\eta_{ceil}$, with the lowest adatom concentration on the surface that occurs next to the islands, $\eta_{edge}$. When $\eta_{edge}$ is greater than $\eta_{ceil}$ pitting is precluded. From Eq. (2.38) and (2.45) this condition is equivalent to

$$\tanh\frac{\ell}{L} > \Lambda \frac{\eta_{ceil} - \eta_e \exp\left(\frac{\Delta E_i}{kT}\right)}{\eta_\infty - \eta_{ceil}}. \qquad (2.46)$$

The next step is to distinguish between instances in which pits can form at any arbitrary location between islands or only adjacent to the islands. When $\eta_{ceil}$ is greater than even the highest adatom concentration on the surface pitting is delocalized. Since the maximum adatom concentration is always smaller than $\eta_\infty$, pitting can occur irrespective of the presence of islands when $\eta_\infty$ is below $\eta_{ceil}$. From Eq. (2.38) and our definition of $\eta_\infty$ this condition is equivalent to $F\tau/\eta_e < e^{\varepsilon/kT} - 1$. Note that $F\tau/\eta_e$ is the supersaturation, i.e. the relative increase of the adatom concentration due to deposition. When $\eta_\infty$ is higher than $\eta_{ceil}$, islands in proximity to each other may still decrease the maximum adatom concentration in the region between them below $\eta_{ceil}$. Using Eq. (2.42) and (2.45) the condition for delocalized pitting is then equivalent to

$$\cosh\frac{\ell}{L} + \frac{1}{\Lambda}\sinh\frac{\ell}{L} < \frac{\eta_\infty - \eta_e \exp\left(\frac{\Delta E_i}{kT}\right)}{\eta_\infty - \eta_{ceil}} \qquad (2.47)$$

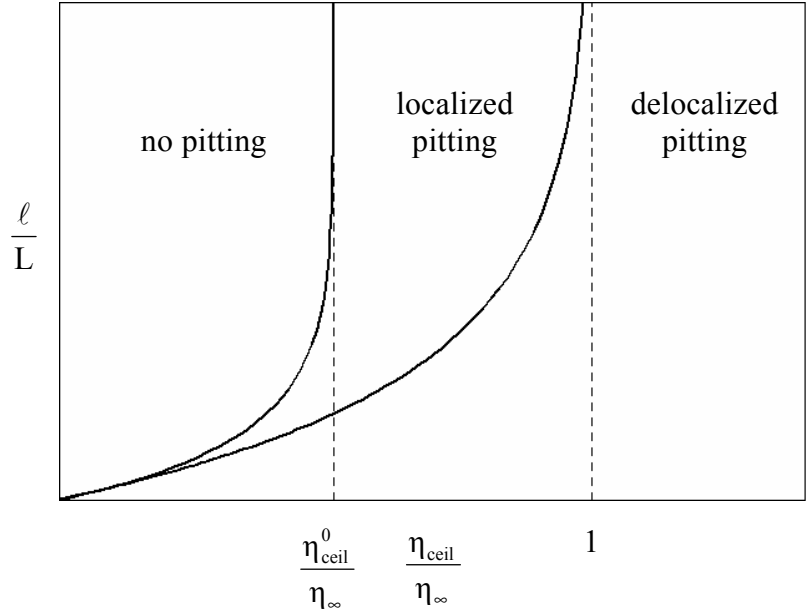

FIG. 2.17: Transitions between no pitting, localized pitting and delocalized pitting as a function of both the ratio of the critical adatom concentration for pit formation to the adatom concentration on a nominally flat film, $\eta_{ceil}/\eta_\infty$ and of the ratio of the island separation to the diffusion length, $\ell/L$. Drawn for $\Lambda = 1$. The dashed lines are asymptotes.

Thus when $\eta_{ceil} < \eta_\infty$, the transition between delocalized and localized pitting depends on the distance between islands, $\ell$.

Figure 2.17 illustrates the transition from absence of pitting to localized and then delocalized pitting as a function of $\eta_{ceil}/\eta_\infty$ and of $\ell/L$. $\eta_{ceil}$ depends on the growth rate and $\eta_\infty$ on the mismatch and temperature. Therefore the x-axis of Fig. 2.17 depends on the material system under certain growth conditions, which will be defined as an "experimental regime". The y-axis depends on the distance between two islands, i.e. the stage of growth. Assuming that islands have already nucleated, $\ell$ decreases as they approach each other due to further growth. In Fig. 2.17 the growth process subsequent to 3D islanding can be conceived as a downward-pointing vertical arrow indicative of the time evolution of the system due to the decrease of $\ell$. Therefore, from Fig. 2.17 one can infer the morphological evolution for a given experimental regime. If $\eta_{ceil}$ is always greater than $\eta_\infty$ pitting is delocalized. This experimental regime will be called *delocalized* pitting, or "ID" for *i*slands followed by *d*elocalized pits. If $\eta_{ceil}$ is less than $\eta_\infty$ but greater than a second transition value, $\eta^0_{ceil}$ pitting is initially possible only near islands until the islands reach a critical separation. This is the *adjacent* pitting experimental regime which will be called "IA". If $\eta_{ceil}$ is lower than $\eta^0_{ceil}$ pitting is precluded until islands are within a critical separation distance. Thus the systems with the lowest values of $\eta_{ceil}/\eta_\infty$ are designated as exhibiting pits *between* islands and are denoted "IB".

Figure 2.18 shows these three experimental regimes as a function of $\mathcal{E}/kT$, the elastic energy as compared to the thermal energy, and $F\tau/\eta_e$, the supersaturation. The transition between experimental regimes IB and IA corresponds to the transition between pitting in between islands and pitting adjacent to islands. If $\Lambda$ is small, this transition is close to the IA-ID transition, the region of phase space associated with experimental regimes of type IA is small. If $\Lambda$ is large on the other hand, $F\tau/\eta_e$ must be much larger than $\mathcal{E}/kT$; limiting the phase space of experimental regimes of type IB. Note that Fig. 2.18 assumes that islands have already nucleated. This criterion will be relaxed when kinetic effects are taken into account to determine if islands nucleate prior to pits, if at all.

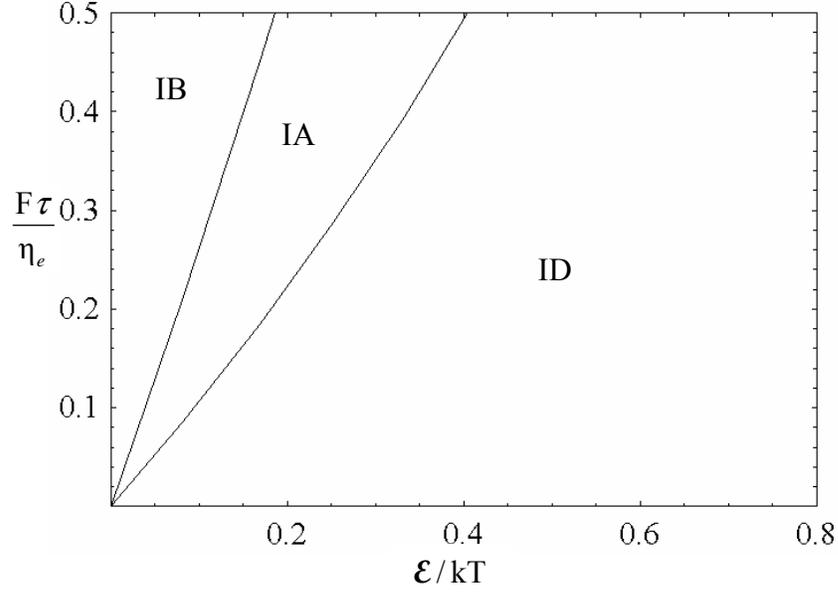

FIG. 2.18: Equilibrium phase diagram showing the domains of the experimental regimes where pits nucleate adjacent to islands (IA), pits nucleate between islands (IB) and pits are delocalized (ID) as a function of the ratio of the elastic energy to the thermal energy, $\mathcal{E}^{pit}/kT$, and of the supersaturation induced by the beam, $F\tau/\eta_e$. Drawn for $\Lambda = 1$.

**6) The kinetics of island and pit nucleation**

The previous section detailed the thermodynamics of pit nucleation. However, the epitaxial growth process occurs on experimentally-determined time scales. Pits will not be experimentally observed unless they nucleate on time scales that are comparable to or faster than this time. In order to predict the experimental observation of pits or lack thereof, it is necessary to incorporate the effect of kinetics into our theoretical analysis.

The rate at which pits nucleate is proportional to the rate at which pits of critical size are generated. The statistics of pit populations on the surface — in particular the number of pits of critical size — can be obtained using the formalism introduced by Walton

(1962). Assuming detailed balance, an expression for the number of pits of size p can be derived

$$N_p = N_1 \, e^{-\frac{E_p - E_1}{kT}} \left(\frac{\eta_e}{\eta}\right)^{p-1} \quad (2.48)$$

where $N_1$ is the number of pits of size 1 and $E_1$ is the energy of a pit of size 1. The nucleation rate is proportional to the number of pits of critical size. Because $N_1 = \exp\left(-\frac{E_1 + \mu}{kT}\right)$, the rate at which pits nucleate is

$$R = R_0 \, \eta \, \exp\left(-\frac{\gamma'^3}{2 \ln^2 \eta/\eta_{ceil}}\right) \quad (2.49)$$

where $\gamma' = \frac{2 \gamma^{pit}}{3 \, kT}$ and $R_0$ is a rate constant related to the attempt frequency. The nucleation rate is high at low $\eta/\eta_{ceil}$ and decreases rapidly with increasing $\eta/\eta_{ceil}$. The nucleation rate decreases when the surface energy increases and, for large values of $\gamma'$, only at low adatom concentrations can pits nucleate at a non-negligible rate. Moreover the maximum nucleation rate is lower at larger $\gamma'$. When $\eta/\eta_{ceil}$ is less than $e^{-\gamma'}$, $p_{crit}$ would be less than 1 and the model breaks down.

The previous section showed that when islands are arbitrarily far apart experimental regimes of type ID show pitting regardless of the presence of islands whereas those of type IA can nucleate pits only adjacent to the islands. An experimental regime that is thermodynamically of type ID would appear to be kinetically of type IA if the probability for a pit to nucleate far from an island is small compared to the probability that it nucleate close to an island. Therefore the ratio of the nucleation rate close to the island to the nucleation rate far from the island will be used to kinetically discriminate experimental regimes of types IA and ID. The pit nucleation rate given by Eq. (2.49) is highest at the islands, where the adatom concentration is the lowest,

$$R_{edge} = R_0 \, \eta_{edge} \, \exp\left(-\frac{\gamma'^3}{2 \ln^2 \eta_{edge}/\eta_{ceil}}\right) \tag{2.50}$$

and it is lowest at the mid-point between the islands, where η is the highest,

$$R_{mid} = R_0 \, \eta_{mid} \, \exp\left(-\frac{\gamma'^3}{2 \ln^2 \eta_{mid}/\eta_{ceil}}\right). \tag{2.51}$$

If the ratio of these two rates $R_{edge} / R_{mid}$ is high, pits nucleate primarily adjacent to the islands as in IA. If this ratio is close to one, pits can nucleate everywhere at almost the same rate and the experimental regime is kinetically of type ID. The *kinetic* IA-ID transition is therefore defined as $R_{edge}/R_{mid}$ much greater than one, which will be 100 for our purposes.

While some experiments showing pits were described in the previous section, most systems do not exhibit any pitting. This implies that there is an important experimental regime for which pitting is precluded for any value of $\ell$. In that case, the maximum nucleation rate on the surface which occurs next to the islands must be negligible regardless of island-island separation. The experimental regimes where pits are never observed to nucleate, i.e. where there are only islands, will be denoted as being of type "I0". The adatom density at the island, $\eta_{edge}$, is at its lowest when $\ell = 0$. The condition for the I0-IB transition can be found by setting $\eta_{edge}$ to $\eta_e \exp\left(\frac{\Delta E_i}{kT}\right)$ in Eq. (2.50).

In Eq. (2.50), $R_{edge}$ gives the nucleation rate in number of pits per site per second. To compare our predictions to experiments, it is more convenient to express this condition in number of pits per sample. The cut-off value is again arbitrary; experimental regimes will be considered to be kinetically of type IB, in which pits nucleate only between islands, when fewer than 1 pit nucleates per minute per 100 μm$^2$; this gives $R_{edge}$ ($\ell \rightarrow \infty$)/$R_0$ < 10$^{-23}$ as a criterion. Experimental regimes are kinetically of type IA when pits can nucleate only close to islands when the latter are far apart. This corresponds to the condition that $R_{edge}$ ($\ell \rightarrow \infty$)/$R_0$ > 10$^{-23}$ and $R_{edge}/R_{mid}$ > 10$^2$. Experimental regimes are kinetically of type ID when pits can nucleate everywhere even if islands are far apart. This corresponds to the condition that $R_{edge}/R_{mid}$ < 10$^2$. Experimental regimes are kinetically of type I0 when pits cannot nucleate even when islands are close together; it is defined as $R_{edge}$ ($\ell = 0$)/$R_0$ < 10$^{-23}$. Figure 2.19 shows the kinetic phase diagram obtained from these conditions. The dashed lines represent the thermodynamics results for the IB-IA and IA-ID transitions from Fig. 2.18 for comparison. When the surface energy γ' is small, the kinetic phase diagram is very close to that from thermodynamics. But when γ' is larger, as in Fig. 2.19, the boundaries shift to higher values of $\mathcal{E}/kT$. This is because when the surface energy is larger, the nucleation process is slower. Experimental regimes of type ID which dominated in Fig. 2.18 may not exist at all on the observable time scale when γ' is large and the discrepancy between thermodynamics and kinetics is most significant. Regimes of type I0 on the other hand, which have no analog when kinetics are not considered, account for most low-misfit material systems.

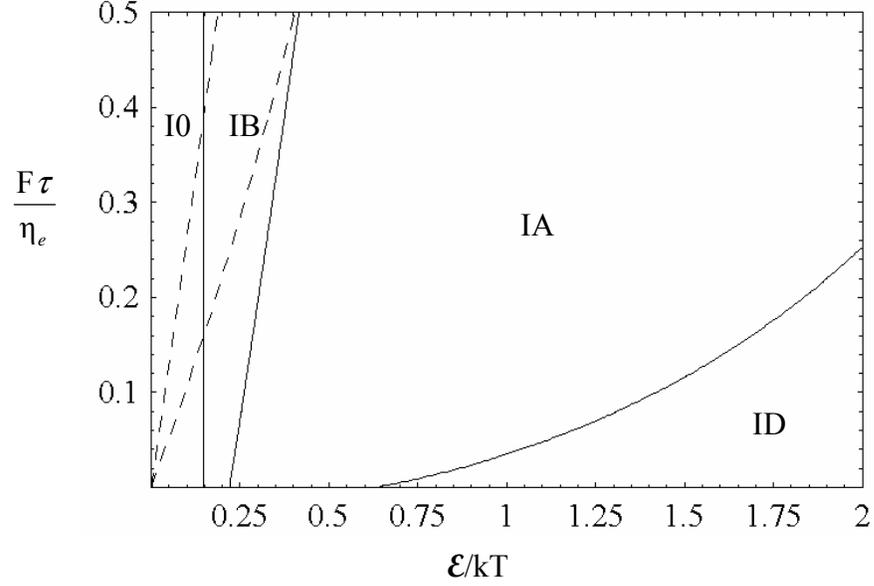

FIG. 2.19: Kinetic phase diagram showing the domains of experimental regimes where only islands nucleate (I0), pits nucleate adjacent to islands (IA), pits nucleate between islands (IB) and pits are delocalized (ID) as a function of the ratio of the elastic energy to the thermal energy, $\mathcal{E}^{pit}/kT$, and of the supersaturation induced by the beam $F\tau/\eta_e$. The dashed lines correspond to the thermodynamic results shown in Fig. 2.17. In this graph, we have assumed $\Lambda = 1$ and $\gamma' = 2$.

The four experimental regimes defined so far assume that islands have already nucleated. To account for the possible absence of island nucleation and the fact that pit may nucleate before islands, two more experimental regimes are added: "P" when pits nucleate before islands and "0" when both islands and pits are kinetically prevented. The nucleation rate of islands is similar to that of pits, Eq. (2.49),

$$R_{isl} = R_0 \frac{\eta_e^2}{\eta_\infty} \exp\left(-\frac{\gamma'^3}{2\ln^2 \eta_{floor}/\eta_\infty}\right) \quad (2.52)$$

If $R_{isl}/R_0 > 10^{-23}$, islands can nucleate on a flat film. If $R_{pit}/R_0 > 10^{-23}$, pits can nucleate on a flat film. If both rates are small, then neither islands nor pits nucleate and the film remains planar, i.e. regime "0". If both can nucleate, it is necessary to determine which one nucleates at a higher rate to discriminate between "I" and "P" regimes. To this end, their nucleation rates on a flat film at $\eta = \eta_\infty$ are compared.

**7) Discussion**

Experimental observations show a wide range of morphologies: planar films, islands alone, islands nucleation followed by pit nucleation, pits alone. Materials systems and experimental procedures also are very diverse; the mismatch, surface energy, temperature can all vary. This section discusses the effects of the changes of such parameters on the surface morphology.

*a) The effect of materials system*

Figure 2.20 shows the experimental regimes as a function of the surface and strain energies. The two diagrams are for two different values of the supersaturation, $F\tau/\eta_e$. At low mismatch and high surface energy, neither islands nor pits can nucleate and the film remains planar (regime 0). At the highest strain energies, pits can nucleate before islands (regime P). If both surface and strain energy are low, islands nucleate (regime I). There are four sub-regimes in regime I. At very low surface and strain energies only islands nucleate (regime I0). For higher strain energies, pits can nucleate but are limited to nucleation between islands (regime IB). For even higher strain energies, they nucleate adjacent to islands (regime IA) or pit nucleation is delocalized between the islands (regime ID).

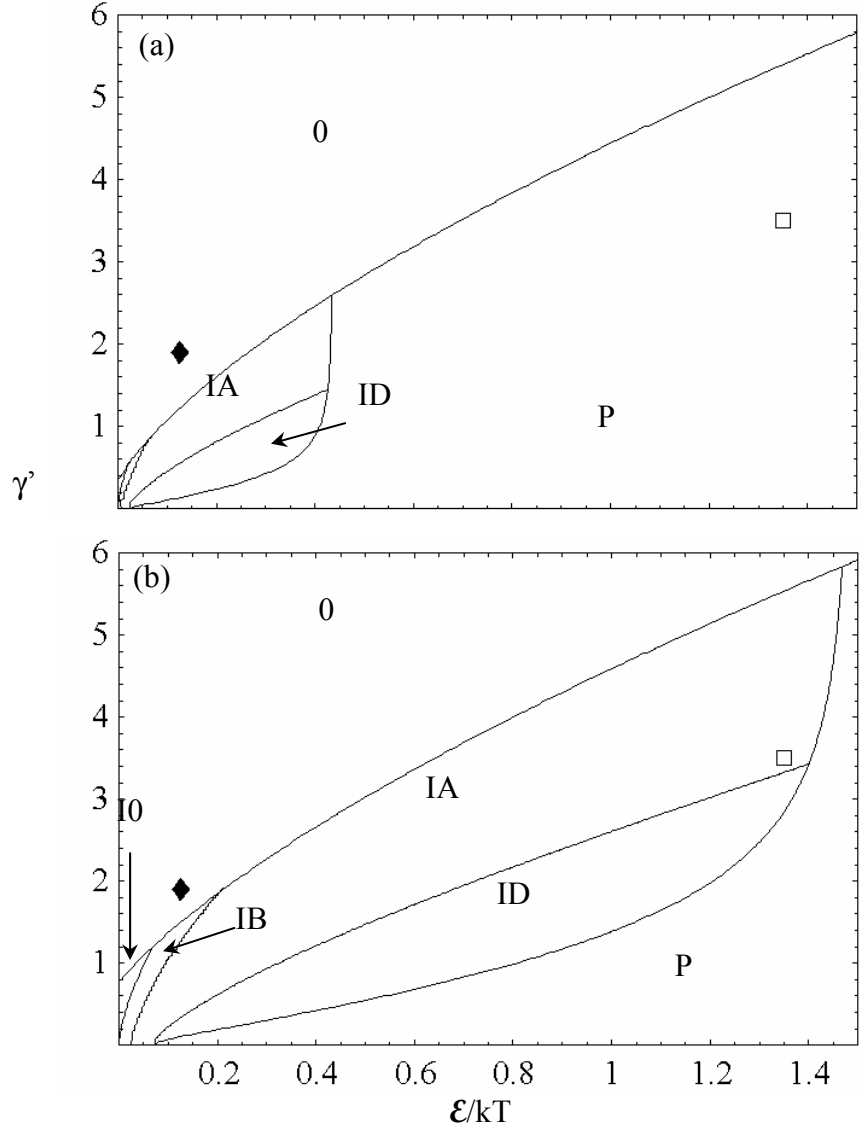

FIG. 2.20: Domains of the various experimental regimes as a function of strain and surface energies, drawn assuming $\eta_e = 0.1$ and $\Lambda = 1$, for two different values of the supersaturation (a) $F\tau/\eta_e = 0.02$, (b) $F\tau/\eta_e = 0.07$. Two experimental systems are denoted for comparison, $In_{0.27}Ga_{0.73}As/GaAs$[7-10] (♦) and $InSb/InAs$[15] (□). Note: the geometry of pits is accounted for in the evaluation of the surface energy.

These six experimental regimes exist at all supersaturations, but which regime dominates depends on the value of the supersaturation. Equation (2.11) predicts that low adatom concentrations promote pit formation. As a result, at lower supersaturations, Fig. 2.20(a) for instance, there are essentially two cases: either pits nucleate before islands or neither islands nor pits nucleate. At high supersaturations, Fig. 2.20(b), the islanding regions dominate regions where pits nucleate first. Previous sections have shown that a high adatom concentration promotes islands formation and prevents the formation of pits; Fig. 2.20 indeed shows that the islanding regions expand at high supersaturations, while pitting regions shrink.

*b) The effect of deposition rate and temperature*

In addition to studying pit nucleation for different strain and surface energies, we have examined the effect of parameters such as deposition rate and temperature for a single materials system. Figure 2.21 shows the predicted experimental regimes as a function of temperature and deposition flux for a film where $\mathcal{E}/kT = 0.13$ and $\gamma' = 1.10$.

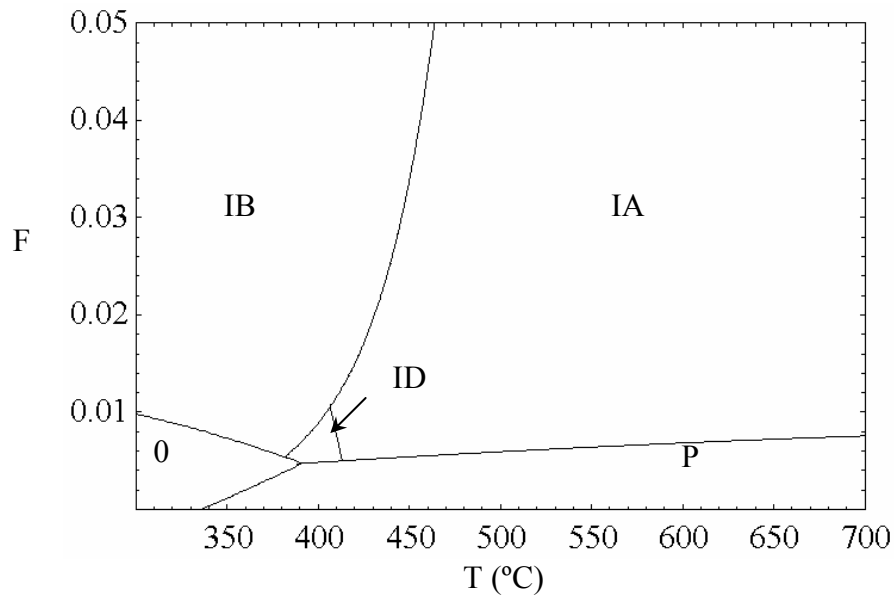

FIG. 2.21: Domains of the various experimental regimes as a function of temperature and deposition rate (in arbitrary units), assuming that at 500°C, $\mathcal{E}/kT = 0.13$, $\Lambda = 1$ and $\gamma' = 1.10$.

The model predicts several different possible morphologies for a given set of growth conditions. For the lowest growth rates and temperatures, the film is predicted to remain planar, with neither islands nor pits able to nucleate on the surface. With increasing growth temperatures, pit nucleation alone is expected at low growth rates. As the growth rate increases, islands are expected to nucleate. However, the model invariably predicts that pitting occurs subsequent to islanding. As we will discuss in the next section, this is consistent with experimental observations in III-V systems.

**8) Comparison to experiments**

The predictions of the model agree well with experimental observations. However, an important *caveat* should be noted in that there exist significant uncertainties in the values of the materials parameters due to the lack of experimental data. The strain energy relieved by the pits, $\mathcal{E}$, should depend on the pit geometry. The surface energy is not well known for all of the materials systems. In our model, the nucleation of pits depends on $\gamma\, p^{2/3}$, the increase of energy due to the surface created when a pit nucleates and grows. This is the net change in surface energy of the pit minus that of a flat film. In the case of growth on (001) surfaces, this gives

$$\gamma\, p^{2/3} = \gamma_p\, A_p - \gamma_{(001)}\, A_{(001)} \tag{2.53}$$

where $A_p$ is the surface area of the pit and $A_{(001)}$ is its basal area in the (001) plane. $\gamma_p$ and $\gamma_{(001)}$ are the surface energies per unit area on the surface of the pit and (001) respectively. As surface energies for planes other than (001) are generally not known, we will assume that $\gamma_p \approx \gamma_{(001)}$ in order to provide an estimate of $\gamma$. This gives

$$\frac{\gamma}{\gamma_{(001)}} = \frac{A_{(001)}}{p^{2/3}}\left(\frac{A_p}{A_{(001)}} - 1\right). \tag{2.54}$$

Here $A_p / A_{(001)}$ and $A_{(001)}/p^{2/3}$ come from the geometry of the pits, which are obtained from experimental observations. Analytical shape optimization has been studied for islands [e.g. Tersoff and Tromp 1993] but not for pits and is beyond the scope of this study. For the purposes of our analysis, we will use the empirically observed pit shapes to determine these factors.

*a) The effect of materials system*

At intermediate mismatch, there is a small region of phase space where pits can nucleate adjacent to islands, as shown in Fig. 2.20(a). Figure 2.22(a) shows a 20 monolayers-thick $In_{0.27}Ga_{0.73}As/GaAs$ film grown at T = 500ºC and F = 2.2 Å/s the surface of which is covered with islands. Next to the islands pits are also observed. These pits form over a large range of growth conditions, but only after a significant number of islands have nucleated [Chokshi *et al.* 2000, 2002, Riposan *et al.* 2002, 2003]. For these growth conditions, $\mathcal{E}/kT \approx 0.13$, and the geometry of the pits indicates $\gamma \approx \gamma_{(001)}/6$. Using the value for $\gamma_{(001)}$ calculated by *ab initio* methods [Pelke *et al.* 1997, Moll *et al.* 1996], $\gamma' \approx 1.9$. Assuming that $\eta_e = 0.1$ and $\Lambda = 1$, this film is expected to reside in the type 0 region, where neither islands nor pits can nucleate and grow, very near the boundary of type IB, where pits may nucleate between islands, as shown by the diamond in Fig. 2.20(b). This small discrepancy may come from the uncertainty in the elastic and surface energies. It may also arise due to the fact that the stress inhomogeneities induced by the islands lead to a local increase in the strain at the island edge. In fact, finite element calculations show that the strain close to an island [Benabbas *et al.* 1999, Meixner *et al.* 2001] can be twice as high as the nominal mismatch. This strain concentration at the island edge would increase $\mathcal{E}$, pushing the system into regime IB.

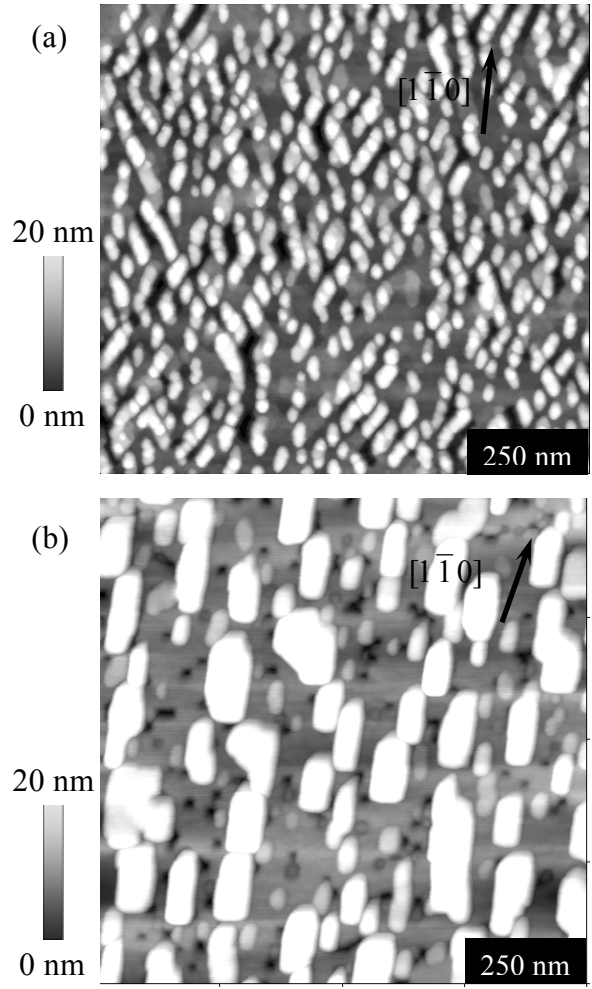

FIG. 2.22: AFM images of (a) a 22 ML thick In$_{.27}$Ga$_{.73}$As film grown on GaAs at 500°C and (b) a 7 ML thick InSb/InAs film grown at 400°C.

At higher mismatch the model predicts that pit nucleation may occur more readily, which has been observed experimentally. Figure 2.22(b) shows a 7 monolayers-thick InSb/InAs film grown at T = 400°C and F ≈ 1.3 Å/s the surface of which is covered with large rectangular islands, with small pits visible between them [Seshadri *et al.* 2000, 2002]. For these growth conditions, $\mathcal{E}/kT \approx 1.35$. Surface energies are not known for InSb but we estimate that the value of γ' for InSb/InAs is around 3.5 based upon the lower melting temperature of antimonides compared to arsenides. Thus, we predict that this film should reside in regime IA close to the IA-ID border, as denoted by the open square in Fig. 2.20(b), while experimentally the morphology is observed to be of type ID. This discrepancy is small considering the uncertainty on some of the parameters. Furthermore, the observed trend is correct, pitting is more favorable in InSb/InAs than in InGaAs/GaAs.

The model predicts that material systems with a high mismatch are more likely to form pits as the driving force (elastic energy relaxation) is higher. However, a high mismatch also implies a very low critical thickness. Pitting may thus be prevented by the lack of material to support a pit [Jesson *et al.* 2000]. As the growth mode in these films is Stransky-Krastanov, denuding the substrate is not energetically favorable. For this reason, a system such as InAs/GaAs would not show any pitting in spite of its high misfit, because the critical thickness is on the order of 1 or 2 monolayers. InSb/InAs can support pit nucleation despite its high mismatch because the film and substrate differ in their group-V species. When the InAs substrate is exposed as a pit grows, the volatile arsenic atoms have a high probability of desorbing. Since the overpressure consists of Sb vapor, the layer is converted to InSb. This results in an effective wetting layer that is infinitely thick, thus allowing for unhampered pit nucleation and growth [Seshadri *et al.* 2000, 2002].

*b) The effect of deposition rate and temperature*

In addition to studying pit nucleation in different materials systems, we can also choose one materials system and study the effect of parameters such as deposition rate, temperature and arsenic overpressure. Figure 2.23 shows the expected experimental regimes as a function of growth rate and deposition temperature for $\gamma' = 1.9$ and $\mathcal{E}/kT = 0.13$, which are close to the nominal values for $In_{0.27}Ga_{0.73}As/GaAs$. For this set of parameters, planar films are expected at low growth rates over a typical range of growth temperatures. At higher growth rates the nucleation of pits between islands is observed at lower temperatures (IB), while pit nucleation adjacent to islands is seen at higher temperatures (IA). These predictions are consistent with experimental results for In$_{0.27}$Ga$_{0.73}$As/GaAs [Chokshi *et al.* 2000, 2002, Riposan *et al.* 2002, 2003]. Figures 2.23(b) and (c) show a pair of atomic force micrographs of In$_{0.27}$Ga$_{0.73}$As/GaAs. The sample in Fig. 2.23(b) is 15 ML thick and was deposited at T = 505°C, F = 5 Å/s, and an As$_4$ overpressure of $12 \times 10^{-6}$ torr. The sample in Fig. 2.23(c) is 21 ML thick and was grown at T = 505°C, F = 0.7Å/s, and an As$_4$ overpressure of $16 \times 10^{-6}$ torr. For the high growth rate sample, pits are observed to nucleate between the islands. When the growth rate is decreased at the same temperature, pits are observed to nucleate adjacent to islands, consistent with the predictions of Fig. 2.23(a).

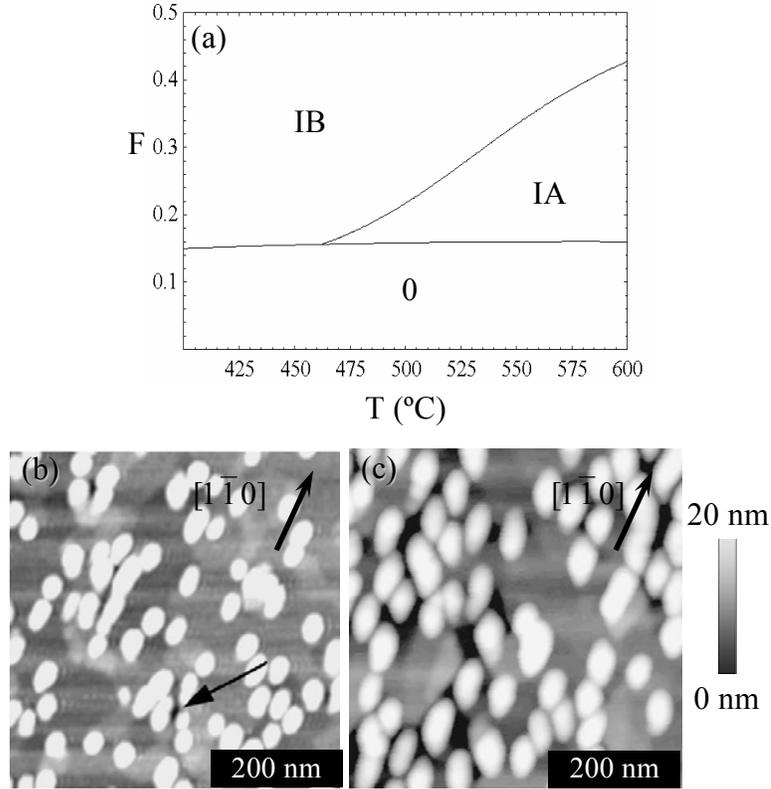

FIG. 2.23: (a) Domains of the various experimental regimes as a function of temperature and growth rate in arbitrary units for $\gamma' = 1.9$, $\mathcal{E}/kT = 0.13$, $\eta_e = 0.1$ and $\Lambda = 1$. AFM micrographs of $In_{0.27}Ga_{0.73}As/GaAs$ grown at T = 505°C and As overpressure = $12 \cdot 10^{-6}$ torr at (b) F = 1.75 ML/s, h = 15 ML and (c) F = 0.25 ML/s, h = 21 ML [Chokshi et al. 2002]. The arrow in (b) points to a pit that has nucleated inside a cluster of islands.

Experimentally, it has also been reported that island and pit nucleation depends on the arsenic overpressure such that at high arsenic overpressure, island and pit nucleation is delayed [Chokshi *et al.* 2000, 2002, Riposan *et al.* 2002, 2003]. The arsenic overpressure is known to have an effect on the chemical potential $\mu$, the surface energy and the diffusivity, however, the dependence of the diffusivity on As overpressure is not well established. It is therefore not possible to compare our model to experiments in terms of the effect of arsenic overpressure as changes in diffusivity cannot be taken into account with any precision. The fact that high arsenic overpressure delays island and pit nucleation suggests that changes in the diffusivity are the primary effect of arsenic overpressure.

*c) Summary*

This model agrees well with the observations in compound semiconductors, but pits have also been observed in SiGe systems. Jesson *et al.* for example observe cooperative nucleation of islands and pits in Si$_{0.5}$Ge$_{0.5}$ grown on Si(001) substrates at low temperature and annealed for 5 min at a T = 590ºC [Jesson *et al.* 1996]. Our model predicts that for these conditions the film should remain planar, in apparent contradiction with the experimental results. However, in our theoretical treatment the assumption was made that pits and islands nucleate independently. Jesson suggested a cooperative nucleation mechanism, i.e. a simultaneous nucleation of an island-pit pair. Such a mechanism is not taken into account in our model. Gray *et al.* show that for $Si_{1-x}Ge_x$ grown on Si(001) at T = 550°C, F = 1Å/s and 25 % < x < 50 %, shallow pits are observed to nucleate prior to the nucleation of islands [Gray *et al.* 2001, 2002, Vandervelde *et al.* 2003]. In contrast, our model predicts that islands should nucleate prior to pit nucleation assuming a symmetry in the aspect ratio of these features. The observations of Gray *et al.* is consistent with a morphology that may arise as a result of a localized surface instability [Shanahan and Spencer 2002] as opposed to a nucleation event. In neither of these systems does our model predict the observed morphologies, suggesting that the mechanisms which dominate in SiGe systems are different from those in compound semiconductors.

**9) Morphological correlations in III-V thin films**

In regimes IA and IB, where pits nucleate is related to islands, the nucleate adjacent to islands and between islands respectively. The data analysis in this section characterizes the emergence of a length scale in experiments by analyzing surface correlations in AFM images.

If we call h(**r**) the height of the film at a position **r** on the surface, then the correlation function C(**r**) is defined as

$$C(\mathbf{r}) = \frac{1}{A} \int_{surface} h(\mathbf{r}') \, h(\mathbf{r}' + \mathbf{r}) \, d\mathbf{r}' \qquad (2.55)$$

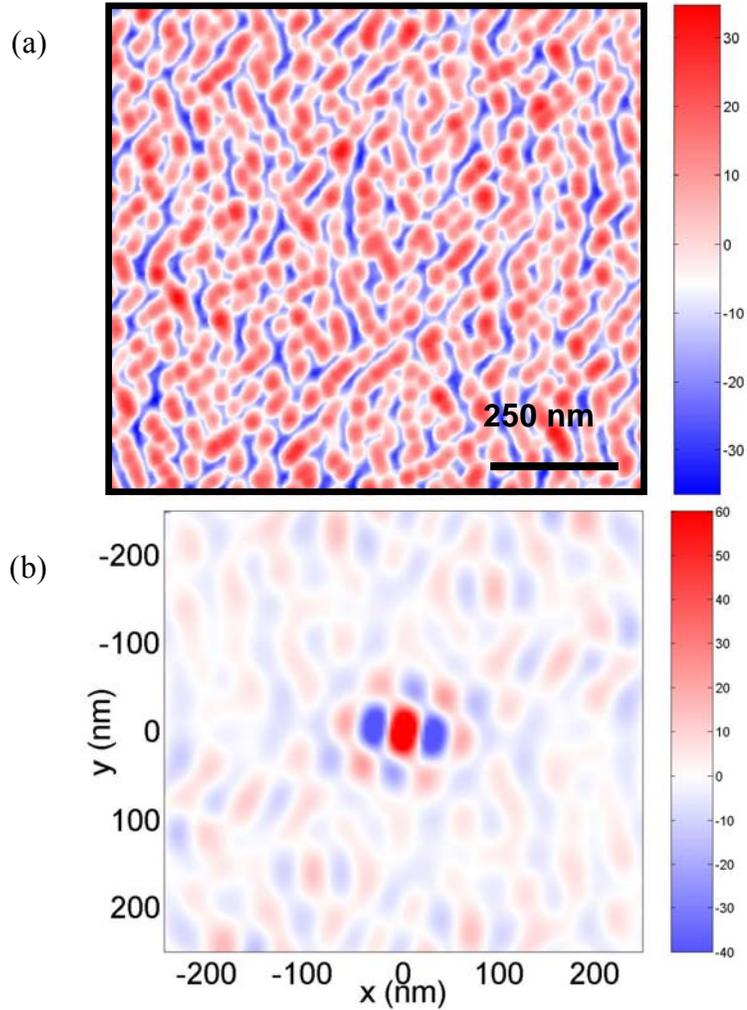

FIG. 2.24: (a) AFM image and (b) correlation map of a 22 ML thick In$_{.27}$Ga$_{.73}$As/GaAs film grown at 505°C, As BEP = 7.9 10$^{-6}$ torr.

where A is the surface area. For a given vector **r**, the height at **r'** and **r'** + **r** are compared. These points are "**r** apart" from each other. If both **r'** and **r'** + **r** are high or both **r'** and **r'** + **r** are low, they are correlated, in this case C is positive. If one is high and the other is low, they are anti-correlated (which is not synonymous with being uncorrelated), in this case C is negative. Plots of C will be referred to as correlation maps below. From AFM micrographs of In$_{.27}$Ga$_{.73}$As grown on GaAs(001) films, we can plot C(x, y) as a function of the position (x, y). For each (x, y) point, blue is negative and red is positive. Figure 2.24(b) gives an example of correlation map for an $In_{.27}Ga_{.73}As$GaAs film.

The correlation map in Fig. 2.24(b) shows a large correlated region at the center. This is a self-correlation which is indicative of the roughness of the film. Indeed at r = 0,

$$C(\mathbf{0}) = \frac{1}{A} \int_{surface} h^2(\mathbf{r'}) \, d\mathbf{r'} \qquad (2.56)$$

which is the square of the RMS roughness. The four blue regions around it are anticorrelated: they correspond to island-pit pairs and therefore show the presence of pits next to or between islands or vice-versa. Around them, forming a diamond are eight red regions; they are island-island or pit-pit correlations. This correlation map clearly shows a number of patterns: along directions close to <110> (the axes of both pictures in Fig. 2.24), red and blue alternate. This ordering is visible up to about 100-200 nm. Fig. 2.24(b) is anisotropic, with patterns aligned along the <110> directions (axes of the graph.) This is related to the anisotropy of surface diffusion.

Figure 2.25 shows AFM micrographs and correlation maps of 22 ML thick $In_{.27}Ga_{.73}As$/GaAs films grown at 505°C for 3 different arsenic overpressures: $12 \times 10^{-6}$ torr (a), $10 \times 10^{-6}$ torr (b) and $7.9 \times 10^{-6}$ torr (c). At high arsenic overpressure, Fig. 2.25 (a), the micrograph shows only isolated decorrelated islands and, apart from the central maximum, the correlation map shows little contrast. At lower arsenic overpressure, Fig. 2.25 (b), the micrograph shows a higher island density and some pits. Consistent with this, the correlation map shows more contrast and correlated and anticorrelated features alternate at a short scale. In the case of the lowest arsenic overpressure, Fig. 2.25 (c), the AFM picture shows closely packed islands and pits. In the correlation map, one can see patterns appearing as already seen in Fig. 2.24: alternating correlated and anticorrelated features. The contrast is stronger than in (b) and the extent of the pattern is greater. The AFM images, Fig. 2.25 (b) and (c), show that features are more organized at lower arsenic overpressure.

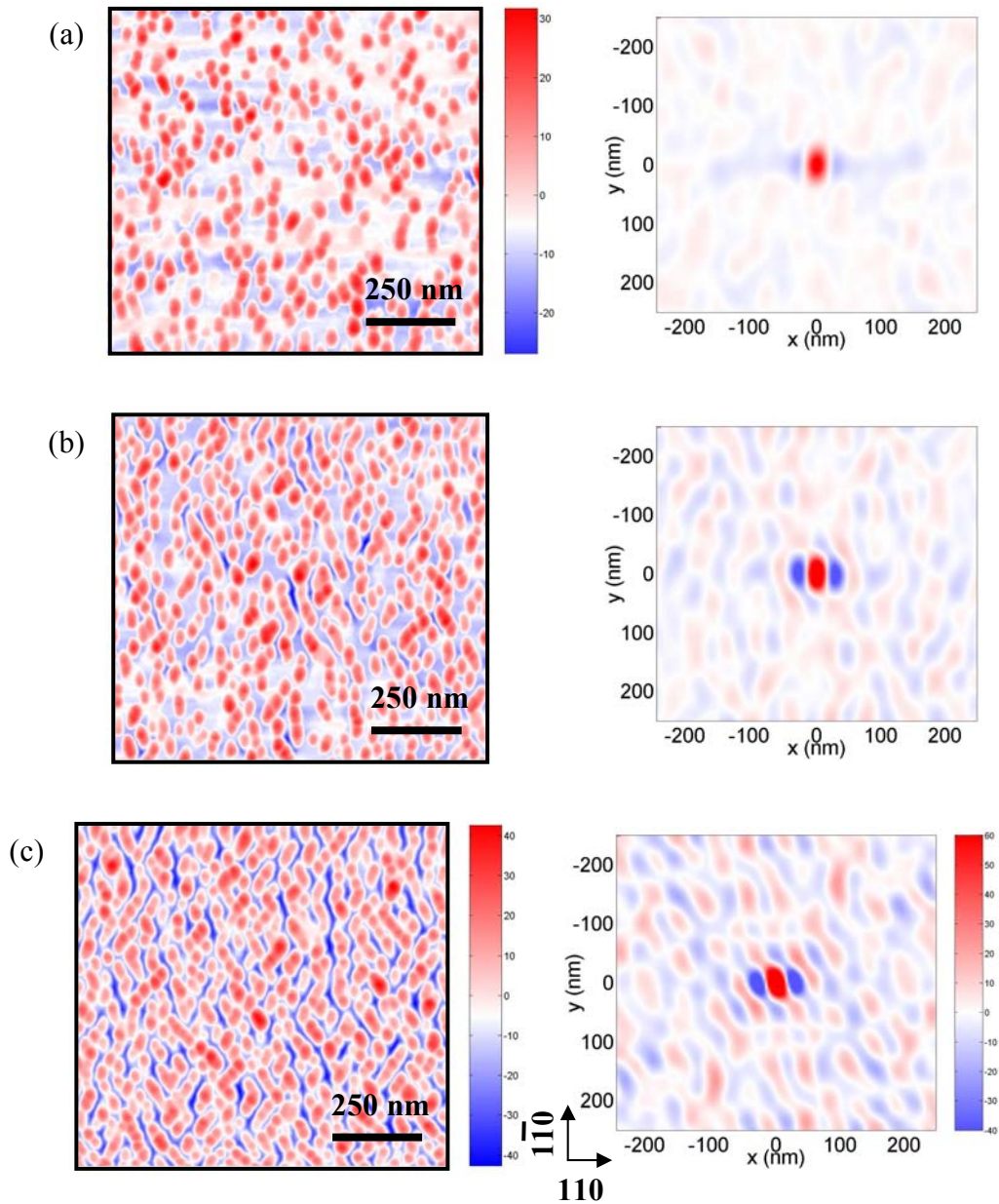

FIG. 2.25: AFM images and correlation maps of 22 ML thick In$_{.27}$Ga$_{.73}$As/GaAs films grown at 505ºC. As BEP = 12x10$^{-6}$ torr (a), 10x10$^{-6}$ torr (b), 7.9x10$^{-6}$ torr (c).

The two small blue regions in the correlation map of figure 2.25 (a) come from the definition of the 0 height. The average height is taken as a reference. This can be problematic because when there are islands and no pits, the average height is above the height of the flat areas. Thus the light blue regions in the correlation map correspond to the absence of correlation: next to an island there is not another island. Further away, heights are uncorrelated leading to a constant value of the correlation function.

Figure 2.26 shows linescans of the correlation maps of Fig. 2.25. It indicates how the surface changes as a function of arsenic overpressure. At low arsenic there is a deep minimum around 25-30 nm, indicative of pits, which is confirmed by the AFM micrographs of Fig. 2.25. There is no such minimum for the sample grown at $12 \times 10^{-6}$ torr and we can see that indeed there are no pits in Fig. 2.25 (a). We predicted that the island-pit distance should be proportional to the diffusion length. Arsenic overpressure can change the diffusivity and thus the diffusion length, but this dependence is not well known quantitatively. If the diffusivity at lower arsenic is higher, then the island-pit distance at low arsenic should also be higher. It is not possible to say if this result supports or contradicts the model since no quantitative theory exists that relates diffusivity and arsenic overpressure.

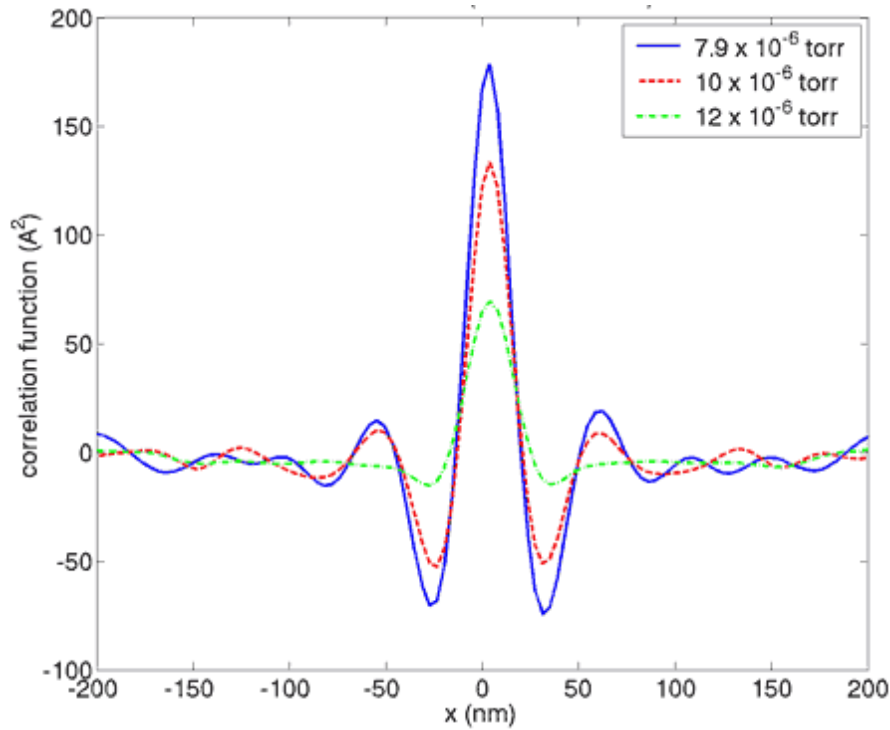

FIG. 2.26: Linescans of the correlation maps of Fig. 2.25 showing the evolution of the surface of 22 ML thick In$_{.27}$Ga$_{.73}$As/GaAs films grown at 505°C as a function of arsenic overpressure. The minimum around 30 nm is indicative of the presence of pits.

The comparison of this model with experiment shows qualitative agreement upon the role of $\mathcal{E}/kT$, the ratio of the volume energy to kT. When $\mathcal{E}/kT$ is low pits nucleate only at high island density — Fig. 2.22(a) — whereas at higher $\mathcal{E}/kT$ pits can nucleate even far from islands — Fig. 2.22(b). The correlation function has then been used to study $In_{.27}Ga_{.73}As/GaAs$. The presence or absence of minima in the correlation maps agrees with the presence or absence of pits in the AFM micrographs.

**10) Summary**

We have studied the nucleation of islands and pits during heteroepitaxial growth of semiconductors. In this model pit nucleation arises from a near-equilibrium nucleation process where the adatom concentration plays a major role. Elastic calculations show that pits can relieve elastic energy more efficiently than islands. However their nucleation is sensitive to the adatom concentration and can be altogether prevented by a high adatom concentration. Also the inhomogeneity of the adatom concentration due to diffusion favors pit nucleation close to the islands where the adatom concentration is lower. We found that while energetic arguments indicate that pits should dominate, they are typically kinetically prevented. Taking kinetics into account, we identified six experimental regimes depending on the growth rate and the elastic energy due to the misfit: pits can nucleate far from islands, adjacent to isolated islands, in between islands, or in the absence of islands. The film can also remain planar or islands alone can nucleate. There is reasonable agreement of the theory with experiments in III-V systems given the uncertainties in quantifying experimental parameters. Correlation functions show the emergence of ripple patterns due to the nucleation of pits close to islands.

# CHAPTER 3 — AN ATOMISTIC-CONTINUUM STUDY OF POINT DEFECTS IN SILICON

## 1) Introduction

Accurate modeling of coupled stress-diffusion problems requires that the effect of stress on the diffusivity and chemical potential of defects and dopants be quantified. Although the aggregate effects of stress on diffusion are readily observable, it is difficult to experimentally measure stress-induced changes in diffusivity and chemical potential. Despite these difficulties a number of careful measurements have been made regarding the effect of stress on diffusivities in model semiconductor systems [Zhao *et al.* 1999A, Zhao *et al.* 1999B], and the formation energies of vacancies have been measured in metals [Simmons and Balluffy 1960]. Due to the experimental challenges, an extensive literature has emerged regarding the numerical calculation of the formation energies of these defects using atomistic simulation [Antonelli *et al.* 1998, Antonelli and Bernholc 1989, Puska *et al.* 1998, Zywietz *et al.* 1998, Song *et al.* 1993, Tang *et al.* 1997, Al-Mushadani and Needs 2003]. Although early work used empirical potentials, more recent work has focused on the application of tight-binding and *ab initio* methods which are more accurate in modeling the alterations in bonding that occur at the defect. These calculations have been limited to a few hundred atoms due to the computational requirements of these methods.

This chapter addresses a number of unresolved issues in the application of atomistic simulations to accurately extract formation volumes and stress fields of point defects. In order to illustrate the methods that can be used to calculate the appropriate thermodynamic and elastic parameters from atomistic data we have performed calculations regarding a simple model point-defect, a vacancy in the Stillinger Weber [Stillinger and Weber 1985] model of silicon. An empirical model of silicon bonding was employed because it allows the exploration of a much larger range of system sizes than would have been possible using a more accurate model. Using an empirical potential precludes a quantitatively accurate measure of, for example, the formation volume of a vacancy in silicon since this model does not properly model the change in bonding that occurs at the vacancy. However the larger system sizes accessible *via* such a method are necessary to demonstrate a new technique for accurately calculating the prediction that does arise from the Stillinger Weber model of silicon and, by extension, in other atomistic potentials. This is critical since our goal is to make firm connections between the atomistic data and continuum concepts that, as we shall show, are not yet convergent on the scale of current *ab initio* calculations. This work paves the way for a multiscale modeling technique in which *ab intio*, atomistic and continuum concepts are used together to extract such quantities with predictive accuracy.

**2) Formation volume**

Chapter 1 introduced the free energy of activation which quantifies the effect of an external stress on the formation and migration of a defect in a crystal. This section focuses on the formation energy which is a part of the activation energy. The formation free energy determines the number of defects in the crystal. It comes from a change in internal energy $E^f$, a change in entropy $S^f$ (usually small) and a work term,

$$G^f = E^f - TS^f - \boldsymbol{\sigma}:\mathbf{V}^f = E^f - TS^f - W_{ext}, \tag{3.1}$$

where $\mathbf{V}^f$ is a tensor describing the change in volume and shape of the system and $W_{ext}$ is the work done by $\sigma$ on the system. The derivative of the free energy with respect to the externally applied stress provides the fundamental definition of this volume term.

Equation (3.1) shows that the free energy depends on the pressure through the work. However it may also be indirectly pressure-dependent if the internal energy depends on the pressure. The internal energy of formation can be split into two parts, an elastic part, $E^f_{LE}$, accounting for the elastic energy related to the crystal relaxation around the vacancy and a core energy, $E^f_{core}$, arising from broken bonds.

$$H^f = E^f_{LE} + E^f_{core} - \sigma : \mathbf{V}^f \qquad (3.2)$$

While the elastic part can be treated using linear elasticity, the core energy part must be treated atomistically. In linear elasticity, there is no interaction between internal and external stresses [Eshelby 1961] therefore $E_{LE}$ does not depend on $\sigma$. The core energy comes from the broken bonds and is therefore expected to be independent of the pressure. Therefore the only dependence of H upon $\sigma$ is from the $\sigma V$ term.

The formation volume is the change of volume of a system upon introduction of a defect. Let system 1 be a perfect crystal under some external stress $\sigma$ and system 2 the same crystal under the same external stress $\sigma$ to which a defect was added. The formation volume $V^f$ is the difference of volumes of the two systems. Similarly $E^f$ is the difference in internal energy between the two systems. The external stress contributes to the internal energy through the elastic energy $E^f_{LE}$, but since these two systems are under the same stress these contributions cancel out.

As described in chapter 1, the stress dependence of the formation of defects is of technological importance. This dependence is captured by the formation volume. If, for a given defect, $\mathbf{V^f}$ is 0 the concentration of this defect does not depend on the external stress. If a defect A has a positive formation volume and a defect B has a negative formation volume, under compression, the number of B defects increases and the number of A defects decreases. Under tension the number of A defects increases and the number of B defects decreases. If part of a film or of a device is under tension and another part is compressive, a segregation of species A and B can result. Therefore the behavior of the dopants/defects under stress depends upon the sign and magnitude of $V^f$.

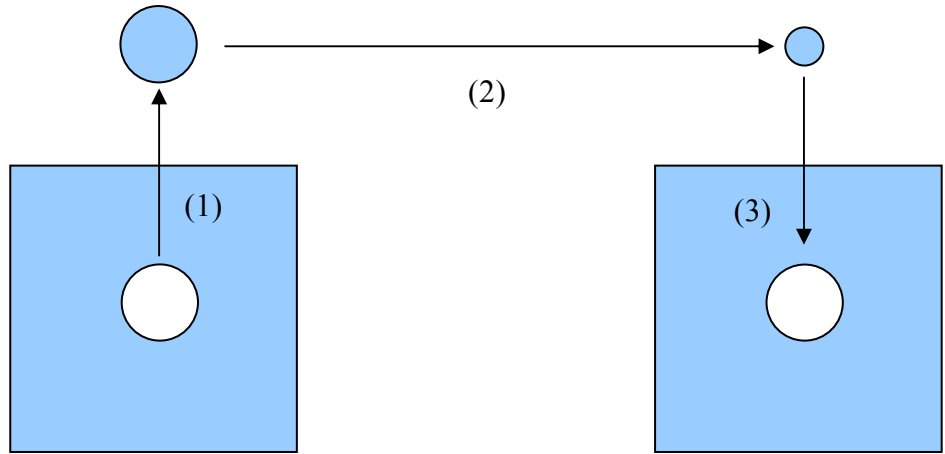

FIG. 3.1: Vacancy as an Eshelby inclusion. Part of the medium is removed (1). Its volume is decreased by $\mathbf{V^t}$ (2). It is reinserted into the medium (3).

Although the origin of the formation volume is atomistic in nature, a formulation in the context of continuum elasticity has also been adapted to interpret $\mathbf{V^f}$ in terms of an internal transformation of the material. This picture assumes the existence of a continuum defect that has a reference state independent of the surrounding crystal. Figure 3.1 shows the theoretical construction that would create such an "Eshelby inclusion" [Eshelby 1961]: material is removed from a continuous medium, the removed material undergoes a transformation described by a tensor $\mathbf{V^t}$, then it is reinserted into the medium. Upon reinsertion into the medium there will be elastic distortions both of the inclusion and of the crystal around it. It is worth noting that the change in volume (and potentially of shape) described by $\mathbf{V}^t$ is the change in shape and volume of the inclusion when not interacting elastically with the surrounding material. Thus $\mathbf{V}^t$ is not equal to the distortion of the inclusion because this distortion is affected by the elasticity of the medium. If the volume of the part of the medium which is removed decreases in step 2, upon reinsertion it will make the medium shrink. It is therefore called a center of contraction. If an external stress is applied, there will be an interaction between the center of contraction and the crystal.

The tensor, $\mathbf{V}^t$, is calculated by assuming a homogeneous strain over the transformed material, and multiplying this strain by the initial, scalar volume of this region. When a vacancy is to be represented by this continuum analogue, the scalar volume is often assumed to be the atomic volume, $\Omega$. In this interpretation the external work is exactly balanced by the work done to transform the inclusion against the external stress, $\sigma$, and can be shown to result in an external work $W_{ext} = \sigma : \mathbf{V}^t$. When evaluated at the boundary the strain field results in the change in volume and shape, $\mathbf{V}^t$, that must be equivalent to $\mathbf{V}^f$ to be consistent with the thermodynamic formulation. However, the arguments leading to $W_{ext} = \sigma : \mathbf{V}^t$ are meaningful only within continuum elasticity [Eshelby 1961], a theory that loses validity in the neighborhood of the defect. While the interpretation of $\mathbf{V}^t$ as a continuum transformation is not physically relevant for a point defect, this transformation can be used to calculate the elastic strain and stress fields in the vicinity of the transformation if $\mathbf{V}^f$ is known.

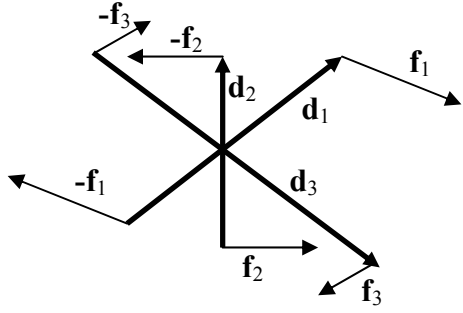

FIG. 3.2: Dipole representation of the point defect.

It is possible to extend the Eshelby inclusion model to point defects such as vacancies by shrinking the inclusion to a point. The elastic field of a vacancy is then modeled using a force dipole. This dipole is similar to an electric dipole. It is composed of forces an infinitesimally small distance apart. Since forces can act in any direction and be separated by a displacement in any direction the dipole is most generally represented by a tensor. This model can work even for a defect with an anisotropic stress field. When the dipole is proportional to the identity matrix it represents an isotropic center of contraction or expansion.

If three force pairs $\mathbf{f}_1$, $\mathbf{f}_2$ and $\mathbf{f}_3$ are applied at three points $\mathbf{d}_1$, $\mathbf{d}_2$ and $\mathbf{d}_3$ away from the vacancy (Fig. 3.2), the dipole is defined as [de Graeve 2002]

$$\mathbf{D} = \sum_i \mathbf{f}_i \otimes \mathbf{d}_i . \qquad (3.3)$$

These three vectors do not have to be orthogonal, they only have to be a basis of 3D space. Far away, when r >> d, the force field is

$$\mathbf{F}(\mathbf{r}) = -\mathbf{D}.\nabla_r [\delta(\mathbf{r})]. \qquad (3.4)$$

For the sake of simplicity, the origin of $\mathbf{r}$ is taken to be at the defect. The analytical form of Eq. (3.4) ensures that the sum of forces is 0. The sum of moments is

$$\int \mathbf{r} \times \mathbf{F}(\mathbf{r}) = (D_{32} - D_{23})\mathbf{e}_1 + (D_{13} - D_{31})\mathbf{e}_2 + (D_{21} - D_{12})\mathbf{e}_3 \qquad (3.5)$$

where $D_{ij}$ is the (i, j) component of the tensor **D** and ($e_i$) are unit vectors. Equilibrium requires that Eq. (3.5) be equal to 0 and thus that **D** be symmetric. The eigenvalues of a symmetric matrix are real and its eigenvectors can be chosen to be orthonormal. The dipole tensor can thus be written as

$$\mathbf{D} = \mathbf{R}.\mathbf{D'}.\mathbf{R}^{-1} = \mathbf{R}.\begin{pmatrix} f'_1 d'_1 & 0 & 0 \\ 0 & f'_2 d'_2 & 0 \\ 0 & 0 & f'_3 d'_3 \end{pmatrix}.\mathbf{R}^{-1} \qquad (3.6)$$

where **R** is a rotation matrix.

Having characterized the force distribution associated with the defect the displacement field can be derived from the elastic solution in a generalized elastic medium. The most general derivation of this kind is to compose the solution from the Green's function that satisfies the equation

$$C_{ijkl} \frac{\partial^2 G_{km}(\mathbf{r})}{\partial x_j \partial x_i} + \delta_{im} \delta(\mathbf{r}) = 0. \qquad (3.7)$$

Here $C_{ijkl}$ is the elastic modulus tensor of the solid and $G_{km}$ is the tensorial elastic Green's function. Once the Green's function is derived from Eq. (3.7), the displacement field can be expressed

$$\mathbf{u}(\mathbf{r}) = -\nabla_r [\mathbf{G}(\mathbf{r}).\mathbf{D}]. \qquad (3.8)$$

The resulting solution can be calculated from the expression [Barnett 1972, de Graeve 2002]

$$\mathbf{u}(\mathbf{r}) = \frac{1}{4\pi^2 \|\mathbf{r}\|^2} \int_0^\pi \left[ -(\mathbf{M}^{-1}.\mathbf{D}.\hat{\mathbf{r}}) + (\mathbf{J}.\mathbf{D}.\mathbf{z}) \right] d\psi. \qquad (3.9)$$

Here $\mathbf{M}^{-1}$ is the inverse of the matrix **M**, where **M** is defined by

$$M_{ir}(\mathbf{z}) = C_{ijrs} z_j z_s, \qquad (3.10)$$

$\hat{\mathbf{r}}$ is given by

$$\hat{\mathbf{r}} = \frac{\mathbf{r}}{\|\mathbf{r}\|}, \tag{3.11}$$

**J** is such that[2]

$$J_{ij} = C_{kp\ell n} M_{ik}^{-1} M_{\ell j}^{-1} \left(z_p \hat{r}_n + z_n \hat{r}_p\right) \tag{3.12}$$

and **z** is

$$\mathbf{z} = \begin{pmatrix} \cos\psi \sin\theta + \sin\psi \cos\theta \cos\varphi \\ -\cos\psi \cos\theta + \sin\psi \sin\theta \cos\varphi \\ -\sin\psi \sin\varphi \end{pmatrix} \tag{3.13}$$

where $\theta$ and $\varphi$ are the polar and azimuthal angles of **r**. The strain is

$$\varepsilon(\mathbf{r}) = \frac{1}{4\pi^2 \|\mathbf{r}\|^3} \int_0^\pi \left[ 2\left(\mathbf{M}^{-1}.\mathbf{D}.\hat{\mathbf{r}}\right)^s \otimes \hat{\mathbf{r}} - 2\left(\mathbf{J}.\mathbf{D}.\hat{\mathbf{r}} \otimes \mathbf{z} + \mathbf{J}.\mathbf{D}.\mathbf{z} \otimes \hat{\mathbf{r}}\right)^s + \left(\mathbf{A}.\mathbf{D}.\mathbf{z} \otimes \mathbf{z}\right)^s \right] d\psi$$

$$\tag{3.14}$$

where the "s" stands for symmetric, i.e.

$$\mathbf{A}^s = \frac{\mathbf{A} + \mathbf{A}^t}{2}. \tag{3.15}$$

If the medium is isotropic, Eq. (3.9) can be written in a closed form [Hirth and Lothe 1982]

$$\mathbf{u}(\mathbf{r}) = -\frac{\mathbf{D}.\hat{\mathbf{r}}}{4\pi C_{11} \|\mathbf{r}\|^2} \tag{3.16}$$

and the strains are

$$\varepsilon_{rr} = \frac{\partial u_r}{\partial \|\mathbf{r}\|} = \frac{(\mathbf{D}.\hat{\mathbf{r}}).\hat{\mathbf{r}}}{2\pi C_{11} \|\mathbf{r}\|^3} \tag{3.17}$$

---

[2] **J** is used here instead of **F** (notation used by Barnett) to avoid confusion with forces.

and

$$\varepsilon_{\theta\theta} = \varepsilon_{\varphi\varphi} = \frac{u_r}{\|\mathbf{r}\|} = -\frac{(\mathbf{D}.\hat{\mathbf{r}}).\hat{\mathbf{r}}}{4\pi C_{11} \|\mathbf{r}\|^3} . \tag{3.18}$$

The stresses then are

$$\sigma_{rr} = C_{11} \varepsilon_{rr} + C_{12} \varepsilon_{\theta\theta} + C_{12} \varepsilon_{\varphi\varphi} = \frac{C_{11} - C_{12}}{2} \frac{(\mathbf{D}.\hat{\mathbf{r}}).\hat{\mathbf{r}}}{\pi C_{11} \|\mathbf{r}\|^3} \tag{3.19}$$

and

$$\sigma_{\theta\theta} = \sigma_{\varphi\varphi} = C_{12} \varepsilon_{rr} + (C_{11} + C_{12}) \varepsilon_{\theta\theta} = -\frac{C_{11} - C_{12}}{2} \frac{(\mathbf{D}.\hat{\mathbf{r}}).\hat{\mathbf{r}}}{2\pi C_{11} \|\mathbf{r}\|^3} . \tag{3.20}$$

As we assume isotropy, $\frac{C_{11} - C_{12}}{2}$ is equal to $C_{44}$. So (keeping in mind that $C_{44}$ actually means "isotropic $C_{44}$", i.e. $\frac{C_{11} - C_{12}}{2}$), we can write

$$\sigma_{rr} = \frac{C_{44}}{C_{11}} \frac{(\mathbf{D}.\hat{\mathbf{r}}).\hat{\mathbf{r}}}{\pi \|\mathbf{r}\|^3} \tag{3.21}$$

and

$$\sigma_{\theta\theta} = \sigma_{\varphi\varphi} = -\frac{C_{44}}{C_{11}} \frac{(\mathbf{D}.\hat{\mathbf{r}}).\hat{\mathbf{r}}}{2\pi \|\mathbf{r}\|^3} . \tag{3.22}$$

The radial force on an area A a distance r from the defect is then

$$F = \sigma_{rr} A = \frac{A}{\pi} \frac{C_{44}}{C_{11}} \frac{(\mathbf{D}.\hat{\mathbf{r}}).\hat{\mathbf{r}}}{\|\mathbf{r}\|^3} \tag{3.23}$$

where A is the surface of the atom on which the force applies.

*A priori*, the dipole may not be enough to represent any point defect and higher order terms, such as a quadrupole, may be necessary. However, results for the vacancy show that the dipole is a good description of this point defect. It is possible that more complicated defects or clusters require a quadrupole term. In any case, the contribution of higher order terms to the stress field should die off faster than the dipole and may be noticeable only close to the defect.

**3) Calculating the formation volume**

*a) Change of volume of the simulation cell*

The most common method used to extract the formation volumes of defects has been the direct measurement of the change in volume of the relaxed supercell upon the introduction of the defect [Zhao *et al.* 1999A, Zhao *et al.* 1999B]. This is a rigorously correct method of calculating the formation volume given two assumptions: that the core energy, defined in Eq. (3.2), is not pressure dependent and that the supercell size is sufficiently large such that defect-defect interactions have a negligible effect on the elastic relaxation of the cell. The former is typically a good assumption. The latter may not always be a good assumption for the small supercell sizes typically simulated by *ab initio* calculation. The vacancy-vacancy interaction will be shown to have a negligible effect even for small systems; however this may not be the case for other defects, in particular the anisotropic ones.

*b) Obtaining the dipole from positions and forces*

Although the above elastic analysis provides a means to calculate the displacement and stress fields around a defect of elastic dipole D, it does not provide a means to extract this dipole value. The dipole value can however be extracted from the forces on the atoms surrounding the defect. In an isotropic medium, the radial force expected on atom n from the dipole is

$$\mathbf{F'}^n = A^n \frac{C_{44}}{C_{11}} \frac{\mathbf{D}.\mathbf{r}^n}{\pi \|\mathbf{r}^n\|^4} \tag{3.24}$$

where $\mathbf{r}^n$ is the position of atom n relative to the center of the defect. This provides the forces as a function of the dipole. In fact the dipole is unknown and the forces can be obtained form atomistics. Equation (3.24) must be somehow inverted to have the dipole as a function of the forces. We define the vector $\mathbf{\Delta}^n$ as the difference between the actual force on atom n, $\mathbf{F}^n$ (obtained from atomistic simulations) and the radial force expected from the dipole in an isotropic medium, $\mathbf{F'}^n$,

$$\mathbf{\Delta}^n = \mathbf{F}^n - \mathbf{F'}^n = \mathbf{F}^n - A^n \frac{C_{44}}{C_{11}} \frac{\mathbf{D}.\mathbf{r}^n}{\pi \|\mathbf{r}^n\|^4}. \tag{3.25}$$

We then define the scalar $\Delta$ by

$$\Delta^2 = \sum_n (\mathbf{\Delta}^n)^2 = \sum_n \left( \mathbf{F}^n - A^n \frac{C_{44}}{C_{11}} \frac{\mathbf{D}.\mathbf{r}^n}{\pi \|\mathbf{r}^n\|^4} \right)^2. \tag{3.26}$$

If the representation of a vacancy (as a center of contraction) in elasticity and its atomistic counterpart were in perfect correspondence, $\Delta$ would be 0. But since $\mathbf{D}$ has 6 components, while there are 3n forces and 3n positions it is not generally possible to find a $\mathbf{D}$ that satisfies the condition $\Delta = 0$. We therefore pick the tensor $\mathbf{D}$ which minimizes $\Delta^2$. To this end, we calculate the derivatives of $\Delta^2$ with respect to the components of $\mathbf{D}$

$$\frac{\partial \Delta^2}{\partial D_{ij}} = \frac{2}{\pi} \frac{C_{44}}{C_{11}} \sum_n A^n \left( -\frac{F_i^n r_j^n}{\|\mathbf{r}^n\|^4} + \sum_k A^n \frac{C_{44}}{C_{11}} \frac{D_{ik} r_k^n r_j^n}{\pi \|\mathbf{r}^n\|^8} \right) = 0. \tag{3.27}$$

This gives

$$\sum_n A^n \frac{F_i^n \, r_j^n}{\|\mathbf{r}^n\|^4} = \sum_n \sum_k \left(A^n\right)^2 \frac{C_{44}}{C_{11}} \frac{D_{ik} \, r_k^n \, r_j^n}{\pi \|\mathbf{r}^n\|^8} \, . \tag{3.28}$$

Letting

$$\mathbf{X} = \sum_n A^n \frac{\mathbf{F}^n \otimes \mathbf{r}^n}{\|\mathbf{r}^n\|^4} \tag{3.29}$$

and

$$\mathbf{Y} = \frac{1}{\pi} \frac{C_{44}}{C_{11}} \sum_n \left(A^n\right)^2 \frac{\mathbf{r}^n \otimes \mathbf{r}^n}{\|\mathbf{r}^n\|^8}, \tag{3.30}$$

Eq. (3.28) can be rewritten as [de Graeve 2002]

$$\mathbf{D} = \mathbf{X}.\mathbf{Y}^{-1}. \tag{3.31}$$

Equation (3.31) provides a means to calculate the value of the dipole using the positions of and the forces on the atoms from atomistics. It is a closed form solution for a generalized defect in an isotropic medium. We will take the sum in Eqs. (3.29) and (3.30) to be over atoms on a cubic shell, as shown in Fig. 3.3.

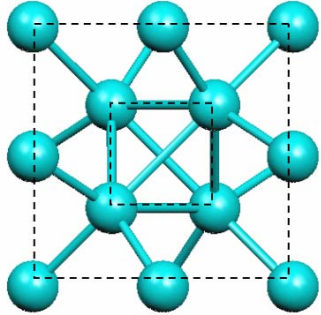

FIG. 3.3: Cubic shells used to calculate the dipole.

From Eqs. (3.29) and (3.30) the forces on atoms are needed to calculate the dipole. However, at equilibrium the net force on any atom is zero. Figure 3.4 shows, in black, an atom belonging to the shell. If all atoms within the shell (white atoms) were removed the only force remaining would be the force from the atoms outside the shell (gray atoms). For the black atom to be at equilibrium, the traction due to atoms inside the shell (wide arrow) must cancel out the traction from the atoms outside the shell. Thus the traction across the surface of the shell due to the vacancy is negative the force on this atom from atoms outside the shell.

*c) Simulation techniques*

So far an expression was obtained for the dipole as a function of positions and forces extracted from atomic simulations. The question of the choice of the technique to use in atomic simulations to obtain forces remains. We now introduce several atomic simulation techniques and compare their strengths and weaknesses.

The families of representations of materials are *ab initio*, tight-binding and empirical potentials. In *ab initio* simulations, the Schrödinger equation is solved (under some assumptions) [Kohn and Sham 1965, Kohn 1999]. These calculations are intrinsically quantum mechanical, which makes them very accurate. However they are computationally intensive which prevents the simulation of large systems. Empirical potentials eliminate the electronic degrees of freedom. The force from one atom on another atom is calculated as a function of their separation distance and the location of surrounding atoms. The expression for this potential is not theoretically derived but some insight from quantum mechanics may be used in motivating these expressions. Empirical potentials have parameters which are fitted to the established properties of the material of interest (known experimentally or from *ab initio* calculations). As a result, while the lattice parameters and cohesive energies are outputs of *ab initio* calculations, they are inputs for empirical potentials. Empirical potentials can be considered to be mostly interpolations between known properties of the material in question (relative energies of crystal structures, elastic properties, etc.) Therefore predictions which rely on aspects of the potential far from the fitting regime are not quantitatively reliable. Tight-binding [Slater and Koster 1954, Goodwin *et al.* 1989] is another simulation technique, it uses a very simplified quantum mechanical description of the atoms. This makes these simulations simpler to implement, less computationally-intensive but also less accurate than *ab initio* calculations. Their relative simplicity also allows for larger systems than *ab initio*. Therefore, both in terms of system size and of accuracy, tight-binding (TB) is intermediate between empirical potentials and *ab initio*.

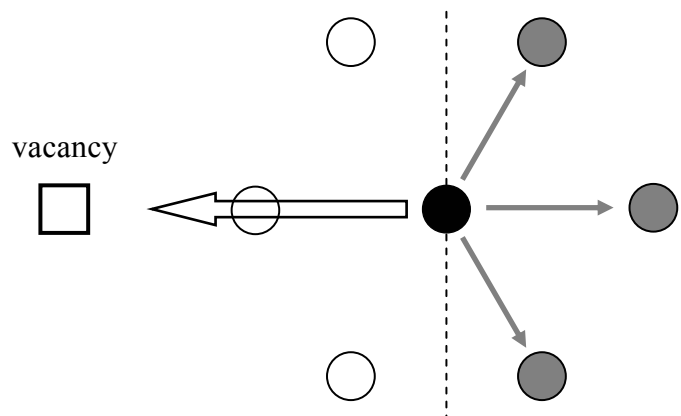

FIG. 3.4: Force on an atom belonging to the shell (black atom) from the atoms outside the shell (gray atoms) and "force from the vacancy" (wide arrow).

Stillinger and Weber [Stillinger and Weber 1985] designed an empirical potential to study the melting of silicon. Due to the covalent nature of silicon bonds, a mere two-body term does not suffice because the energy would then be proportional to the number of bonds which would drive the system to a close-packed structure. Stillinger Weber (SW) potential uses both two-body and three body terms:

$$\Phi = \sum_{i<j} v_2(r_{ij}) + \sum_{i<j<k} v_3(r_{ij}, r_{jk}, \theta_{jik}). \qquad (3.32)$$

The first summation is over pairs of atoms and the second is over triples. For a given pair of atoms, the two-body term depends only on the distance r between the atoms,

$$v_2(r) = \varepsilon A \left[ B\left(\frac{r}{\sigma}\right)^{-4} - 1 \right] \exp\left(\frac{1}{r/\sigma - a}\right) \qquad (3.33)$$

where $\varepsilon$, A, B and $\sigma$ are positive constants. The exponential term drives $v_2$ to 0 when r/$\sigma$ approaches the constant a from below. $\sigma$a is therefore a cut-off distance. Equation (3.33) applies when r/$\sigma$ < a and $v_2$ is set to 0 when r/$\sigma$ > a. The value of a is chosen such that the cut-off occurs between first and second nearest neighbors, as a consequence there is no two-body interaction between second nearest neighbors. Whereas, physically, atoms further apart contribute to the energy limiting two-body interactions to first nearest-neighbors simplifies the relationship between the model parameters and many properties such as lattice parameters and bond lengths for various crystal structures, elastic constants. This greatly simplifies the routine to optimize the parameters.

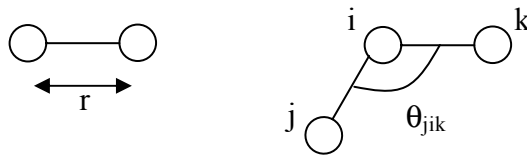

FIG. 3.5: The two-body terms only depend on the interatomic distance while the three-body terms also account for the angle between the bonds.

The three-body term models aspects of the sp$^3$ bonding that cannot be adequately described by two-body interactions. In this term the energy depends both on distances and angles, as shown in Fig. 3.5,

$$v_3(r_{ij}, r_{ik}, r_{jk}, \theta_{ijk}) = \varepsilon\, h(r_{ij}, r_{ik}, \theta_{jik}) + \varepsilon\, h(r_{ji}, r_{jk}, \theta_{ijk}) + \varepsilon\, h(r_{ki}, r_{kj}, \theta_{ikj}) \qquad (3.34)$$

where

$$h(r_1, r_2, \theta) = \lambda \exp\left(\frac{\gamma}{r_1/\sigma - a} + \frac{\gamma}{r_2/\sigma - a}\right)\left(\cos\theta + \frac{1}{3}\right)^2 \qquad (3.35)$$

and λ and γ are positive constants. Again the exponential plays the role of a cut-off and h is zero when $r_1$ or $r_2$ is greater than σa. Notice that in the case of a perfect diamond cubic crystal, cos θ = -1/3 due to the tetrahedral symmetry and the three-body terms do not contribute to the energy.

When an atom is removed to form a vacancy, its first nearest-neighbors can relax (generally inward). Different simulation techniques predict different amounts of relaxation and different formation energy. Table 3.1 shows the range of formation energy of a vacancy and of the radial component of the displacement of the first nearest neighbors from experiment, *ab initio,* tight binding, Stillinger Weber and other empirical potentials. Space group $T_d$ corresponds to a radial displacement of the first nearest-neighbors while in $D_{2d}$ there is a pairing of nearest-neighbors which form two dimers with the distance between the two atoms of a dimer smaller than the distance from atoms of the other dimer. The formation energy obtained by *ab initio* calculations is not very wide-ranged and is consistent with experimental results. The displacement of the first nearest neighbors, on the other hand, can vary greatly (by a factor of two.) In *ab initio* simulations for instance it varies between -0.48 Å and -0.22 Å. Pushka and coworkers also found the symmetry to be either $D_{2d}$ or $T_d$ depending on the size of their system [Pushka *et al.* 1998]. This indicates that energy converges faster than geometry and that geometric data, such as formation volumes, cannot be obtained with small systems. According to empirical potentials, the first nearest neighbors may move inwards or outwards. These simulations are the least reliable because the potential are fitted to perfect crystals and therefore poorly model the changes in bonding near a defect.

| technique | references | space group | energy (eV) | displacement (Å) |
|---|---|---|---|---|
| experiment | Watkins 1964; Dannefaer 1986 | | 3.6 ± 0.2 | |
| *ab initio* | Antonelli 1989, 1998; Zhu 1996; Puska 1998; Zywietz 1998 | $D_{2d}$* | 3.3 → 3.65 | -0.48 → -0.22 |
| tight binding | Song 1993; Lenosky 1997; Tang 1997; Munro 1999 | $D_{2d}$ | 3.68 → 5.24 | -0.50 → -0.42 |
| SW | Stillinger and Weber 1985 | $T_d$ | 2.82 | - 0.56 |
| other potentials | Balamane 1992 | $T_d$ | 2.82 → 3.70 | -0.51 → +0.24 |

Table 3.1: Formation energy of a vacancy and displacement of the first nearest neighbors from experiment, ab initio, tight binding, Stillinger Weber and other empirical potentials. *: there exist a few reports of $T_d$ symmetry.

Any technique, be it experimental or computational, has limitations. It is therefore not always possible to use only one technique. Simulation techniques can be limited in two ways: accuracy and computational cost. The most accurate techniques being the most computationally intensive, they are limited to small systems. Computationally less intensive techniques on the other hand are not efficient far from equilibrium, in particular where the lattice is distorted (defects, surfaces.) Multiscale modeling of materials aims to bring two (or more) different techniques together, each providing its specific strength(s) and compensating for the weakness(es) of the other technique. One possibility is to use several techniques within the same simulation: *ab initio* is used where accuracy is needed and an empirical potential is used where structural changes are not expected to occur. This provides a means to increase the system size without increasing the computational cost significantly. A slightly different kind of simulation uses atomistics close to a singularity (crack tip, defect, indenter) and continuum mechanics for the rest of the system [Shilkrot *et al.* 2002].

## 4) Results

*a) Atomistic results*

The dipole tensor, **D**, gives the magnitude and anisotropy of the center of contraction and cannot be obtained by elasticity, but must be determined by the microscopic structure of the point defect. A number of different techniques have been used to characterize the relaxation around a point defect. One typical method is to note the relaxation of the nearest neighbor atoms. However this method is not effective for describing the asymptotic elastic relaxation in the vicinity of the defect, which is important for accurately calculating the relaxation volume, i.e. the quantity necessary to predict the thermodynamic response of the defect to stress. We detail here a systematic method for extracting the relaxation around the vacancy. One method to obtain **D** would be to fit the displacement curve as a whole to the asymptotic elastic solution. While this is feasible it is not an efficient way to proceed and involves fitting the curve in regions close to the defect and close to the periodic boundary where the solution in an infinite medium cannot be expected to apply. Rather we obtain **D** from Eq. (3.31), i.e. we find the value of the dipole that provides a best fit to the forces obtained form atomistics.

*b) Isotropy of the vacancy in silicon*

Equations (3.9) to (3.31) make no assumptions as to the isotropy of the dipole although (3.16) to (3.31) do assume an isotropic elastic medium. Equilibrium only requires that the tensor be symmetric to ensure that there is no net moment. However, conjugate gradient (CG) calculations show that the actual dipole of a vacancy, as may be expected, is nearly isotropic. Figure 3.6(a) shows the ratio of off-diagonal term of **D** to diagonal terms of **D**. Far from both the vacancy and the boundaries the dipole is very close to being diagonal. At any shell the non-diagonal terms are never more than a few percent of the diagonal terms. Figure 3.6(b) shows the standard deviation for the diagonal terms of **D** normalized by the trace of **D** as a function of the shell where **D** is calculated. When **D** is calculated far from the vacancy the standard deviation is less than 0.1 % of the trace and the three diagonal terms are essentially equal. Thus for shells far enough from the vacancy, the dipole is nearly proportional to the identity tensor. An example of such a tensor (in eV) is

$$\mathbf{D} = \begin{pmatrix} 8.506 & 3.2 \times 10^{-3} & -1.7 \times 10^{-3} \\ -9.4 \times 10^{-3} & 8.501 & -5.4 \times 10^{-3} \\ -0.2 \times 10^{-3} & -2.4 \times 10^{-3} & 8.506 \end{pmatrix}. \tag{3.36}$$

Therefore, we can write the dipole as

$$\mathbf{D} = D\,\mathbf{I} \qquad (3.37)$$

where D is a scalar and **I** is the identity tensor. In what follows, when we refer to the dipole, we will be referring to the scalar D.

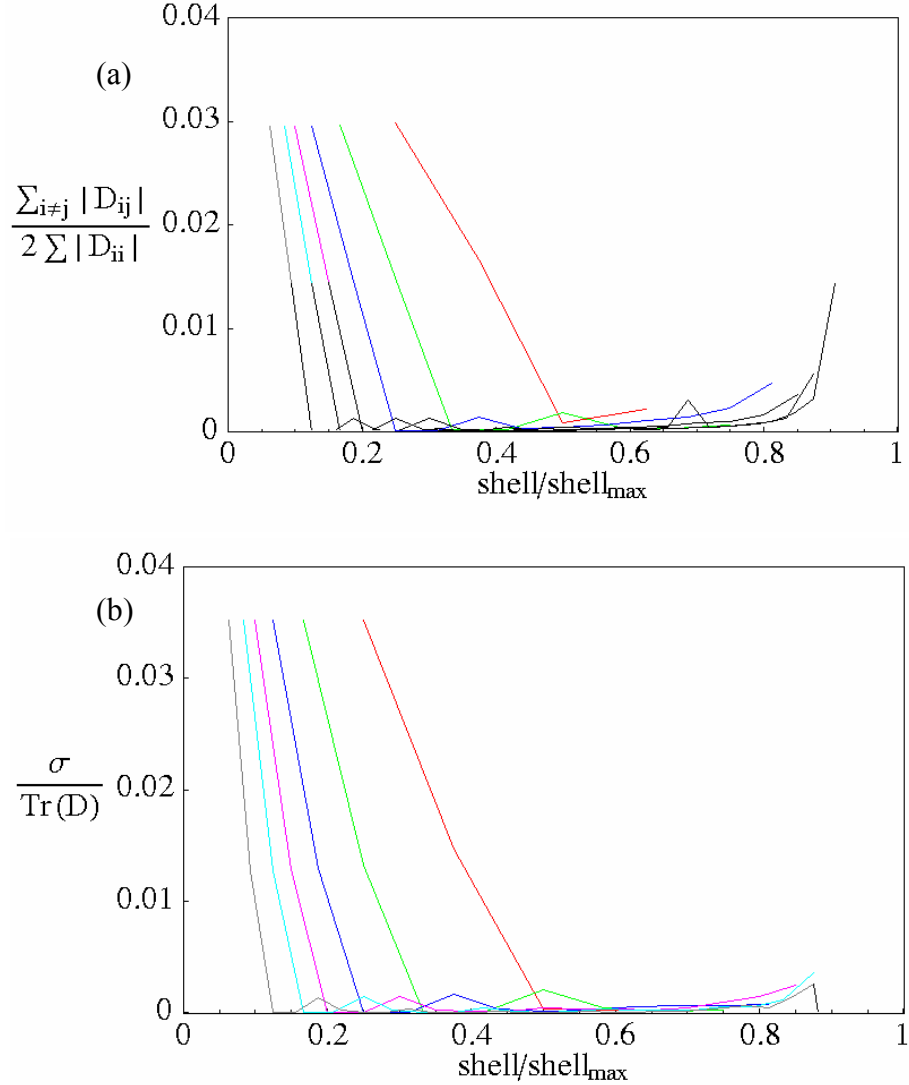

FIG. 3.6: The ratio of off-diagonal terms of **D** to diagonal terms (a) and the standard deviation for the diagonal terms of **D** (b) as a function of the shell where **D** is calculated. Plotted for 512, 1 728, 4 096, 8 000, 13 824 and 32 768 atoms.

*c) Displacement field and formation volume*

Once extracted the value of the dipole can be used to calculate the formation volume. The simplest case to calculate is an isotropic elastic sphere of radius R, where the radial displacement is given by the expression:

$$u_r(r) = \frac{D}{4\pi r^2 C_{11}}\left[1 + 2\frac{C_{11} - C_{12}}{C_{11} + 2C_{12}}\left(\frac{r}{R}\right)^3\right]. \tag{3.38}$$

While the first term arises directly from the asymptotic elastic field from Eq. (3.16), the second term is imposed by the free boundary at R. From Eq. (3.38) it follows that the measured formation volume is related directly to the displacement at the outer boundary

$$V^f = 4\pi R^2 u_r(R) = \frac{3D}{C_{11} + 2C_{12}}. \tag{3.39}$$

Note that $V^f$ is independent of R. For a large system, where continuum elasticity applies, the formation volume is independent of the size of the system. Since a large cube should not be different from a large sphere, Eq. (3.39) is expected to hold for any isotropic system where finite size effects can be neglected, independent of geometry.

The dipole values that were obtained from a series of conjugate gradient calculations using the Stillinger-Weber model ranging in size from 512 atoms to 32 768 atoms. The value of D was calculated on concentric shells around the defect. The shell of first nearest neighbors of the vacancy is not used: since there is nothing strictly inside this shell, the external force is 0 at equilibrium. Shells too close to the vacancy show evidence of discreteness effects. This is to be expected since continuum elasticity does not apply down to the atomic scale. Ten samples were used for each system size except the larger ones since they were more computationally-intensive. In some cases the simulations converge to distinct vacancy structures with different formation energies, different formation volumes and different dipole values. Figure 3.7 shows the value of the dipole as a function of the shell where it is calculated for 30 samples made of 4 096 atoms. There are two kinds of curves corresponding to two structures of the vacancy. Within each structure there exists some variation of the properties. Only simulations leading to the lowest energy structure were considered to plot the figures (other than Fig. 3.7) in this chapter. In all figures bearing error bars, the error bars are sample-to-sample variations among the samples of the lowest-energy structure. Therefore they do not account for systematic errors due to system size effects. Table 3.2 shows the number of lowest energy samples obtained for each system size.

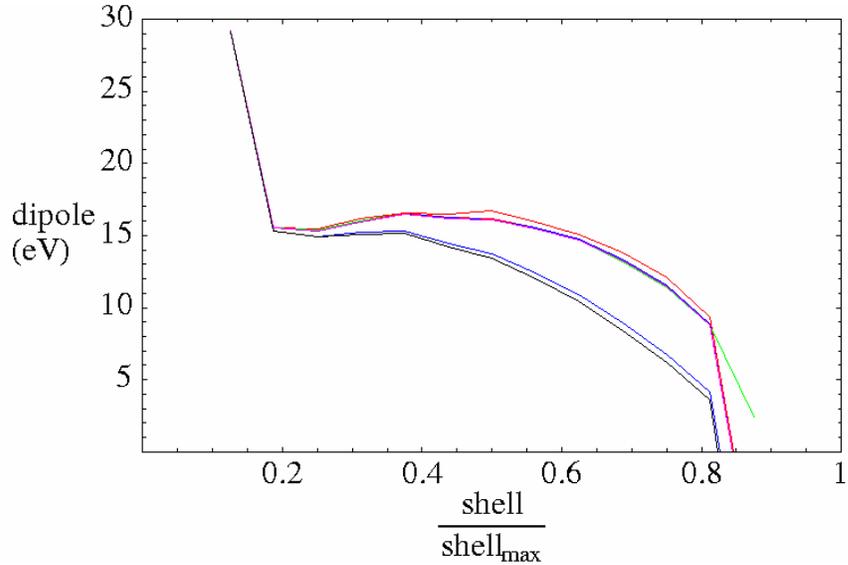

FIG. 3.7: Dipole values as a function of the cubic shell at which the force data is extracted for 30 samples made of 4 096 atoms.

| system size | 512 | 1 728 | 4 096 | 8 000 | 13 824 | 32 768 |
|---|---|---|---|---|---|---|
| number of samples | 10 | 10 | 10 | 10 | 6 | 4 |

Table 3.2: Number of sample used for each system size.

Figure 3.8(a) shows the dipole values as a function of the shell at which it is calculated. Figure 3.8(b) shows the dipole as a function of the shell over shell$_{max}$, where shell$_{max}$ is half the vacancy-vacancy distance. Thus shell/shell$_{max}$ varies between 0 at the

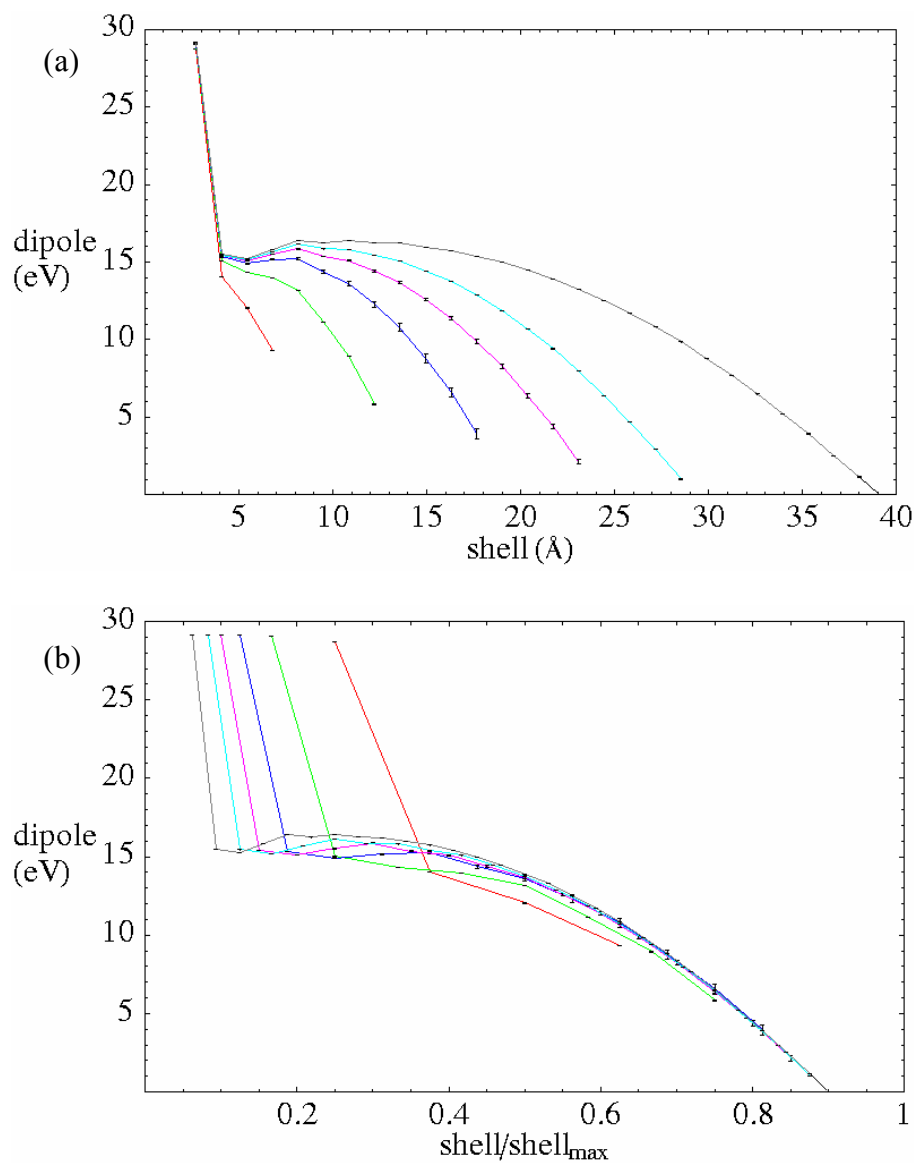

FIG. 3.8: Dipole values as a function of the cubic shell at which the data are extracted (a) and as a function of the ratio of the shell to the largest shell (b). Each shell is numbered by the distance that separates the closest atom in the shell from the vacancy. The error bars are from sample-to-sample standard deviation. Plotted for 512, 1 728, 4 096, 8 000, 13 824 and 32 768 atoms.

vacancy and 1 at the boundary of the simulation cell. The fact, shown in Fig. 3.8(b), that the curves for different systems sizes are close together far from the vacancy is an indication that linear elasticity applies there. The error bars correspond to sample-to-sample standard deviation. The shell of the first nearest neighbors was not plotted as indicated above and the shell of second nearest neighbors gives a very high dipole due to finite size effects. The third to fifth shells give a fairly low dipole value, again a finite size effect. The sixth shell and above form a plateau where the dipole is almost constant. For shells further out, the boundary has an increasingly important influence and the dipole decreases. The sixth and seventh shells will be used to extract the dipole because they are the smaller shells without finite size effects. For small systems, 512 and 1 728 atoms, there is no evident plateau since there is no region far enough from both the defect and the boundary.

We can now use the dipole extracted from the atomistic simulations to obtain the formation volume from Eq. (3.39) and Stillinger Weber elastic constants. This volume is plotted in Fig. 3.9 along with the direct measurements of the change of volume of the simulation supercell. The calculated formation volume does not match the relaxation of the simulation cell. The reason for this discrepancy is that the calculated formation volume assumes that the system is isotropic. Since there is no closed-form expression for the stress field in the anisotropic case a fully anisotropic calculation would be much more complicated. In the next sections a method will be discussed to correct for anisotropy when calculating the dipole value from isotropic equations.

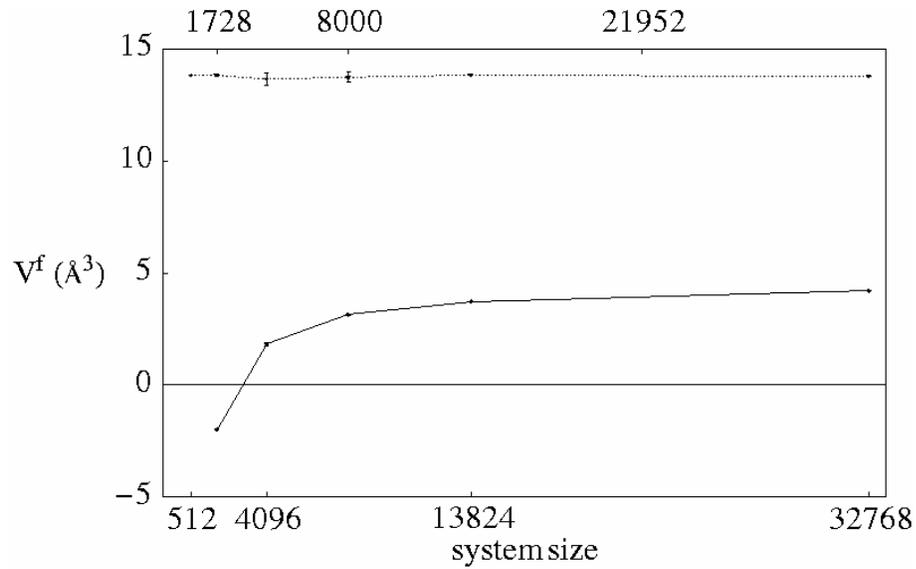

FIG. 3.9: The formation volume versus the system size measured both using Eq. (3.39) (solid line) and from direct measurements of the change of volume of the simulation supercell (dashed line). The error bars correspond to sample-to-sample standard deviation; they do not account for systematic errors.

*d) Finite element calculations*

In order to correct for the assumption of isotropy made in Eqs. (3.29) and (3.30) it is necessary to calculate a value of $D/V^f$ appropriate for determining the formation volume in the anisotropic medium given the dipole extracted assuming an isotropic medium. This has been addressed by a series of finite element (FE) calculations in which the stress field around the defect was obtained and related to the volumetric relaxation of the box [Bouville *et al.* 2004D]. Obtaining the relationship between the extracted dipole and the formation volume required a convergence study of the solution with respect to the refinement of the discretization. The constitutive behavior of the mesh was taken from the anisotropic (cubic) elastic moduli of the Stilinger Weber potential [Balamane *et al.* 1992].

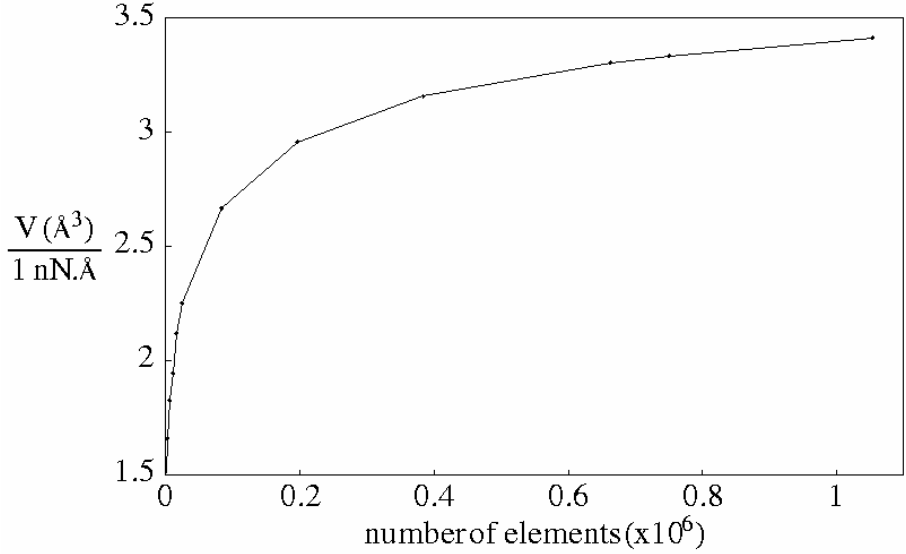

FIG. 3.10: Relaxation volume as a function of the number of elements.

The vacancy was modeled by a cube-shaped hollow region of dimensions 1/48 x 1/48 x 1/48 of the system size located at the centroid of the mesh. The dipole was represented by point forces, directed toward the origin, applied at the centers of the inner faces of this cube. With the points of application of these forces being known, their magnitudes were specified such that the dipole strength was 1 nN.Å (= 0.624 eV). For the dipole to be equivalent to forces applied at the first nearest neighbors of the vacancy, this corresponds to a system of dimensions 12 x 12 x 12 unit cells. The outer surfaces of the cube were allowed to relax inward while maintaining the planarity of the surfaces. The extent of this relaxation was varied until corresponding normal force on each outer face vanished. These boundary conditions were easier to implement than periodic ones, and were therefore preferred. They resulted in displacement fields for which the relaxation volume differed by less than $10^{-2}$ Å$^3$ from the fields for periodic boundary conditions.

Since the dipole is an elastic singularity and the cubic shape introduces further stress concentrations, the finite element solutions were slow to converge with mesh refinement. This necessitated considerably fine meshes. Figure 3.10 shows the relaxation volume as a function of the number of elements. If the number of elements is $6N^3$, the number of nodes is $6(N+1)^3 - 12(N+1)^2$.

The three-dimensional stress tensor obtained at element quadrature points with each mesh was projected to the nodes of the mesh using a least-squares formulation. The radial stress component at each node was then obtained. The slow convergence rate applies to these stresses also. Finite element error analysis predicts that the stress projected to the nodes converges at the rate

$$|\sigma_{node} - \sigma_{exact}| \leq C\, h^2, \tag{3.40}$$

where h is the element size and C is a constant [Hughes 2000]. The same is true of the volume. Thus using the results from two mesh sizes the asymptotic value can be extrapolated:

$$V \approx \frac{\left(\dfrac{h_1}{h_2}\right)^2 V_2 - V_1}{\left(\dfrac{h_1}{h_2}\right)^2 - 1}. \tag{3.41}$$

Figure 3.11 shows the volume obtained from Eq. (3.41) where the size of mesh 2 is constant ($6 \times 56^3$ elements) and the size of mesh 1 is on the x axis. This shows that, unexpectedly, Eq. (3.41) does not provide an asymptotic value independent of the choice of the meshes. This is because convergence is very slow for finite element calculations with a singularity. The stresses have the same problem.

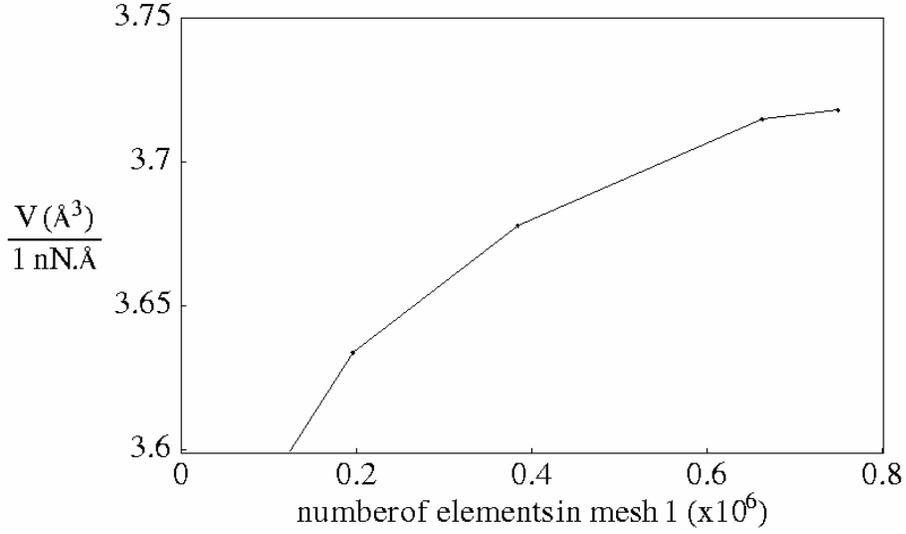

FIG. 3.11: The volume obtained from Eq. (3.41) where the size of mesh 1 is the x axis and the size of mesh 2 is $6 \times 56^3$.

Figure 3.12(a) shows the output dipole per unit input dipole as a function of the shell where it is calculated for five finite element meshes. The curves for the finer meshes have similar shapes and they are similar to what was observed atomistically (Fig. 3.8). However the magnitude of the dipole is different for the different meshes due to the lack of convergence. The high values close to the defect are due to finite size effects and the fact that the dipole was implemented as force pairs a finite distance apart.

Equation (3.39) provides a relationship between the formation volume and the dipole for a sphere of radius R made of an isotropic material. However the proportionality constant applies only to an isotropic medium. In an anisotropic medium the formation volume is also of the form $V^f = K D$ but in this case the proportionality constant K is unknown. Since both atomistic and finite elements results follow this relationship,

$$\frac{V^f_{FE}}{D_{FE}} = \frac{V^f_{at}}{D_{at}}. \qquad (3.42)$$

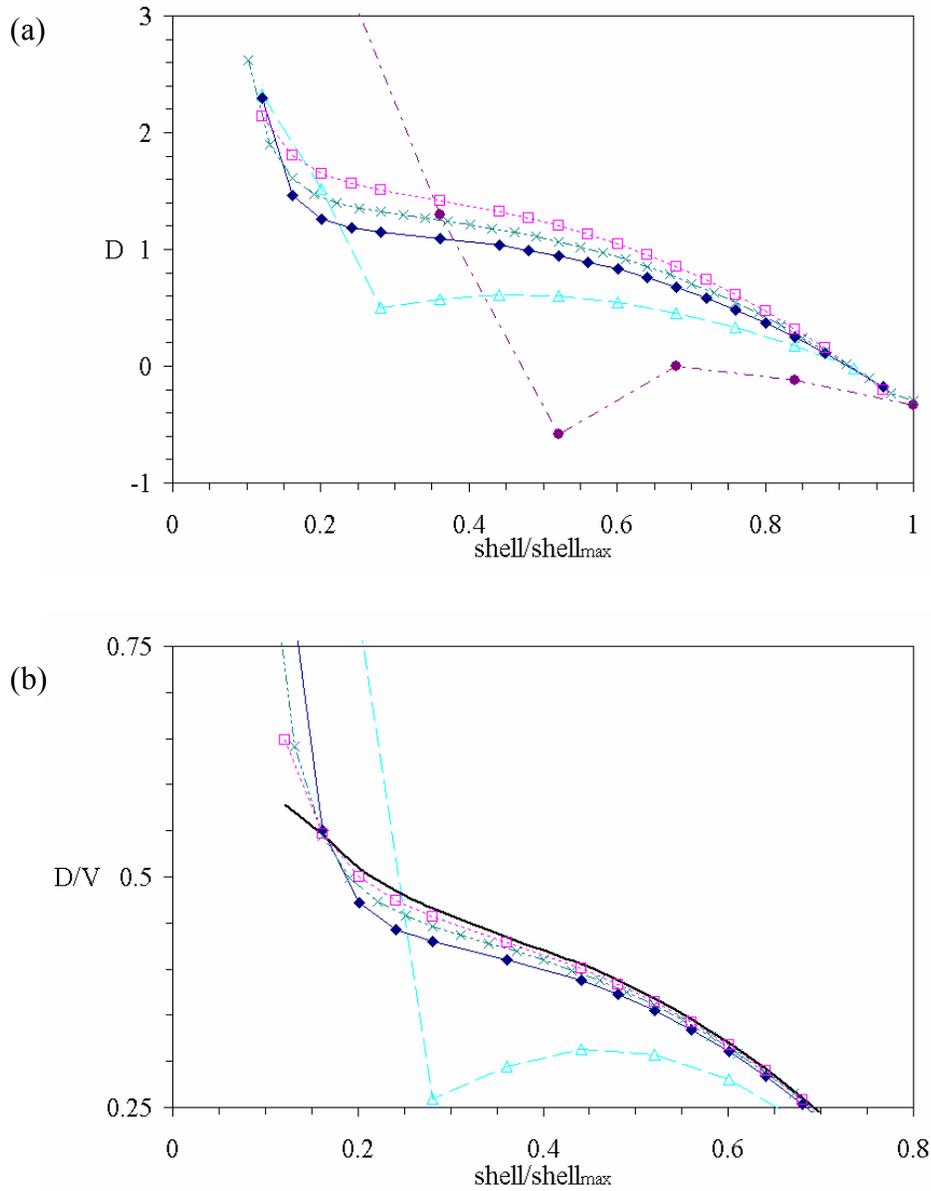

FIG. 3.12: Ratio of the output dipole to the input dipole (a) and to the relaxation volume in eV/Å$^3$ (b) as a function of the cubic shell at which the data is extracted from finite element calculations. Closed circles: 6x6$^3$ elements, open triangles: 6x12$^3$, closed diamonds: 6x24$^3$ elements, crosses: 6x36$^3$ elements and open squares: 6x48$^3$ elements. The thick solid line in (b) is an extrapolation.

In order to obtain $V_{at}^{f}$ from $D_{at}$, only the ratio $V_{FE}^{f}/D_{FE}$ is needed. Although $V_{FE}^{f}$ and $D_{FE}$ converge slowly their ratio may not. Figure 3.12(b) shows the ratio of the output dipole to the volume, $D_{FE}/V_{FE}^{f}$, as a function of the shell. Unlike the volume and the dipole taken separately, the ratio is nearly converged. The two curves are closer together than those in Fig. 3.12(a). Figure 3.13 shows $D_{FE}$ as a function of $V_{FE}^{f}$ for five different mesh sizes. $D_{FE}/V_{FE}^{f}$ is almost independent of the mesh although $V_{FE}^{f}$ and $D_{FE}$ are not converged yet.

The medium used in the FE calculations is anisotropic. The dipole was extracted from the FE calculations using Eqs. (3.29) through (3.31). Thus the error introduced in the results shown in Fig. 3.8 by the use of Eqs. (3.29) through (3.31), which assume an isotropic medium, also exists in the results relating the dipole value to the formation volume shown in Fig. 3.10. Since the FE results are used to derive a formation volume from the dipole extracted in this way it is reasonable to expect that the errors cancel out and the formation volume obtained no longer includes a systematic error arising from an assumption of isotropy.

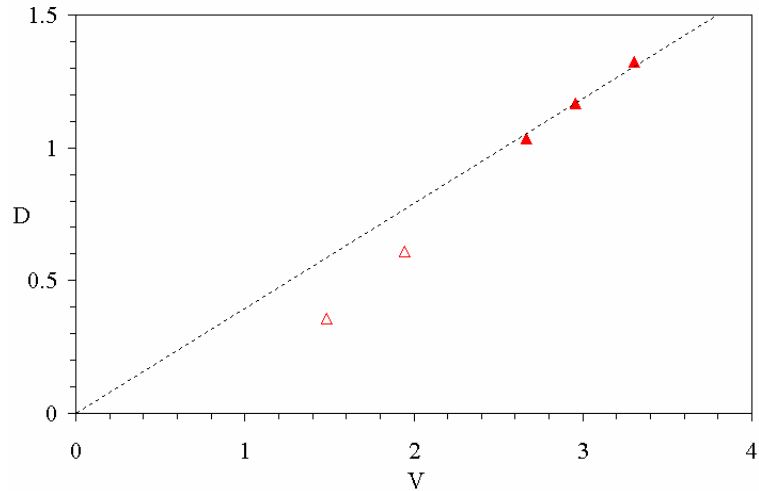

FIG. 3.13: $D_{FE}$ (calculated at the shell situated at 0.45) in eV as a function of $V_{FE}^{f}$ in Å$^3$ for five different mesh sizes: $6\times6^3$, $6\times12^3$, $6\times24^3$, $6\times32^3$ and $6\times48^3$. The dotted line shows that $D_{FE}/V_{FE}^{f}$ is almost constant for the larger shells $6\times24^3$, $6\times32^3$ and $6\times48^3$ (filled symbols).

Since $D_{FE}/V_{FE}^f$ is close to convergence, Eq. (3.41) can be applied to it. Figure 3.14 shows the result for $D_{FE}/V_{FE}^f$ (in eV/Å$^3$) thus obtained. The dotted line is a power law fit to the part of the data far enough from the vacancy for finite size effects to be neglected. Its equation is

$$D_{FE}/V_{FE}^f = 0.47 - 0.59\left(\frac{\text{shell}}{\text{shell}_{max}}\right)^{2.73}. \tag{3.43}$$

At the defect, $D_{FE}/V_{FE}^f \approx 0.47$ eV/Å$^3$, as opposed to 0.67 eV/Å$^3$ in the isotropic case.

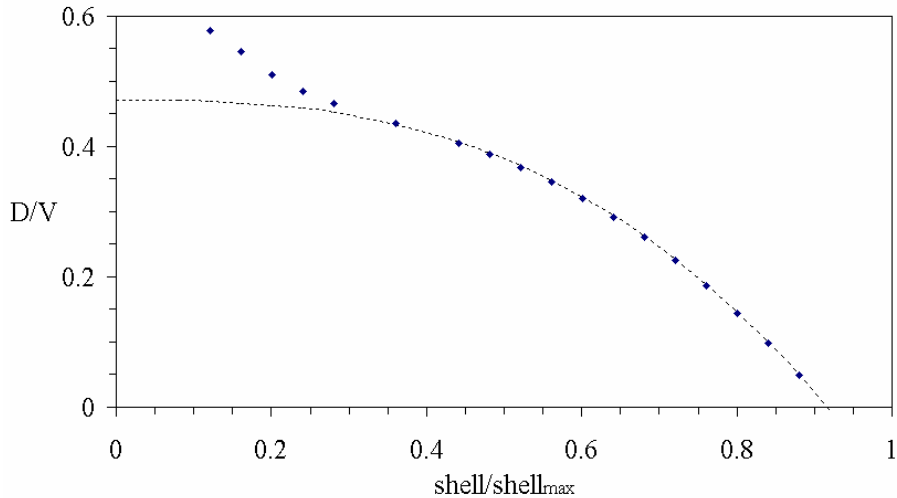

FIG. 3.14: $D_{FE}/V_{FE}^f$ (in eV/Å$^3$) from finite elements as a function of the shell at which the data is extracted. The dotted line is a power law fitted to the data far from the vacancy.

*e) System size effects*

Figure 3.15(a) shows as a function of the system size the values of the formation volume obtained from the dipole and of the formation volume calculated by directly measuring the change in volume of the supercell upon the introduction of the defect to a

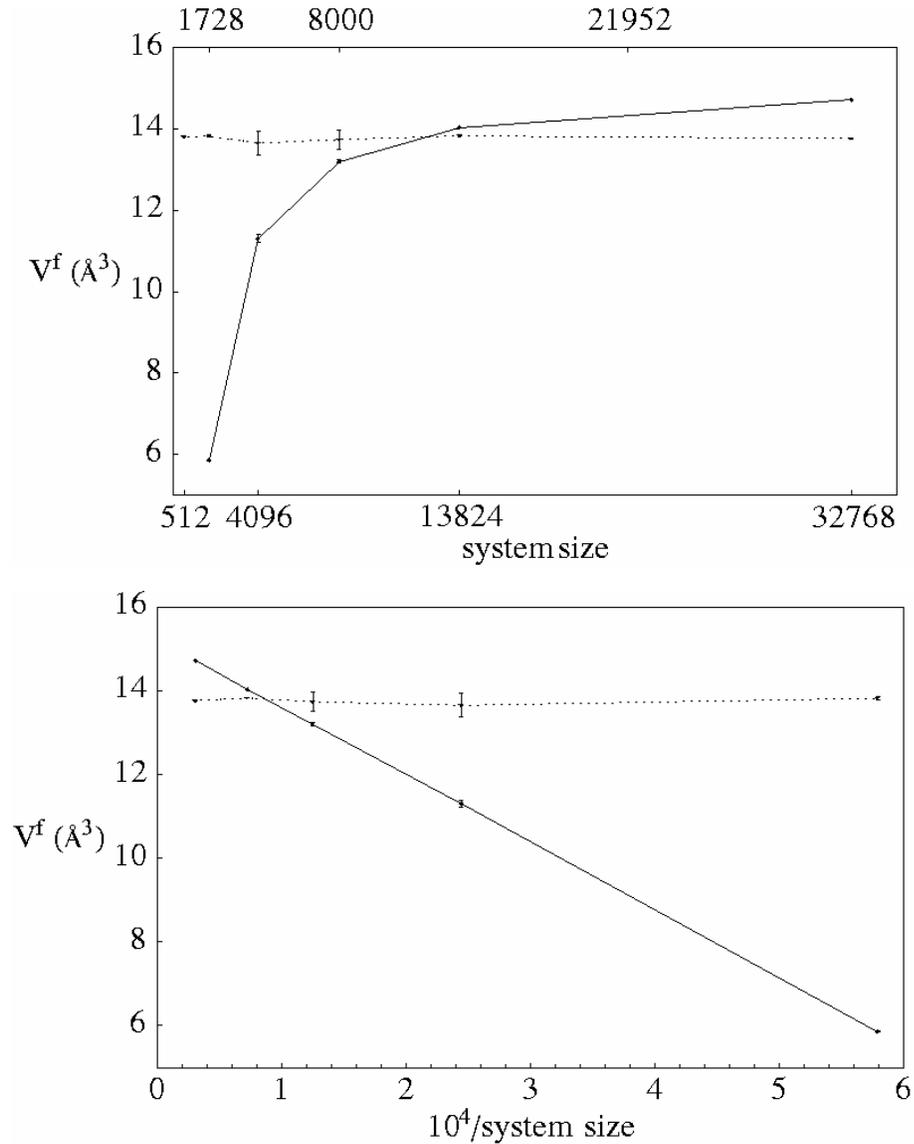

FIG. 3.15: The formation volume as a function of the system size (a) and of $10^4$ over the system size (b). Solid line: volume calculated using the dipole; dotted line: direct extraction from atomistic simulations. The error bars correspond to sample-to-sample standard deviation; they do not account for systematic errors.

system held at zero pressure. Figure 3.15(b) shows the formation volume as a function of 10 000 over the system size. If there is convergence, the formation volume for an infinitely large system can be read at the intersection between the curve and the y-axis. The volume obtained through the dipole converges to a value of 15 Å$^3$ while the direct measurement gives 13.8 Å$^3$.

**5) Summary**

Accurate calculation of formation volumes from atomistic models is important for modeling stress-defect interactions during diffusive processes. The Stillinger Weber potential was used because it allows for the simulation of larger systems than quantum mechanical methods. We presented a new method which calculates the formation volume by matching stresses near the defect to the asymptotic elastic prediction. This method has been shown to converge with system size to a value close to that obtained by measuring the change in volume of the simulation cell. This validates the new method presented in this chapter. It is now possible to find the elastic field around the defect given V$^f$. This will enable the simulation of real systems by superposing the stress field surrounding the individual defects.

As shown in table 3.1, the Stillinger Weber description of the vacancy is not quantitatively accurate. In order to obtain quantitatively accurate results a better description of silicon is necessary close to the vacancy. However, this improvement in model accuracy should not happen at the cost of a dramatic shrinkage of the system size. Two possibilities are available. One could use tight-binding which is more accurate than empirical potentials but not as computationally intensive as *ab initio*, the description of the vacancy is fair and the system can be simulated using thousands of atoms. This approach is not efficient because tight-binding is used far from the vacancy where an empirical potential would be good enough since only the correct elastic properties are needed. Another solution is then to use *ab initio* (or tight-binding) methods close to the vacancy, where the system is far from the equilibrium structure, and an empirical potential further away where computationally-intensive methods are not necessary.

A full-scale simulation of diffusion in semiconductors would require data on other point defects, interstitials, substitutionals, vacancy-interstitial pairs and other defects [Goedecker *et al.* 2002] since all of these defects can exist and interact in the devices. The methodology we developed can be applied to these defects, with some modification. These methods may also be applicable to other kinds of crystal defects, such as dislocations [Shilkrot *et al.* 2002] or more generally to any material inhomogeneity leading to a singularity in the stress field. We applied this methodology to silicon, but it is general enough to be applied to other materials.

# CHAPTER 4 — CONCLUSIONS

This dissertation addresses two aspects of the interaction between stress and diffusion in semiconductors. Chapter 2 presents a study of the role of kinetics in the formation of pits in stressed thin films and chapter 3 is a description of how atomic-scale calculations can be used to extract the thermodynamic and elastic properties of point-defects. For both surface features in heteroepitaxy and bulk point-defects, there is an interaction between phenomena at the atomic and macroscopic scales. In heteroepitaxy the nucleation of surface features is controlled by rate equations that arise from atomic scale considerations and continuum concentration and stress fields. Thus predictions regarding the formation of surface features is obtained from the interaction between atomistic and continuum effects. For point-defects, data obtained from atomistic simulations are used as input into finite element calculations to extract the formation volume of the vacancy. The relaxation of the atoms surrounding the vacancy is treated atomistically and this result is used as an input into finite element calculations to provide a formation volume at the macroscopic scale.

The formation of both point-defects and surface features depends on the stress state of the system. The lattice mismatch is an important factor that determines the morphology of the surface during epitaxial growth. In homoepitaxy films generally remain smooth while heteroepitaxial films tend to form ripples, islands or pits. The magnitude of the mismatch also determines whether islands will be coherent or dislocated. Likewise the presence of stress in the bulk can determine the kinds of defects that dominate and how diffusion occurs. Under compression, defects with a negative formation volume will dominate while those with a positive formation volume will dominate under tension. This can have an impact on diffusion mechanisms, if vacancies dominate vacancy-assisted diffusion is more likely to be the main diffusion mechanism.

Elastic calculations predict that pits can relieve elastic energy more efficiently than islands in heteroepitaxial films. Pits have commonly been observed only in the presence of impurities, but recently new evidence has arisen that pits can form during heteroepitaxial growth in the absence of contamination. To understand this phenomenon the nucleation of islands and pits was studied assuming a near-equilibrium nucleation process where the adatom concentration plays a major role. This analysis predicts that pit nucleation can be prevented by a high adatom concentration, but that inhomogeneities of the adatom concentration arising from diffusion favor pit nucleation close to the islands where the adatom concentration is lower. While energetic arguments indicate that pitting should occur in most systems, their nucleation is usually kinetically prevented. Accounting for kinetics, six experimental regimes are predicted to exist depending on the growth rate and the elastic energy arising from misfit: pits can nucleate far from islands, adjacent to isolated islands, in between islands, or in the absence of islands. The film can also remain planar or islands alone can nucleate. The theoretical arguments presented here lead to predictions for the circumstances under which these various regimes should arise. These have been mapped on to a non-equilibrium "phase diagram" that predicts the morphology as a function of surface energy, temperature, deposition rate, lattice mismatch and surface diffusivity.

These findings are in reasonable agreement with experiments in III-V semi-conductors. More precise predictions would require that surface energies be better known and that the effect of the shape of the features be accounted for. In addition, strain inhomogeneities arising from the presence of islands could be incorporated into the model in order to study whether strain and adatom concentration inhomogeneities tend to reinforce each other. Since no analytical form is known for the strain field around an island, these effects could be only addressed qualitatively.

In III-V systems, pits have been observed to nucleate after islands [Chokshi *et al.* 2000, 2002, Riposan *et al.* 2002, 2003, Seshadri *et al.* 2000, Lacombe 1999]. However pits have also been observed before islands in SiGe/Si systems [Gray *et al.* 2001, 2002, Vandervelde *et al.* 2003]. Pits could then be used as nucleation sites for quantum dots, which leads to a high density and even spatial distribution of the dots [Deng and Krishnamurthy 1998, Songmuang *et al.* 2003]. They can also give birth to more complex structures where a "wall" surrounds a pit [Gray *et al.* 2001, 2002, Vandervelde *et al.* 2003]. The study of pit nucleation presented here provides a means of understanding the nucleation of islands subsequent to that of pits in addition to the nucleation of pits subsequent to islanding. This makes possible a complete study of island and pit nucleation which would include the cooperative kinetics of these processes. Chapter 2 also included a preliminary investigation of the late stages of the growth morphology by analyzing ripple arrays which appear to arise from such processes.

Chapter 3 focused on the effect of stress on the formation of point defects. Accurate calculation of formation volumes is important for modeling stress-defect interactions in diffusive processes. The most common methods for calculating these quantities are to measure the relaxations of the first nearest neighbors and the change in volume of the simulation cell. In the new method presented in this dissertation the formation volume is calculated by matching stresses near the defect to the asymptotic elastic prediction. This method has been shown to converge to about the same volume as direct measurements for large systems.

The methodology presented in this dissertation was applied to vacancies but it could be adapted for other defects. The stress field in a crystal arises due to external stress plus the superposition of the individual stress fields due to the defects. A full-scale simulation of diffusion in semiconductors would therefore require data on vacancies, interstitials, substitutionals and vacancy-interstitial pairs since all of these defects can exist and interact in devices. A natural follow-up work is thus to obtain the dipole tensors for these defects. Then the stress at any point in the system could be known and incorporated into, for example, a kinetic Monte Carlo simulation of the defect kinetics. Although chapter 3 described a methodology which can be applied to other defects, some modifications would be necessary. In the case of defects involving several atoms for instance the "position" of the defect is not obvious, there is an additional degree of freedom in the determination of the dipole value. Also when the dipole is not isotropic the dipole tensor has more than one degree of freedom, which would require a large set of finite element calculations.

Shilkrot and coworkers (2002) studied indentation and dislocation dynamics on a multiscale: close to the indenter the system was modeled atomistically while a discrete dislocation model was used further out. A similar study is possible for point defects if their stress fields are known. The boundary conditions of the atomistic zone must account for the stress fields of defects in the continuum zone. Also if the relationship between stress and formation free energy is known, a chemical potential gradient can be inferred from a stress gradient; this will drive the diffusion of the defects. The diffusion of the defects would lead to a change in the stress state leading to a feedback loop between stress field and local defect concentration. Such simulations could have a significant impact in the semiconductor industry. If the desired dopant profile could be obtained as an equilibrium profile the manufacturing of semiconductors would be eased and would not require as many steps.

The method introduced in chapter 3 was applied to defects in silicon. However it is more general and could theoretically be applied to other materials. As the scale of materials science decreases the fact that the properties of small systems such as thin films can be different from the bulk properties continues to grow in importance. For devices on smaller and smaller scales, voids of smaller and smaller sizes will have increasingly dramatic effects. Although large-scale defects are due to synthesis and are not very likely to arise from diffusion (at least at low temperature), nano-scale voids can form if vacancies diffuse and merge. Electronic properties are very sensitive to defects, thus they are impaired by such mechanisms faster than mechanical properties. However the shrinkage of the scale of mechanical devices will soon result in similar reliability issues. The mechanical and electronic properties of carbon nanotubes are exceptional [Ajayan 1999, Dresselhaus and Dai 2004], which makes them good candidates for future devices [Liu *et al.* 1999, Baughman *et al.* 2002], but they may be very sensitive to the presence of defects. The model developed in 3D could be adapted to 2D and the stress field around a defect in a graphite sheet, or a single-walled carbon nanotube, could be analyzed similarly.

The study of surface features during heteroepitaxial growth, presented in chapter 2, focused on the adatom concentration inhomogeneities. This was enough to obtain several regimes in fair agreement with experimental observations, but it could be completed by accounting for the stress inhomogeneities across the surface. The stress close to an island or a pit is not the nominal stress coming from the mismatch, surface features are sources of stress. Shilkrot and Srolovitz (1997) used elastic dipoles to study the interaction between steps and adatoms. It would be possible to do a similar work for islands. The methodology developed for point-defects in chapter 3 could then be used to extend the work done in chapter 2. If the islands are sufficiently far apart they can be treated as points. Otherwise they can be represented by a series of dipoles at their perimeter instead of only one dipole. This way both stress and adatom concentration inhomogeneities would be accounted for. If there are defects in the film or at its surface, such as the end of a threading dislocation, they may participate to the inhomogeneity of the stress field on the surface. The net stress field can be obtained by the superposition of the individual stress fields from the islands, pits and various defects, albeit in the presence of a free surface.

# APPENDIX — POTENTIALS FOR III-V SEMICONDUCTORS

## 1) Introduction

Molecular dynamics (MD) and Monte Carlo (MC) simulations require a model for the energy of an atomic system to calculate forces (MD) or the energy differences between configurations (MC). This expression of the energy as a function of the position of the atoms is often decomposed into inter-atomic potentials. Many potentials have been proposed, for noble gases, metals, semiconductors, etc. The choice of the potential is critical as the simulations do not probe GaAs but rather the model described by the potential. Therefore it is important that the model accurately represents the salient characteristics of the material bonding.

Potentials are typically designed to reproduce aspects of the physics of the material as accurately as possible, including crystal structure (FCC or BCC for metals, diamond cubic or ZnS for semiconductors, etc.), elastic properties, liquid phase, phonon spectrum, surfaces, etc. The most fundamental description of this energy would require quantum mechanics. Since the potentials do not include any explicit quantum mechanics, their parameters must be fitted to empirical data. The set of properties to which the potential can be fitted is typically much larger than the set of parameters. The appropriate choice of properties depends in part on the purpose of the study in which the potential will be used. To study the growth of thin films, for instance, surfaces are important while the properties of the liquid phase or bulk defect formation energies may be secondary. The opposite is true for studies of ion implantation. Also, not all properties are easy to calculate: while the cohesive energy of a given crystal structure can be readily obtained from the (known) positions of the atoms, surface reconstructions require that atoms be allowed to relax. Fitting the potential to surface reconstruction is therefore more computationally intensive. While it may seem possible to obtain better accuracy using a more complicated potential with more parameters more complicated potentials are harder to implement and debug, more computationally intensive and more likely to have spurious local minima. As a consequence energy minimization methods may not be reliable if the energy is not a smooth function of atomic positions.

This appendix reviews the available potentials for III-V semiconductors. Compound semiconductors are covalent but there is also some ionicity associated with partial charges on the anions and cations. Section 3 is a study of the effects of ionicity that was undertaken to help decide whether the semiconductors are to be simulated as purely covalent materials or partly ionic materials. Adding ionicity adds accuracy and complication. The purpose is to find a trade-off that is simple but accurate.

**2) Review of potentials for III-V semiconductors**

The simplest potentials, such as the Lennard-Jones potential, only take into account the distance between atoms, not the coordination number or angles. In this case, the energy is directly proportional to the number of bonds and such potentials can only find a close-packed structure as the ground state. Therefore they are not suitable for semiconductors. The diamond cubic (DC) and zinc-blende (ZnS) structures are open structures with a coordination number Z of 4. Their stability comes from the fact that, although less numerous, the bonds are stronger. The energy of an atom is Z $E_{bond}$, so a lower Z can lead to a lower energy as long as this strengthens the bonds. Close-packed structures would have much weaker bonds and therefore a higher energy. Thus, it is not possible to study covalently-bonded materials with a pair (two-body) potential.

|        | Smith | Sayed | Ashu    | Nordlund | Murugan | Nakamura | Albe |
|--------|-------|-------|---------|----------|---------|----------|------|
| Ga-As  |       |       | Sayed   | Sayed    |         | Sayed[*] |      |
| Ga-Ga  |       | Smith | Smith   | Smith    | ??      | Smith    |      |
| As-As  |       | Smith | Smith   | Smith    | ??      | Smith    |      |
| In-As  |       |       |         |          |         | Ashu     |      |
| In-In  |       |       |         | Ashu     |         | ??       |      |
| In-Ga  |       |       | average | Ashu     |         | ??       |      |
| Al-As  |       |       |         | Sayed    |         | ??       |      |
| Al-Al  |       |       |         | Sayed    |         | Sayed    |      |
| Al-Ga  |       |       |         | average  |         | ??       |      |
| Al-In  |       |       |         |          |         |          |      |

Table A.1: Potentials for III-V semiconductors and the parameters they use. White cells correspond to new parameters and dark gray cells correspond to parameters which were not used. The light gray cells are for parameters which were used in the paper but which were not original: the name of another author means that the parameters from this author were used and "average" means that the III-III' parameters were obtained as an average of the III and III' parameters. Murugan and Ramachandran did not publish parameters for the Ga-Ga and As-As bonds and did not cite any other work. Although Nakamura and coworkers cite the works of Smith, Sayed and Ashu, they do not explicitly state what was drawn from these works and what they fitted themselves. "Sayed[*]" refers to the fact that some of the parameters were taken from Sayed.

The two main categories of potentials for semiconductors are bond-order potentials (Tersoff [Tersoff 1986, Tersoff 1988a, Tersoff 1988b, Tersoff 1989] and Brenner [Brenner 1990] developed potentials with a similar form for Si and C respectively) and potentials with an explicit three-body term [Stillinger and Weber 1985]. Since there are few studies which give parameters for III-V semiconductors under the analytical form of Stillinger and Weber, this appendix will focus on GaAs potentials based on the Tersoff potential. Papers generally present their potential and the results of the simulations in which it was used (often implantation damage), but few of them test the potentials abilities to model basic properties. In order to evaluate the published potentials, we analytically tested (at 0 K) five Tersoff-based potentials for GaAs [Smith 1992, Sayed *at al.* 1995, Albe *et al.* 2002, Nakamura *et al.* 2000, Murugan and Ramachandran 1999], two for InAs [Ashu *et al.* 1995, Nordlund 2000] and one for AlAs [Sayed *at al.* 1995]. Table A.1 summarizes the interdependence between these works. This table illustrates the fact that many authors fit Tersoff potential themselves for one material system and use published parameters for some other material systems. Nordlund for instance simulated GaAs, InAs and AlAs but fitted new parameters for In-As bonds only. Although many investigators simulated GaAs, until Albe and coworkers's work (2002), all authors used Smith's parameters [Smith 1992] for pure As and Ga.

*a) Analytical form of Tersoff potential*

Unlike the Stillinger Weber potential, there is no explicit three-body term in the analytical form designed by Tersoff. Instead the two-body energy depends on the environment of the atom. The energy is given by

$$E = \frac{1}{2} \sum_{i \neq j} V_{ij} \qquad (A.1)$$

with

$$V_{ij} = f_c(r_{ij})[f_R(r_{ij}) - f_A(r_{ij})] \quad (A.2)$$

where $f_R$ is the repulsive part and $f_A$ is the attractive part of the potential. $f_c$ is a smooth cut-off function defined by

$$f_c(r) = \begin{cases} 1, & r < R_{ij} \\ \frac{1}{2} + \frac{1}{2}\cos\left(\pi \frac{r - R_{ij}}{S_{ij} - R_{ij}}\right), & R_{ij} < r < S_{ij} \\ 0, & r > S_{ij} \end{cases} \quad (A.3)$$

with R and S constants. Thus when the distance between the atoms is less than R, $f_c = 1$ and the cut-off does not play a role while for r > S $f_c = 0$ and there is no interaction between the atoms. This accounts for the short-ranged nature of the bonds and reduces the number of bonds for which an energy must be computed, making the simulations more efficient. Between R and S, $f_c$ goes from 1 to 0; due to the analytical form of $f_c$, $f_c$ and its derivatives are continuous at R and S. Although the cut-off is an efficient way of speeding up simulations, it can introduce artifacts. When the distance between two atoms is between R and S, the energy of the bonds is due in part to $f_c$ which has no physical meaning. In the Tersoff potential (and most other potentials) the cut-off is set between first and second nearest neighbors. This minimizes the number of bonds for which an energy or forces must be computed and it also simplifies the fitting procedure by decreasing the number of degrees of freedom and by making the properties of bulk III-V semiconductors independent of III-III and V-V bonds between second nearest neighbors.

$f_R$, the repulsive part, is given by

$$f_R(r) = A_{ij} e^{-\lambda_{ij} r} \quad (A.4)$$

where A and λ are positive constants. The attractive part is of the same form

$$f_A(r) = B_{ij} b_{ij} e^{-\mu_{ij} r} \tag{A.5}$$

with B and μ positive constants. R, S, A, B, λ and μ depend on the species involved, e.g. $A_{Ga\text{-}As}$ may be different from $A_{Ga\text{-}Ga}$ and $A_{As\text{-}As}$. While the repulsive part depends on the interatomic distance only, the attractive part also depends on the function b defined as

$$b_{ij} = \left[1 + (\beta_i \zeta_{ij})^{n_i}\right]^{\frac{-1}{2n_i}} \tag{A.6}$$

where ζ is a "pseudo" coordination number:

$$\zeta_{ij} = \sum_{k \neq i,j} f_c(r_{ik}) g(\theta_{ijk}) \tag{A.7}$$

with

$$g(\theta) = 1 + \frac{c_i^2}{d_i^2} - \frac{c_i^2}{d_i^2 + (\cos\theta - h_i)^2} . \tag{A.8}$$

If $c = 0$ (or $c^2 \ll d^2$), $g(\theta) = 1$ and then $\zeta = Z$, the coordination number. As a consequence b is constant and the energy depends only on bond lengths and coordination number, not on angles. Therefore a potential where $c^2 \ll d^2$ can reproduce the lattice constant, the cohesive energy and the bulk modulus of the ZnS structure and can give the ZnS structure as the ground state but the Young's modulus and $c_{44}$ are almost 0 because of the lack of angular dependence of the energy. Tersoff designed this simple potential form [Tersoff 1988a] for pedagogical reasons and Smith [Smith 1992] and Ashu [Ashu *et al.* 1995] have inadvertently recreated the same artifact.

There are 9 parameters (A, B, λ, μ, β, n, c, d, h) plus two for the cut-off R and S. The latter are not systematically optimized. The 9 parameters are generally fitted to the lattice parameter, the cohesive energy and the bulk modulus of the zinc-blende bulk, and sometimes to the elastic constants $c_{11}$, $c_{12}$ and $c_{44}^0$ defined below. The parameters were not fully fitted to other phases, though it was checked that ZnS was the ground state.

The eight sets of parameters are presented in Table A.2. Parameters are summarized in this table along with Tersoff 2 [Tersoff 1988a] and Tersoff 3 [Tersoff 1988b] for Si as a comparison. One can see that very different sets of parameters are used. Some use large values for c along with small $\gamma$ (for instance T3 and Murugan and Ramachandran).

|  | Si | | GaAs | | | | | InAs | | AlAs |
|---|---|---|---|---|---|---|---|---|---|---|
|  | T2 | T3 | Smith | Sayed | Albe | Naka-mura | Murugan | Ashu | Nord-lund | Sayed |
| A | 3264.7 | 1830.8 | 3088.5 | 2543.3 | 3306.2 | 13287.6 | 1702.2 | 2246.6 | 1968.3 | 2307.9 |
| B | 95.37 | 471.18 | 469.97 | 314.46 | 1929.30 | 13.19 | 381.50 | 417.67 | 266.57 | 219.14 |
| *A/B* | *34.231* | *3.886* | *6.572* | *8.088* | *1.714* | *1007.6* | *4.462* | *5.379* | *7.384* | *10.532* |
| l1 | 3.239 | 2.480 | 2.828 | 2.828 | 2.301 | 4.599 | 1.540 | 2.530 | 2.598 | 2.809 |
| l2 | 1.326 | 1.732 | 1.842 | 1.723 | 2.015 | 0.249 | 0.770 | 1.671 | 1.422 | 1.558 |
| De | 2.624 | 2.666 | 2.180 | 2.180 | 2.100 | 7.100 | 21.376 | 2.398 | 5.173 | 2.493 |
| Re | 2.313 | 2.295 | 2.345 | 2.340 | 2.350 | 2.260 | 2.842 | 2.441 | 2.214 | 2.353 |
| S = l1/l2 | 2.443 | 1.432 | 1.535 | 1.641 | 1.142 | 18.447 | 2.000 | 1.514 | 1.826 | 1.803 |
| b | 1.465 | 1.466 | 1.614 | 1.561 | 1.523 | 0.757 | 0.770 | 1.454 | 1.359 | 1.479 |
| c | 4.838 | 100390 | 0.078 | 1.226 | 1.290 | 1.226 | 18223 | 1.307 | 5.172 | 1.450 |
| d | 2.042 | 16.622 | 4.505 | 0.790 | 0.560 | 0.790 | 12.384 | 91.553 | 1.666 | 0.829 |
| (c/d)^2 | 5.615 | 4E+07 | 3.0E-04 | 2.407 | 5.306 | 2.407 | 2E+06 | 2E-04 | 9.640 | 3.060 |
| h | 0.000 | -0.598 | -3.411 | -0.518 | -0.237 | -0.518 | -0.500 | -0.570 | -0.541 | -0.521 |
| *arccos h* | *90.0* | *126.7* | ---- | *121.2* | *103.7* | *121.2* | *120.0* | *124.7* | *122.8* | *121.4* |
| n | 22.956 | 0.787 | 5.504 | 6.317 | 1.000 | 6.317 | 0.310 | 6.332 | 0.756 | 4.048 |
| g | 0.337 | 1.1E-06 | 0.381 | 0.357 | 0.017 | 0.357 | 4E-07 | 0.387 | 0.319 | 0.331 |
| *g * (c/d)^2* | *1.891* | *40.122* | *1.2E-04* | *0.860* | *0.088* | *0.860* | *0.805* | *8E-05* | *3.072* | *1.013* |
|  | *0.727* | *0.153* | *1.000* | *0.975* | *0.959* | *0.975* | *0.345* | *1.000* | *0.451* | *0.915* |
| cut-off | 2.8 | 2.7 | 3.4 | 3.4 | 3 | 3.4 | 2.34 | 3.4 | 3.5 | 3.4 |
|  | 3.2 | 3 | 3.6 | 3.6 | 3.2 | 3.6 | 2.65 | 3.6 | 3.7 | 3.6 |

Table A.2: The eight sets of parameters for Ga-As, In-As and Al-As potentials. The Tersoff 2 and Tersoff 3 potentials for Si are also presented for comparison.

*b) Elastic constants*

Cubic crystals have 3 independent elastic constants $c_{11}$, $c_{12}$ and $c_{44}$; the stiffness matrix is of the form

$$\begin{bmatrix} c_{11} & c_{12} & c_{12} & 0 & 0 & 0 \\ c_{12} & c_{11} & c_{12} & 0 & 0 & 0 \\ c_{12} & c_{12} & c_{11} & 0 & 0 & 0 \\ 0 & 0 & 0 & c_{44} & 0 & 0 \\ 0 & 0 & 0 & 0 & c_{44} & 0 \\ 0 & 0 & 0 & 0 & 0 & c_{44} \end{bmatrix}. \quad (A.9)$$

Bulk modulus B, Young's modulus along (001), $E_{(001)}$, and Poisson's ratio $\nu$ are functions of $c_{11}$ and $c_{12}$. Thus only two of these five constants plus $c_{44}$ are necessary. We directly calculated $c_{11}$, $c_{44}$ and B. From those three numbers $E_{(110)}$ (that is along the direction of surface dimers), the anisotropy ratio and Cauchy ratio were also calculated. $c_{44}^0$ and $\zeta$ (and also the pressure derivative of the bulk modulus) were also separately calculated. When the unit cell is sheared in the x-z plane by an amount $\gamma$ all atoms move along x by $\gamma z$. Once they have moved they can relax to decrease their energy: instead of moving by ($\gamma z$, 0, 0) they move by ($\gamma z + \delta x$, $\delta y$, $\delta z$). In the first case (no relaxation), the energy of the crystal under the shear strain $\gamma$ will be overestimated, therefore the stiffness will also be overestimated. The constant obtained this way is not $c_{44}$, it is another constant called $c_{44}^0$. When atoms are allowed to relax the real elastic constant $c_{44}$ is obtained. In such a case there is a shift of the group III sublattice along y by $\zeta \gamma$ (this is the definition of $\zeta$)[3]. $c_{44}^0$ is easier to determine, since there is no degree of freedom there is no need to optimize the energy by searching for a root. However having the correct $c_{44}^0$ does not mean that $c_{44}$ will be correct too; Smith and Ashu have the right magnitude for $c_{44}^0$ but their $c_{44}$ is 0.

---

[3] This $\zeta$ is completely different from the $\zeta_{ij}$ used in the expression of the potential. $\zeta$ is the usual letter to refer to this relaxation and $\zeta_{ij}$ is the notation used by Tersoff.

Results for GaAs are presented in Table A.3 and results for InAs are in Table A.4. Results for AlAs are not presented because there was only one reported potential at the time of this writing [Sayed *at al.* 1995]. Bond lengths and cohesive energies for dimer, "graphite", zinc-blende (ZnS), rock-salt (NaCl) and CsCl structures are given. Bulk modulus B and its first derivative with respect to pressure, B', are also given for ZnS, NaCl and CsCl structures. For ZnS, the elastic properties and the melting point are also presented. The two "experimental/*ab initio*" columns give a lower and an upper bound as found in the literature. For each potential the first column is the value of the property as calculated with the potential, the second column gives the relative error (value$_{potential}$ – value$_{exp}$)/value$_{exp}$. The last column assigns demerit points when this error is too large. In the case of GaAs, Smith does poorly because $c_{11} = c_{12}$ (leading to a Young's modulus of 0) and $c_{44} = 0$. Sayed finds the "graphitic" structure (sheet of GaAs were each atom has bonds with three atoms of the other species) barely higher in energy than the ZnS ground state. Albe's main weakness is that $c_{44}$ is a bit too low. For InAs, Ashu's potential has the same problems as Smith's GaAs (E = 0 and $c_{44}$ = 0). Nordlund's "graphite" is low in energy (similar to Sayed.)

| | **GaAs** | Exp/*ab initio* | | Smith | | | Sayed | | | Albe | | |
|---|---|---|---|---|---|---|---|---|---|---|---|---|
| Dimer | Bond length (A) | 2.510 | 2.550 | 2.35 | -6.57% | 0.0 | 2.35 | -6.57% | 0.0 | 2.35 | -6.37% | 0.0 |
| | Δ Bond length (A) | 0.062 | 0.102 | -0.096 | -194% | 9.0 | -0.104 | -202% | 9.4 | -0.098 | -196% | 9.1 |
| 1 | Cohesive energy (eV/at) | -1.055 | -1.005 | -1.09 | 3.32% | 0.0 | -1.09 | 3.32% | 0.0 | -1.05 | 0.00% | 0.0 |
| | Δ Cohesive energy (eV/at) | 2.299 | 2.349 | 2.229 | -3.03% | 0.0 | 2.161 | -6.01% | 0.1 | 2.304 | 0.00% | 0.0 |
| | ω (cm-1) | 215 | 215 | 215 | 0.00% | 0.0 | 215 | 0.00% | 0.0 | 278 | 29.30% | 0.0 |
| | **Total** | | | | **4%** | **9.1** | | **12%** | **9.5** | | **19%** | **9.1** |
| Graph. | Bond length (A) | **2.342** | **2.342** | 2.35 | 0.13% | 0.0 | 2.353 | 0.48% | 0.0 | 2.460 | 5.05% | 0.0 |
| 3 | Δ Bond length (A) | -0.106 | -0.106 | -0.096 | -8.94% | 0.0 | -0.096 | -9.68% | 0.0 | 0.013 | -112% | 3.8 |
| | Cohesive energy (eV/at) | **-2.785** | **-2.785** | -3.150 | 13.10% | 0.0 | -3.196 | 14.74% | 0.0 | -2.444 | -12.25% | 0.0 |
| | Δ Cohesive energy (eV/at) | 0.569 | 0.569 | 0.169 | -70.22% | 21.5 | 0.055 | -90.30% | 35.3 | 0.911 | 60.02% | 1.3 |
| | **Total** | | | | **10%** | **21.5** | | **44%** | **35.3** | | **11%** | **5.1** |
| ZnS | Bond length (A) | 2.448 | 2.448 | 2.441 | -0.26% | 1.7 | 2.449 | 0.04% | 0.0 | 2.448 | -0.01% | 0.0 |
| 4 | Lattice parameter (A) | 5.6532 | 5.6533 | 5.638 | -0.26% | 0.0 | 5.655 | 0.04% | 0.0 | 5.653 | -0.01% | 0.0 |
| | Cohesive energy (eV/at) | -3.354 | -3.354 | -3.319 | -1.04% | 0.0 | -3.251 | -3.08% | 0.0 | -3.354 | 0.01% | 0.0 |
| | T m (K) | 1513 | 1513 | 1250 | -17.38% | 0.7 | 1050 | -30.60% | 2.2 | 1900 | 25.58% | 1.6 |
| | C 11 (GPa) | 120.7 | 121.5 | 81.8 | -32.27% | 9.1 | 117.9 | -2.31% | 0.1 | 123.7 | 1.87% | 0.0 |
| | C 12 (GPa) | 53.3 | 56.42 | 81.8 | 44.91% | 15.9 | 53.0 | -0.65% | 0.0 | 48.2 | -9.60% | 0.9 |
| | ν | 0.305 | 0.318 | 0.500 | 56.99% | 26.7 | 0.310 | 0.00% | 0.0 | 0.280 | -8.10% | 0.7 |
| | E (001) (GPa) | 83.9 | 89.8 | 0.0 | -99.99% | 60.0 | 85.1 | 0.00% | 0.0 | 96.7 | 7.65% | 0.6 |
| | E (110) (GPa) | 114.6 | 133.5 | 0.0 | -100% | 60.0 | 129.9 | 0.00% | 0.0 | 98.9 | -13.66% | 1.8 |
| | anisotropy ratio | 0.53 | 0.57 | 2.69 | 372.75% | 104.8 | 0.47 | -10.82% | 0.3 | 0.96 | 69.67% | 11.2 |
| | C 44 (GPa) | 60.2 | 60.6 | 0.0 | -100% | 60.0 | 68.6 | 13.24% | 1.7 | 39.1 | -34.97% | 11.3 |
| | C 44 0 (GPa) | 72.6 | 75.0 | 81.8 | 9.02% | 0.2 | 95.2 | 26.96% | 1.7 | 73.2 | 0.00% | 0.0 |
| | ζ | 0.76 | 0.76 | 0.99 | 30.26% | 0.6 | 0.54 | -29.47% | 0.5 | 0.55 | -27.89% | 0.5 |
| | Cauchy ratio | 0.87 | 0.94 | 87895 | 9350460% | 37.5 | 0.77 | -11.57% | 0.3 | 1.23 | 31.05% | 2.3 |
| | Born ratio | 1.02 | 1.09 | 1.00 | -1.50% | 0.0 | 1.25 | 15.03% | 0.6 | 0.71 | -30.44% | 2.2 |
| | B (GPa) | 75.7 | 78.1 | 81.8 | 4.69% | 0.2 | 74.6 | -1.40% | 0.0 | 73.4 | -3.05% | 0.1 |
| | B' | 4.00 | 5.00 | 4.80 | 0.00% | 0.0 | 4.71 | 0.00% | 0.0 | 4.50 | 0.00% | 0.0 |
| | *sub-total* | | | | *80%* | *176.5* | | *7%* | *5.3* | | *66%* | *31.2* |
| | **Total** | | | | **81%** | **179.0** | | **9%** | **7.5** | | **70%** | **32.8** |
| NaCl | Bond length (A) | 2.625 | 2.639 | 2.673 | 1.30% | 0.0 | 2.850 | 8.00% | 0.0 | 2.660 | 0.80% | 0.0 |
| 6 | Δ Bond length (A) | 0.177 | 0.191 | 0.232 | 21.29% | 0.2 | 0.401 | 110.03% | 3.7 | 0.212 | 11.14% | 0.0 |
| | Cohesive energy (eV/at) | -3.084 | -3.078 | -2.585 | -16.00% | 0.0 | -1.567 | -49.08% | 0.0 | -3.086 | 0.08% | 0.0 |
| | Δ Cohesive energy (eV/at) | 0.270 | 0.276 | 0.734 | 165.88% | 6.3 | 1.683 | 509.93% | 13.1 | 0.268 | -0.78% | 0.0 |
| | B (GPa) | **77.6** | **77.6** | 89.6 | 15.39% | 0.0 | 47.6 | -38.62% | 0.0 | 95.6 | 23.22% | 0.0 |
| | B' | **5.1** | **5.1** | 5.16 | 1.50% | 0.0 | 5.32 | 4.70% | 0.0 | 4.77 | -6.19% | 0.0 |
| | **Total** | | | | **3%** | **6.5** | | **21%** | **16.8** | | **0%** | **0.1** |
| CsCl | Bond length (A) | 2.837 | 2.837 | 2.843 | 0.20% | 0.0 | 3.000 | 5.75% | 0.0 | 2.830 | -0.24% | 0.0 |
| 8 | Δ Bond length (A) | 0.389 | 0.389 | 0.401 | 3.13% | 0.0 | 0.551 | 41.68% | 0.7 | 0.382 | -1.69% | 0.0 |
| | Cohesive energy (eV/at) | -2.864 | -2.864 | -2.219 | -22.51% | 0.0 | -1.368 | -52.25% | 0.0 | -2.782 | -2.86% | 0.0 |
| | Δ Cohesive energy (eV/at) | 0.490 | 0.490 | 1.100 | 124.49% | 4.4 | 1.883 | 284.31% | 10.2 | 0.572 | 16.76% | 0.1 |
| | B (GPa) | **82.2** | **82.2** | 90.3 | 9.92% | 0.0 | 51.3 | -37.57% | 0.0 | 104.9 | 27.65% | 0.0 |
| | B' | **5.3** | **5.3** | 5.42 | 2.12% | 0.0 | 5.55 | 4.50% | 0.0 | 4.96 | -6.63% | 0.0 |
| | **Total** | | | | **2%** | **4.4** | | **14%** | **10.9** | | **0%** | **0.1** |
| | **GRAND TOTAL** | | | | | **220** | | | **80** | | | **47** |

Table A.3: Results for GaAs

|  | InAs | Exp./ ab initio | | Ashu | | | Nordlund | | |
|---|---|---|---|---|---|---|---|---|---|
| Dimer | Bond length (A) | **2.500** | **2.500** | 2.44 | -2.35% | 0.0 | 2.21 | -11.60% | 0.0 |
| 1 | Δ Bond length (A) | -0.123 | -0.123 | -0.117 | -4.88% | 0.0 | -0.414 | 236.01% | 9.0 |
|  | Cohesive energy (eV/at) | **-1.200** | **-1.200** | -1.20 | -0.08% | 0.0 | -2.59 | 115.83% | 0.0 |
|  | Δ Cohesive energy (eV/at) | 2.350 | 2.350 | 2.366 | 0.70% | 0.0 | 0.975 | -58.49% | 8.1 |
|  | Total |  |  |  | 0% | 0.0 |  | 23% | 17.1 |
| Graph. | Bond length (A) | **2.502** | **2.502** | 2.46 | -1.76% | 0.0 | 2.517 | 0.61% | 0.0 |
| 3 | Δ Bond length (A) | -0.121 | -0.121 | -0.100 | -17.16% | 0.1 | -0.107 | -11.88% | 0.1 |
|  | Cohesive energy (eV/at) | **-3.040** | **-3.040** | -3.449 | 13.46% | 0.0 | -3.529 | 16.08% | 0.0 |
|  | Δ Cohesive energy (eV/at) | 0.510 | 0.510 | 0.116 | -77.20% | 25.8 | 0.036 | -92.84% | 37.0 |
|  | Total |  |  |  | 8% | 25.9 |  | 50% | 37.1 |
| ZnS | Bond length (A) | 2.623 | 2.623 | 2.558 | -2.47% | 116.9 | 2.624 | 0.03% | 0.0 |
| 4 | Lattice parameter (A) | 6.058 | 6.058 | 5.908 | -2.47% | 0.0 | 6.060 | 0.03% | 0.0 |
|  | Cohesive energy (eV/at) | **-3.550** | **-3.550** | -3.565 | 0.43% | 0.0 | -3.565 | 0.43% | 0.0 |
|  | T m (K) | 1215 | 1215 | 1200 | -1.23% | 0.0 | 1100 | -9.47% | 0.2 |
|  | C 11 (GPa) | 83.3 | 83.3 | 68.0 | -18.31% | 3.1 | 83.5 | 0.23% | 0.0 |
|  | C 12 (GPa) | 45.3 | 45.3 | 68.0 | 50.22% | 16.8 | 45.2 | -0.19% | 0.0 |
|  | ν | 0.352 | 0.352 | 0.500 | 41.94% | 15.0 | 0.351 | -0.28% | 0.0 |
|  | E (001) (GPa) | 51.4 | 51.4 | 0.0 | -100% | 50.0 | 51.7 | 0.67% | 0.0 |
|  | E (110) (GPa) | 79.3 | 79.3 | 0.0 | -100% | 50.0 | 79.4 | 0.16% | 0.0 |
|  | anisotropy ratio | 0.48 | 0.48 | 5.00 | 942.11% | 95.7 | 0.48 | 0.95% | 0.0 |
|  | C 44 (GPa) | 39.6 | 39.6 | 0.00001 | -100% | 50.0 | 39.5185 | -0.21% | 0.0 |
|  | C 44 0 (GPa) | **49.5** | **49.5** | 68.05 | 37% | 0.0 | 42.6 | -13.94% | 0.0 |
|  | ζ | **0.65** | **0.65** | 0.99 | 52.31% | 0.0 | 0.65 | 0.00% | 0.0 |
|  | Cauchy ratio | 1.14 | 1.14 | 6804950 | 594869701% | 25.0 | 1.14 | 0.01% | 0.0 |
|  | Born ratio | 1.14 | 1.14 | 1.00 | -11.96% | 0.4 | 1.13 | -0.70% | 0.0 |
|  | B (GPa) | 58.0 | 58.0 | 68.0 | 17.39% | 2.9 | 58.0 | 0.01% | 0.0 |
|  | B' | **4.5** | **4.5** | 4.58 | 0.74% | 0.0 | 4.52 | -0.74% | 0.0 |
|  | *sub-total* |  |  |  | *50%* | *167.9* |  | *0%* | *0.2* |
|  | Total |  |  |  | 85.2% | 284.8 |  | 0.6% | 0.5 |
| NaCl | Bond length (A) | 2.812 | 2.812 | 2.827 | 0.54% | 0.0 | 2.883 | 2.53% | 0.0 |
| 6 | Δ Bond length (A) | 0.189 | 0.189 | 0.269 | 42.32% | 0.7 | 0.259 | 37.31% | 0.6 |
|  | Cohesive energy (eV/at) | -3.280 | -3.280 | -2.711 | -17.36% | 0.0 | -2.728 | -16.82% | 0.0 |
|  | Δ Cohesive energy (eV/at) | 0.270 | 0.270 | 0.855 | 216.59% | 18.8 | 0.837 | 210.08% | 17.7 |
|  | B (GPa) | **67.1** | **67.1** | 72.1 | 7.39% | 0.0 | 62.2 | -7.39% | 0.0 |
|  | B' | **4.9** | **4.9** | 4.96 | 0.98% | 0.0 | 4.86 | -0.98% | 0.0 |
|  | Total |  |  |  | 5.8% | 19.5 |  | 24.7% | 18.2 |
| CsCl | Bond length (A) | 3.029 | 3.029 | 3.022 | -0.25% | 0.0 | 3.016 | -0.43% | 0.0 |
| 8 | Δ Bond length (A) | 0.406 | 0.406 | 0.463 | 14.13% | 0.1 | 0.392 | -3.41% | 0.0 |
|  | Cohesive energy (eV/at) | -2.880 | -2.880 | -2.209 | -23.31% | 0.0 | -2.576 | -10.56% | 0.0 |
|  | Δ Cohesive energy (eV/at) | 0.670 | 0.670 | 1.357 | 102.48% | 4.2 | 0.989 | 47.68% | 0.9 |
|  | B (GPa) | **72.1** | **72.1** | 71.4 | -1.03% | 0.0 | 72.9 | 1.03% | 0.0 |
|  | B' | **5.1** | **5.1** | 5.23 | 1.85% | 0.0 | 5.04 | -1.85% | 0.0 |
|  | Total |  |  |  | 1.3% | 4.3 |  | 1.2% | 0.9 |
|  | **GRAND TOTAL** |  |  |  |  | **334** |  |  | **74** |

Table A.4: Results for InAs

Elastic properties are the main problem of all potentials (surfaces may be a bigger issue but very little data are available). Smith (GaAs) and Ashu (InAs) have almost no angular dependence of the energy ($c^2 \ll d^2$). This implies that the energy of an atom with four nearest-neighbors at a given distance will be the same whether these four atoms are on ZnS lattice sites or not, many degenerate states topologically close to ZnS have the same energy. Under some conditions (for instance shear or tension) the crystal will be almost infinitely soft, with E = 0 and $c_{44}$ = 0. In these two cases the bulk modulus B = $(c_{11} + 2 c_{12})/3$ is correct but $c_{11}$ = $c_{12}$, which leads to a Young's modulus (proportional to $c_{11} - c_{12}$) of 0. Clearly the potential cannot be fitted to one elastic constant only. B is simple to calculate but it is hydrostatic and therefore it totally ignores the stiffness coming from the angles. The use of B alone as a representative of elastic properties may be why Smith and Ashu did not notice the problem their potentials had with the angular dependence. Sayed *et al.*, Albe *et al.* (GaAs) and Nordlund *et al.* (InAs) have reasonable results for $c_{11}$ and $c_{12}$. The GaAs potential from Albe and coworkers has a low $c_{44}$ (40 GPa instead of 60 GPa) because $c_{44}^0$ was used in the fitting process instead of $c_{44}$. $c_{44}$ was also an issue in Tersoff 2; $c_{44}$ and surfaces seem to be properties empirical potentials do not reproduce well [Balamane *et al.* 1992].

*c) Dimers*

Experiments [Lemire 1990] and *ab initio* calculations [see for instance Al-Laham and Raghavachari 1991, Song *et al.* 1994] agree: the GaAs dimer is longer (2.53 Å) than the GaAs bond length in the ZnS structure (2.44 Å). Figure A.1(a) shows the bond length in several structures over the ZnS (or DC) bond length for Si, GaAs, InAs and AlAs. The only problem comes from the GaAs and Si dimers which have an opposite behavior. Figure A.1(b) shows that the normalized energy does not seem to depend on the material system either. There is a discrepancy for AlAs for highly coordinated structures. Whether this is due to deficiency in the *ab initio* results or whether this is a real phenomenon is unclear.

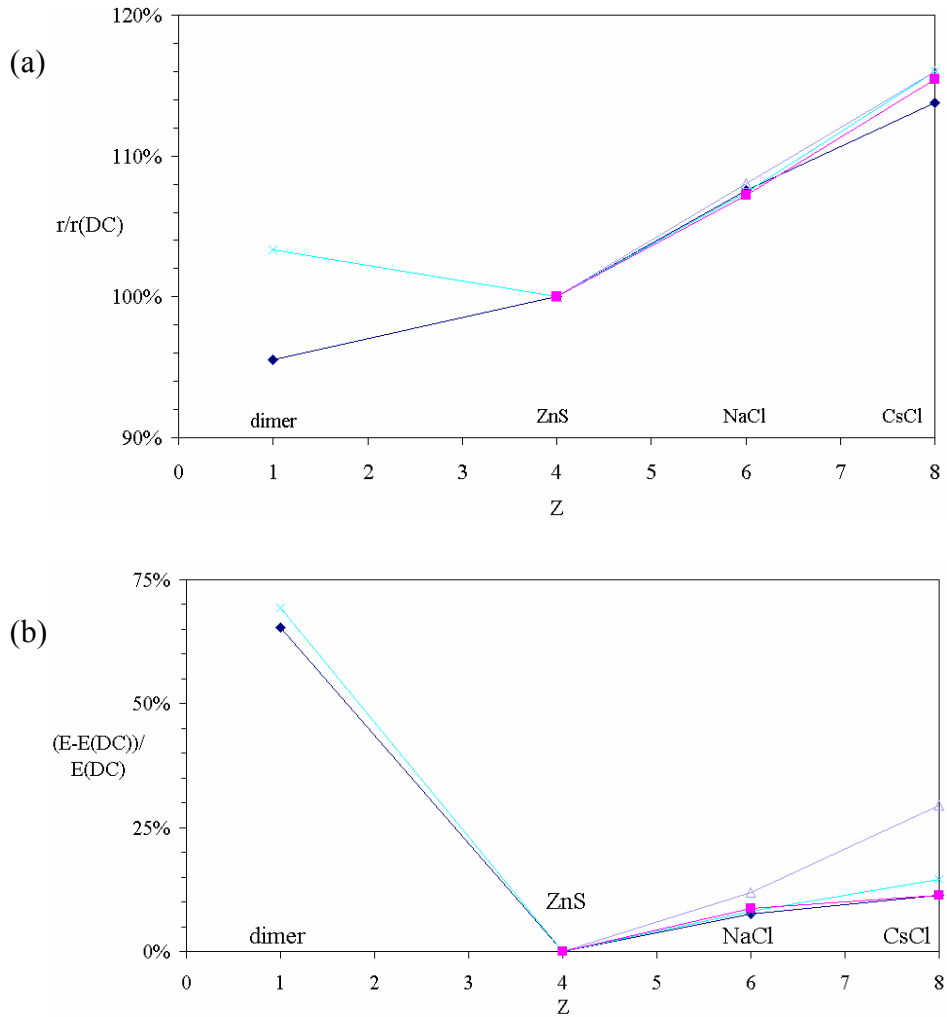

FIG. A.1: Normalized bond length (a) and cohesive energy (b) for Si and III-V. Diamonds: silicon, crosses: GaAs, open triangles: AlAs, squares: InAs.

The silicon dimer is shorter than the bond length in the diamond cubic structure (see Fig. A.1) and no data could be found for C or Ge nor for InAs or AlAs. It is therefore not possible to tell whether there is a global difference between groups IV and III-V or if this applies only to the particular case of Si and GaAs. All potentials give a dimer shorter than the bulk (2.35 Å *vs.* 2.44 Å for GaAs) but they generally get the right energy anyway, except for Nordlund's InAs which describes the dimer very poorly (the dimer properties were not used in the fitting process). This comes from the analytical form given by Tersoff itself: it cannot find a dimer longer than the bulk bond length whatever parameters are used. Keep in mind that this potential was developed for Si for which this is not a problem.

*d) Surfaces*

Isolated dimers are not very interesting *per se* for the study of thin film growth. Unlike isolated dimers, dimers on surfaces are three-fold coordinated which may make a big difference as the coordination number is taken into account in the potential. Hence surface dimers may be correctly described even if isolated ones are not.

Only Albe's potential for GaAs has been tested for surfaces [Albe *et al.* 2002]. Testing surfaces requires a MD simulation or a conjugate gradient as the number of degrees of freedom (relaxation of several atoms) is too high to allow for an analytical treatment. It was found that As-rich structures, $\gamma(2x4)$ and $c(4x4)$, were poorly described mainly because the arsenic dimers were too long or even not stable at all. Albe and coworkers's interest is in implantation damage, where melting and antisite defects are more important than in thin film growth. They did not want to change the As potential too much to conserve the properties of pure As [Nordlund, private communication].

Tersoff modified his potential (nicknamed Tersoff 2 or T2) because its angular dependence was too weak and $c_{44}$ was too low by an order of magnitude [Tersoff 1988a]. The new potential (nicknamed Tersoff 3 or T3) was better at describing elasticity but had relatively poor results on the surface [Tersoff 1988b]. Using the criteria we used (nothing concerning surfaces) T3 would have been evaluated as much better than T2. It is possible that the Sayed potential gives better results on surfaces than Albe even if the latter looks better for the criteria used here.

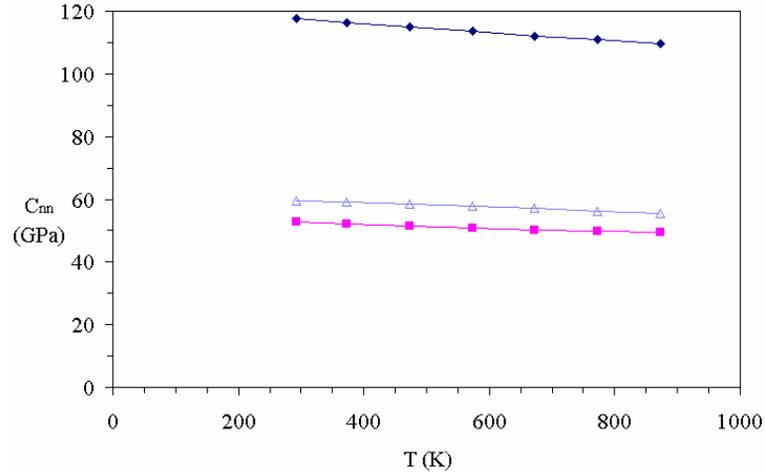

FIG. A.2: Change in the elastic constants with temperature. Diamonds: $c_{11}$, squares: $c_{12}$ and open triangles: $c_{44}$.

*e) Melting point and high temperature behavior*

$T_m$ is very well described for InAs (Ashu gives 1200 K and Nordlund 1100 K both close to the experimental value of 1215 K). For GaAs, $T_m$ ranges from 1050 K for Sayed to 1900 K for Albe (experimental: 1513 K).

All properties presented in this chapter were calculated at 0 K. Therefore the melting temperature may matter because the fact that elasticity is well described at 0 K does not imply that it is well described at growth temperatures. Figure A.2 shows the elastic constants $c_{11}$, $c_{12}$ and $c_{44}$ as a function of temperature [Burenkov 1973] along with trend lines. For instance a growth temperature of 800 K would be .75 $T_m$ according to Sayed but only .4 $T_m$ for Albe. However it is hard to get quantitative results at high temperature without MD simulations, and the implementation of the potential to run MD simulations is beyond the scope of this work. The elastic properties change linearly with temperature (see Fig. A.2) but in the case of GaAs the (experimental) slope is small so the (real) elastic properties at 800 K are not much different from those at 0 K. However it is not possible to know if the slope will be small for the potentials too.

*f) Summary*

Two sets of potential parameters [Nakamura *et al.* 2000, Murugan and Ramachandran 1999] do not get the right bond length and cohesive energy for the ZnS structure. Therefore they were not included in our study. In the 6 remaining potentials, Smith and Ashu committed the same error: c is much smaller than d. This leads to almost no angular dependence and $c_{44}$ and Young's modulus become 0. This makes them useless. For GaAs, Sayed gives the best results for elasticity against Albe as the latter has a weak $c_{44}$. They give the same results for dimers but Sayed does not describe other structures very well, the "graphitic" structure for instance in very close in energy to the ZnS ground state. No comparison being possible for surfaces it is not really possible to decide which potential is best. For InAs, once Ashu is removed only one potential (Nordlund) remains. There are two "useable" potentials for GaAs and one only for InAs and AlAs.

One remaining issue is to determine how relevant the Sayed potentials for GaAs and AlAs and Nordlund potential for InAs are for surfaces as results for surfaces are only available for Albe potential (GaAs). Tersoff potentials for silicon did not reproduce both elasticity and surfaces well: T2 was good at surfaces but the bulk was too soft and for T3 the elasticity was correct but not surfaces. For surfaces to be reproduced correctly, bonds need to be soft which correlates to a bad description of bulk elasticity. It is quite possible that a potential thought to be rather good when judged on results for elasticity may describe surfaces poorly.

Another issue is that in the bulk there are only III-V bonds while on surfaces there are also III-III and V-V bonds. As surfaces do not take only the III-V parameters into account, they also need pure III and pure V parameters. This implies that surfaces could be improved by modifying the pure III and/or pure V parameters without changing the III-V bonds. Albe and coworkers claim that their parametrization for pure gallium and pure arsenic give the right ground states for pure materials (which are rather exotic structures) but their description of surfaces is not very good. Surfaces could be improved by changing the Ga and As potentials but this may lead to a less accurate description of the pure phases. To people concerned with crystal growth (as opposed to Albe, Nordlund and their coworkers who study ion implantation), surface reconstructions should be more important than pure Ga or As.

**3) Ionicity of III-V semiconductors**

*a) Why ionicity in a covalent material?*

III-V semiconductors are mostly covalent but not exclusively: their bonds are partly ionic. Even if in the bulk perfect crystal the ionicity is not very important (which remains to be checked), it may matter for surfaces and defects, when there can be III-III or V-V bonds. The potentials studied in the previous section do not take into account the ionic character of III-V semiconductors. We only found one potential taking this into account [Nakamura *et al.* 2000] but it is not fully developed. Indeed they used the parameters from Smith, Sayed and Ashu and added ionicity to that. This is problematic since when ionicity is added, the bond length and cohesive energy will be wrong because the contribution of ionicity will be double-counted in the fit to the total energy and in the fit to the Coulombic energy. Also it is not clearly stated how their parameters were obtained. Some of them were directly taken from Smith, Sayed or Ashu but this is not explicitly acknowledged. Sometimes they used some of the parameters published by others but not all of the parameters.

*b) Implementation of ionicity: the problem of long-range interactions*

Coulombic interactions die off as one over the distance, whereas covalent bonds are essentially local. A consequence is that the force on an atom has contributions from very remote atoms. Let a spherical shell of radius r and thickness δr. Its volume is $4 \pi r^2 \delta r$, hence the number of atoms in this shell scales as $r^2$. Each of these atoms has a contribution to the ionic energy scaling as 1/r. Hence the total contribution of the shell scales as r. This implies that the further away the shell, the larger its contribution. This is the opposite of covalent bonds for which only nearby atoms are bonded.

In a simulation with periodic boundaries, the simulation cell is repeated infinitely. Therefore the system being simulated is infinite. Since the further away the shell is the larger its contribution, there cannot be a cut-off, i.e. it is not possible to take only atoms within some radius into account. All atoms must contribute to the energy of a given atoms, and calculating the energy of a given atom may take an infinite amount of time. Schemes have been designed to make the implementation of long-range interactions more efficient, the most famous of them is known as the Ewald sphere [Ewald 1921]. The ionicity contribution comes on top of the covalent bonding, which increases the computational cost of the simulation. Therefore ionicity should be included only if it appears necessary to describe the physics.

*c) Change in bond length due to ionicity*

The silicon dimer is shorter than the bond length of the bulk diamond cubic structure (2.24 Å *vs.* 2.35 Å) but it is the opposite for GaAs (2.53 Å *vs.* 2.44 Å) (see also the paragraph "dimers" in the previous section). A possible explanation for such a difference between Si and GaAs is that GaAs is partly ionic and Si is purely covalent.

The total internal energy, U, of an atom is given by

$$U(r) = U_0(r) + U_c(r) \tag{A.10}$$

where $U_0$ is the covalent part of the energy and $U_c$ the coulombic part:

$$U_c(r) = -\frac{\alpha q^2}{4\pi\varepsilon_0 r} = -\frac{Q}{r} \tag{A.11}$$

where $\alpha$ is the Madelung constant (see Table A.5). At equilibrium,

$$\frac{\partial U}{\partial r} = \frac{\partial U_0}{\partial r} + \frac{Q}{r^2} = 0. \tag{A.12}$$

Let $r_0$ the value of the bond length at the minimum of the covalent energy (i.e. the bond length the crystal would adopt if it were purely covalent). We can write that for a mostly covalent material,

$$r = r_0 + \delta r \tag{A.13}$$

with $\delta r \ll r_0$. Expanding the derivative of U at $r_0$ we get

$$\frac{\partial U}{\partial r} \approx \left.\frac{\partial U}{\partial r}\right)_{r_0} + \left.\frac{\partial^2 U}{\partial r^2}\right)_{r_0} \delta r = 0 + \frac{Q}{r_0^2} + \left.\frac{\partial^2 U}{\partial r^2}\right)_{r_0} \delta r. \tag{A.14}$$

Expression (A.14) is equal to 0 at equilibrium, leading to

$$\delta r = \frac{U_c(r_0)}{r_0 \left.\dfrac{\partial^2 U}{\partial r^2}\right)_{r_0}} \tag{A.15}$$

| structure | dimer | ZnS | NaCl | CsCl |
|---|---|---|---|---|
| Madelung constant | 1 | 1.638 | 1.7476 | 1.763 |
| k | — | $\frac{8}{9}\sqrt{3} \approx 1.54$ | 1 | $\frac{4}{9}\sqrt{3} \approx 0.77$ |
| $B_{GaAs}$ (GPa) | — | 76.5 | 90* | 100* |

Table A.5: Madelung constant, geometric constant k and bulk modulus B for various structures. *: from Albe *et al.* 2002

The atomic volume of a crystal $\Omega = kr^3$ where k is a geometric constant which depends on the structure only (see Table A.5 for numerical values). Hence $d\Omega^2 = 9k^2 r^4 dr^2$. From the definition of the bulk modulus,

$$B = \Omega \frac{d^2U}{d\Omega^2} = \frac{1}{9kr} \frac{d^2U}{dr^2}. \tag{A.16}$$

Inserting this in equation (A.15) leads to

$$\delta r = \frac{U_c(r_0)}{9k r_0^2 B}. \tag{A.17}$$

Equation (A.17) shows that δr is always negative ($U_c$ is negative) and that its magnitude is larger for softer materials or structures. Therefore we expect the dimer to contract more than the bulk crystals.

*d) Summary*

The partial charge on Ga and As atoms in $Ga^{+q} As^{-q}$ is q = 0.2 e or q = 0.4 e where e is the charge of the electron. The literature [Blakemore 1982 and references therein] is not clear about this: it is hard to know if 0.4 e is the charge or the difference of charges (q = 0.4 e or 2q = 0.4 e). For GaAs (assuming a charge of 0.2 e for both ZnS and dimers), the dimer would be shortened by less than 0.02 Å. Thus ionicity cannot explain why the dimer is longer than the bond length of the bulk ZnS structure (2.53 Å *vs.* 2.44 Å).

It is worth noticing that ionicity may matter for the electron-counting rule i.e. for reconstructions. However surfaces are generally not well described even for silicon [Balamane *et al.* 1992] for which there is no ionicity at all (this seems to be due to the inaccurate angular dependence).

The issue of ionicity is two-fold: is it important and is it complicated? Only the first question was dealt with above. The second question is related to techniques to implement coulombic interactions, the most famous being the Ewald summation [Ewald 1921]. The Ewald summation is time consuming, so nobody would want to implement it if ionicity is not really needed. However one big problem remains: no potential is fitted taking ionicity into account.

## 4) Summary

For GaAs, only two potentials have fair values for bond length, cohesive energy and elasticity in the ZnS structure. Sayed gives the best results for elasticity against Albe as the latter has too weak a value for $c_{44}$. No comparison being possible for surfaces it is not really possible to decide which potential is best. For InAs and AlAs, only Nordlund's potential gives good enough results. There are two "useable" potentials for GaAs and one only for InAs and AlAs.

There has not been any decent attempt at incorporating coulombic interactions into the potential. It is not clear whether ionicity would be an improvement worth the computational cost of long-range interactions.